\documentclass[lettersize,journal]{IEEEtran}
\usepackage{amsmath,amsfonts,acronym,amssymb,amsthm,svg,dsfont,multicol,csquotes,cite,bbding}

\renewcommand{\P}[2][]{{\mathbb{P}_{#1}}{\left(#2\right)}} 
\newcommand{\E}[2][]{{\mathbb{E}_{#1}}{\left(#2\right)}} 

\newcommand{\abs}[1]{\ensuremath{\left|#1\right|}} 
\newcommand{\norm}[2][]{\ensuremath{{\left\Vert{#2}\right\Vert}_{#1}}} 
\newcommand{\eqdef}{\ensuremath{\triangleq}} 
\newcommand{\set}[1]{\ensuremath{\left\{{#1}\right\}}} 
\newcommand{\indic}[1]{\ensuremath{\mathds{1}\!\left\{#1\right\}}} 
\renewcommand{\leq}{\leqslant} 
\renewcommand{\geq}{\geqslant} 
\newcommand{\argmin}{\mathop{\text{argmin}}} 
\newcommand{\argmax}{\mathop{\text{argmax}}} 

\newcommand{\gcheck}{\textcolor{green}{\checkmark}} 
\newcommand{\rxmark}{\textcolor{red}{\XSolidBrush}}

\DeclareMathAlphabet{\eurm}{U}{eur}{m}{n}
\DeclareMathAlphabet{\mathbsf}{OT1}{cmss}{bx}{n}
\DeclareMathAlphabet{\mathssf}{OT1}{cmss}{m}{sl}
\DeclareMathAlphabet{\mathcsf}{OT1}{cmss}{sbc}{n}

\DeclareSymbolFont{bsfletters}{OT1}{cmss}{bx}{n}  
\DeclareSymbolFont{ssfletters}{OT1}{cmss}{m}{n}
\DeclareMathSymbol{\bsfGamma}{0}{bsfletters}{'000}
\DeclareMathSymbol{\ssfGamma}{0}{ssfletters}{'000}
\DeclareMathSymbol{\bsfDelta}{0}{bsfletters}{'001}
\DeclareMathSymbol{\ssfDelta}{0}{ssfletters}{'001}
\DeclareMathSymbol{\bsfTheta}{0}{bsfletters}{'002}
\DeclareMathSymbol{\ssfTheta}{0}{ssfletters}{'002}
\DeclareMathSymbol{\bsfLambda}{0}{bsfletters}{'003}
\DeclareMathSymbol{\ssfLambda}{0}{ssfletters}{'003}
\DeclareMathSymbol{\bsfXi}{0}{bsfletters}{'004}
\DeclareMathSymbol{\ssfXi}{0}{ssfletters}{'004}
\DeclareMathSymbol{\bsfPi}{0}{bsfletters}{'005}
\DeclareMathSymbol{\ssfPi}{0}{ssfletters}{'005}
\DeclareMathSymbol{\bsfSigma}{0}{bsfletters}{'006}
\DeclareMathSymbol{\ssfSigma}{0}{ssfletters}{'006}
\DeclareMathSymbol{\bsfUpsilon}{0}{bsfletters}{'007}
\DeclareMathSymbol{\ssfUpsilon}{0}{ssfletters}{'007}
\DeclareMathSymbol{\bsfPhi}{0}{bsfletters}{'010}
\DeclareMathSymbol{\ssfPhi}{0}{ssfletters}{'010}
\DeclareMathSymbol{\bsfPsi}{0}{bsfletters}{'011}
\DeclareMathSymbol{\ssfPsi}{0}{ssfletters}{'011}
\DeclareMathSymbol{\bsfOmega}{0}{bsfletters}{'012}
\DeclareMathSymbol{\ssfOmega}{0}{ssfletters}{'012}

\newcommand{\calF}{{\mathcal{F}}}

\newcommand{\calI}{{\mathcal{I}}}

\newcommand{\calR}{{\mathcal{R}}}

\acrodef{3GPP}{3rd Generation Partnership Project}
\acrodef{ABP}{Auxiliary Beam Pair}
\acrodef{ABS}{Adaptive Beam Search}
\acrodef{ABT}{Active Beam Tracking}
\acrodef{AEP}{Asymptotic Equipartition Property}
\acrodef{AGC}{Automatic Gain Control}
\acrodef{AL}{Agile Link}
\acrodef{AoA}{Angle of Arrival}
\acrodef{AoD}{Angle of Departure}
\acrodef{AWGN}{Additive White Gaussian Noise}
\acrodef{AVC}[AVC]{Arbitrarily Varying Channel}
\acrodef{BAI}{Best Arm Identification}
\acrodef{BER}{Bit-Error-Rate}
\acrodef{BEC}{Binary Erasure Channel}
\acrodef{BLER}{Block-Error Rate}
\acrodef{BPSK}{Binary Phase-Shift Keying}
\acrodef{BSC}{Binary Symmetric Channel}
\acrodef{BICM}[BICM]{Bit-Interleaved Coded-Modulation}
\acrodef{BS}{Base Station}
\acrodef{CDF}[CDF]{Cumulative Distribution Function}
\acrodef{CDL}{Clustered Delay Line}
\acrodef{CGF}[CGF]{Cumulant Generating Function}
\acrodef{CLT}[CLT]{Central Limit Theorem}
\acrodef{CS}{Compressed Sensing}
\acrodef{CSI}[CSI]{Channel State Information}
\acrodef{DBZ}{Dynamic Beam Zooming}
\acrodef{DMC}[DMC]{Discrete Memoryless Channel}
\acrodef{DMS}[DMS]{Discrete Memoryless Source}
\acrodef{DNN}{Deep Neural Network}
\acrodef{DoF}{Degrees of Freedom}
\acrodef{DWNA}{Discrete White Noise Acceleration}
\acrodef{EKF}{Extended Kalman Filter}
\acrodef{FC}{Fixed Confidence}
\acrodef{FER}[FER]{Frame Error Rate}
\acrodef{FLOP}{Floating-Point Operation}
\acrodef{gNB}{Next Generation Node B}
\acrodef{GPU}{Graphics Processing Unit}
\acrodef{HAD}{Hybrid Analog-Digital}
\acrodef{HBA}{Hierarchical Beam Alignment}
\acrodef{HOO}{Hierarchical Optimistic Optimization}
\acrodef{HOSUB}{Hierarchical Optimal Sampling of Unimodal Bandits}
\acrodef{HPBW}{Half-Power Beamwidth}
\acrodef{HPM}{Hierarchical Posterior Matching}
\acrodef{HUCB}{Hierarchical Upper-Confidence Bound}
\acrodef{IA}{Initial Alignment}
\acrodef{iid}[i.i.d.]{independent and identically distributed}
\acrodef{IMT-BR}{Integrated Monopulse Tracking with Beam Recovery}
\acrodef{IRS}{Intelligent Reflective Surface}
\acrodef{ISAC}{Integrated Sensing and Communication}
\acrodef{IoT}[IoT]{Internet of Things}
\acrodef{KL}{Kullback-Leibler}
\acrodef{KF}{Kalman Filter}
\acrodef{LCB}{Lower-Confidence Bound}
\acrodef{LPD}[LPD]{Low Probability of Detection}
\acrodef{LDPC}[LDPC]{Low-Density Parity-Check}
\acrodef{LTE}{Long Term Evolution}
\acrodef{LOS}{Line of Sight}
\acrodef{LSTM}{Long Short Term Memory}
\acrodef{LUCB}{Lower-Upper Confidence Bound}
\acrodef{mmWave}{millimeter-Wave}
\acrodef{MAB}{Multi-Armed Bandit}
\acrodef{MAC}[MAC]{multiple-access channel}
\acrodef{MCTS}{Monte-Carlo Tree Search}
\acrodef{ME}{Mobile Entity}
\acrodef{MGF}[MGF]{Moment Generating Function}
\acrodef{MIMO}[MIMO]{Multiple-Input Multiple-Output}
\acrodef{MISO}{Multiple-Input Single-Output}
\acrodef{ML}{Machine Learning}
\acrodef{MLC}[MLC]{Multi-Level Coding}
\acrodef{MLS}[MLS]{Multi-Level Sampling}
\acrodef{MSD}[MSD]{Multi-Stage Decoding}
\acrodef{NR}{New Radio}
\acrodef{NLOS}{Non-LOS}
\acrodef{OFDM}{Orthogonal Frequency Division Multiplexing}
\acrodef{OSUB}{Optimal Sampling for Unimodal Bandits}
\acrodef{PAC}{probably approximately correct}
\acrodef{PDF}[PDF]{Probability Density Function}
\acrodef{PE}{Pure Exploration}
\acrodef{PF}{Particle Filter}
\acrodef{PMF}[PMF]{Probability Mass Function}
\acrodef{PPM}[PPM]{Pulse Position Modulation}
\acrodef{PSD}{Power Spectral Density}
\acrodef{PSK}{Phase Shift Keying}
\acrodef{QKD}{Quantum Key Distribution}
\acrodef{RACH}{Random Access Channel}
\acrodef{RE}{Resource Elements}
\acrodef{RF}{Radio Frequency}
\acrodef{RL}{Reinforcement Learning}
\acrodef{ROC}{Receiver Operating Characteristic}
\acrodef{RS}{Reference Signal}
\acrodef{RSRP}{Reference Signal Received Power}
\acrodef{QoS}{Quality of Service}
\acrodef{QPSK}{Quadrature Phase-Shift Keying}
\acrodef{QuaDRiGa}{Quasi Deterministic Radio Channel Generator}
\acrodef{RNG}{Random Number Generator}
\acrodef{RSS}{Received Signal Strength}
\acrodef{RV}{random variable}
\acrodef{SCMA}{Sparse Code Multiple Access}
\acrodef{SDR}{Semi-Definite Relaxation}
\acrodef{SIMO}{Single-Input Multiple-Output}
\acrodef{SINR}{Signal to Interference plus Noise Ratio}
\acrodef{SNR}{Signal-to-Noise Ratio}
\acrodef{SS}{Synchronization Signal}
\acrodef{SSB}{Synchronization Signal Block}
\acrodef{SSE}{Successive Subtree Elimination}
\acrodef{TDD}{Time-Division Duplexing}
\acrodef{TDM}{Time-Division Multiplexing}
\acrodef{TPCP}{Trace-Preserving Completely-Positive}
\acrodef{2PHTS}{Two Phase Heteroscedastic Track-and-Stop}
\acrodef{TS}{Thompson Sampling}
\acrodef{TAS}{Track-and-Stop}
\acrodef{UAV}{Unmanned Aerial Vehicle}
\acrodef{UCB}{Upper-Confidence Bound}
\acrodef{UCB-V}{Upper-Confidence Bound - Variance}
\acrodef{UBA}{Unimodal Beam Alignment}
\acrodef{UCA}{Uniform Circular Array}
\acrodef{ULA}{Uniform Linear Array}
\acrodef{UPA}{Uniform Planar Array}
\acrodef{UE}{User Equipment}
\acrodef{WINNER}{Wireless World Initiative for New Radio}
\acrodef{wrt}[w.r.t.]{with respect to}
\acrodef{WSS}{Wide Sense Stationary}

\usepackage{array,svg}

\usepackage{textcomp}
\usepackage{stfloats}
\usepackage{hyperref}
\usepackage[only,llbracket,rrbracket]{stmaryrd}

\usepackage{xr}
\IEEEpubid{This work has been submitted to the IEEE for possible publication.    Copyright may be transferred without notice, after which this version may no longer be accessible.}

\hypersetup{
 unicode=false, 
 pdftoolbar=true, 
 pdfmenubar=true, 
 pdffitwindow=true, 
 pdftitle={\@title}, 
 pdfauthor={\@author}, 
 pdfsubject={}, 
 pdfnewwindow=true, 
 pdfkeywords={}, 
 colorlinks=true, 
 linkcolor={darkblue}, 
 citecolor={darkblue}, 
 urlcolor={darkblue}, 
 anchorcolor={black} 
}

\usepackage{verbatim}
\usepackage{graphicx}
\usepackage[ruled,linesnumbered]{algorithm2e}
\usepackage{multicol,multirow,tabularx}
\usepackage{graphicx,float}
\usepackage{subfig}
\usepackage{mathtools}

\DeclarePairedDelimiter\ceil{\lceil}{\rceil}

\DeclareMathOperator\erf{erf}

\theoremstyle{plain}
\newtheorem{lemma}{Lemma}

\newtheorem{assumption}{Assumption}

\newcommand{\mrb}[1]{\textcolor{black}{#1}}

\newcommand{\nbbox}[1]{\hspace{2pt}\raisebox{0.75pt}{\boxed{\mbox{\tiny{#1}}}}}

\begin{document}

\definecolor{darkgray}{rgb}{0.66, 0.66, 0.66}
\definecolor{darkblue}{rgb}{0.26,0.5,0.70}

\title{\acl{MAB} Dynamic Beam Zooming\\ for mmWave Alignment and Tracking}

\author{Nathan Blinn,~\IEEEmembership{Student Member,~IEEE} and Matthieu Bloch,~\IEEEmembership{Senior Member,~IEEE}\\
Emails: nblinn6@gatech.edu and matthieu.bloch@coe.gatech.edu


}



\maketitle

\begin{abstract}

We propose an \ac{ISAC} algorithm that exploits the structure of a hierarchical codebook of beamforming vectors using a best-arm identification \ac{MAB} approach for initial alignment and tracking of a \ac{ME}. The algorithm, called \ac{DBZ}, performs beam adjustments that mitigate the severe outages associated with wireless mmWave systems and allow for adaptive control of the parameters governing communications.  
We analyze the sample complexity of \ac{DBZ} and use it to inform how the algorithm adapts to the non-stationary \ac{MAB} statistics based on \ac{ME} motion and \ac{SNR}. We perform extensive simulations to validate the approach and demonstrate that \ac{DBZ} is competitive against existing Bayesian algorithms, without requiring channel multi-path or fading knowledge. In particular, \ac{DBZ} outperforms other low-complexity algorithms in the low \ac{SNR} regime. We also illustrate the efficacy of \ac{DBZ} in standardized rural and urban scenarios using NYU Sim.

\end{abstract}

\begin{IEEEkeywords}
5G, 6G, millimeter-Wave, MIMO, Beamforming, Multi-Armed Bandits.
\end{IEEEkeywords}

\section{Introduction}\label{sec:intro}
Communication in the  \ac{mmWave} spectrum (30 GHz to 300 GHz) is envisioned as a key enabler of ultra-high-speed data delivery with low latency for next generation wireless systems~\cite{cui2021integrating}. %
The severe path loss inherently associated with \ac{mmWave} frequencies, however, creates unique engineering challenges. First, compensating for the path loss requires transceivers to combine massive \ac{MIMO} arrays to form highly focused, narrow beams~\cite{rappaport2015wideband} together with \ac{HAD} architectures to reduce the otherwise impractical number of associated \ac{RF} paths~\cite{alkhateeb2014channel}. Second, ensuring persistent and reliable communication between entities requires efficient beam refinement and management to initiate alignment and track \ac{ME} movement over time~\cite{wang2022beam}. Beam alignment and tracking can be viewed as \emph{sensing} tasks, leveraging approaches in radar \cite{soumya2023recent}, that support a \emph{communication} task. While the two tasks could be independently addressed, joint designs within the framework of \ac{ISAC} offer opportunities for enabling emerging applications~\cite{liu2022integrated} and efficiently utilize increasingly congested wireless resources and constrained hardware \cite{wei2023integrated}. 

\subsection{Related Works}\label{ssec:relatedworks}
The 5G standard currently only offers basic support for \ac{mmWave} beam alignment and tracking in the form of an exhaustive search for beam directions~\cite[Section 4]{3GPP_TS_38213_R17}. Consequently, several classes of \ac{ISAC} algorithms for \ac{mmWave} beam alignment and tracking have been investigated, each offering different complexity-measurement-performance tradeoffs.  \mrb{The classes are summarized in Table \ref{tab:algcomparisonsT} and discussed next.} 

\begin{table}[H]
\centering
\caption{\mrb{Comparison of Algorithms with \ac{ISAC} Defining Features.}}
\label{tab:algcomparisonsT}
\begin{tabularx}{\columnwidth}{ >{\raggedright\arraybackslash\color{black}}p{0.11\columnwidth}
                               >{\centering\arraybackslash\color{black}}p{0.13\columnwidth}
                               >{\centering\arraybackslash\color{black}}p{0.11\columnwidth}
                               >{\centering\arraybackslash\color{black}}p{0.09\columnwidth}
                               >{\centering\arraybackslash\color{black}}p{0.09\columnwidth}
                               >{\centering\arraybackslash\color{black}}p{0.13\columnwidth} }
\hline
\textbf{Reference}          & \textbf{Alignment}  	&\textbf{Alignment} 		& \textbf{Motion}         & \textbf{CSI}     & \textbf{Computational}\\
                            & \textbf{Complexity}  	&\textbf{Accuracy} 		& \textbf{Adaptive}        & \textbf{Adaptive}  & \textbf{Overhead}\\
\hline\hline
KF \cite{va2016beam}  										&N/A				&\rxmark   	&\gcheck   	&\rxmark   	&Low \\ 
RL \cite{zhang2020beam}, \cite{chiang2021machine}    		    &High			& \rxmark   	&\gcheck 	& \gcheck  	&Low \\
ABP \cite{zhu2018high}							   			&High			& \rxmark  	&\gcheck  	& \gcheck   &Low \\
HPM \cite{chiu2019active}  									&Low				&\gcheck 	&\rxmark 	& \rxmark   &High \\
2PHTS \cite{wei2022fast}										&Low				&\gcheck		&\rxmark		& \gcheck	&High\\
HBA \cite{wu2019fast}										&Low				&\rxmark		&\rxmark		& \gcheck	&Low\\
HOSUB \cite{blinn2021mmwave}									&Fixed			&\rxmark		&\rxmark 	& \gcheck 	&Low \\
DL-IA \cite{sohrabi2021deep}									&Fixed			&\gcheck		&\rxmark 	& \rxmark 	&High* \\
DL \cite{lim2021deep}										&N/A				&\rxmark   	&\gcheck  	& \rxmark   &High* \\
PF \cite{chung2020adaptive} 									&N/A				&\rxmark		&\gcheck		& \rxmark	&High\\
ABT \cite{ronquillo2021active}    							&Low				& \gcheck  	&\gcheck  	& \rxmark 	&High\\
Present Work 												&Med-Low			& \gcheck  	&\gcheck  	& \gcheck   &Med-Low\\
\hline
\multicolumn{6}{l}{\textcolor{black}{\footnotesize * High computational overhead pending hardware implementation.}}
\end{tabularx}
\end{table}

A first class of algorithms uses Bayesian decision-making and leverages hierarchical beamforming codebooks \cite{xiao2016hierarchical}. 
In particular,~\cite{chiu2019active} proposes the \ac{HPM} algorithm that exploits \ac{CSI} and measurements to update the posterior probabilities of the incoming beam direction, choosing increasingly narrower beams as the posteriors identify more precisely the likely beam direction. The approach of \ac{HPM} has also been recently extended to track \ac{ME}s~\cite{ronquillo2023integrated}.

\IEEEpubidadjcol

A second class of algorithms selects beamformer weights using a \ac{DNN}~\cite{sohrabi2021deep} instead of relying on a predefined hierarchical codebook.  
Numerical results show that the performance of this ``codebook-free'' approach without \ac{CSI} matches the performance of \ac{HPM} with full \ac{CSI}. \ac{DNN} approaches, however, may take a significant number of samples to correctly point a beam.  For example, the deep reinforcement learning algorithm in~\cite{tandler2023deep} takes about $10^5$ samples to converge at runtime.  

\IEEEpubidadjcol

A third class of algorithms attempts to circumvent the computational complexity incurred by Bayesian and \ac{DNN} approaches using \ac{CS} techniques.  The idea is to exploit the sparsity associated with \ac{mmWave} channels \cite{rappaport2015wideband} to quickly identify the direction of incoming signal.  To infer user location, the approach in \cite{rasekh2017noncoherent,hassanieh2018fast} is to generate random peaks in multiple beam patterns to quickly identify the optimal combinations to form beamforming weights.  

Particularly relevant to the present work, a fourth class of  algorithms exploits the conceptual analogy between beam steering and arm play in a \acf{MAB} problem to lower complexity without sacrificing performance. In brief, every beam direction, which corresponds to a set of phase shifts applied to array elements, may be viewed as an arm to pull in a \ac{MAB} algorithm and the \ac{RSRP} acquired with every choice of beam direction provides a reward that may be exploited by a \ac{MAB} exploration strategy. 
Experimental results in \cite{hashemi2018efficient} show that the alignment of two users exhibits an approximate unimodal structure \cite{combes2014unimodal,yu2011unimodal} that can be efficiently exploited.  Few reward structures are perfectly unimodal and the algorithm in \cite{hashemi2018efficient,combes2014unimodal} may get stuck in local maxima. To address this, \cite{blinn2021mmwave} adapts \ac{OSUB} \cite{combes2014unimodal} for use with a hierarchical codebook; numerical simulations show a substantial reduction in the number of samples required, with robustness to multi-path effects but no theoretical guarantees.  The \ac{HBA} algorithm~\cite{wu2019fast} adapts the X-arm bandit algorithm~\cite{bubeck2011x}, but only indirectly exploits the hierarchical structure since pencil beams acquire reward signals.   The \ac{2PHTS} algorithm \cite{wei2022fast} uses grouped sums of arms as ``super-arms,'' which are broader beams, to create a two-level hierarchical beamformer and adopts the \ac{TAS} framework of \cite{garivier2016optimal,kaufmann2016complexity}.  With respect to track tasks, \ac{MAB} algorithms in the regret setting provide a low-overhead approach, where for instance \cite{zhang2020beam} chooses arms close to the empirical best to play at each round.   

A final class of algorithms uses tools from adaptive control for \ac{AoA} and/or \ac{AoD} estimation over time. \cite{va2016beam} estimates the fading coefficient along with the angles using an \ac{EKF} for a low computational overhead approach to motion compensation, but no initial alignment. Beamwidth control over time ultimately prevents large outages associated with \ac{mmWave} channels during instances of misalignment. For instance, \cite{chung2020adaptive,lim2019beam} both adapt \ac{PF}s to make dynamic adjustments to the beamwidth over time, at the expense of high computational overhead.

\subsection{Contributions and Outline}\label{ssec:contributions}

The main contributions of the present work are as follows:
\begin{itemize}
 \item We present \acf{DBZ}, an \ac{ISAC} algorithm for \ac{mmWave} beam alignment and tracking \mrb{that offers high alignment accuracy and automatic adaptation to changes in \ac{CSI} and motion while preserving a relatively low complexity. In particular, as summarized in Table~\ref{tab:algcomparisonsT},} \ac{DBZ} strikes competitive performance against Bayesian methods exploiting full channel knowledge and \ac{ME} motion such as \cite{ronquillo2021active,chiu2019active}. 
\item We show that \ac{DBZ} is able to operate in the low SNR regime, avoiding intrinsic numerical issues of competing approaches~\cite{wei2022fast}.\footnote{ \ac{2PHTS} uses an approximation of the actual channel model. The empirical mean may become negative in the low-\ac{SNR} regime, causing numerical issues in the computation of the relative entropy with the \ac{TAS} baseline algorithm.} 
\item \mrb{We guarantee the accuracy of the initial alignment phase by showing that \ac{DBZ} is $\delta$-\ac{PAC} and derive a closed-form expression for the sample complexity.  We also use the sample complexity to inform how the algorithm adapts to the time-varying statistics.}
\item We provide extensive simulations, including realistic environments from NYU Sim \cite{ju2018nyusimmotion, sun2017nyusim}.
\end{itemize}
The remainder of the document is organized as follows. \mrb{In Section \ref{sec:problemstatement}, we introduce the system model used for our \ac{ISAC} scenario. In Section~\ref{sec:hierarchical} we describe the hierarchical codebooks used for beam alignment and tracking.} In Section \ref{sec:algs}, we introduce \ac{DBZ}, which uses a \ac{MAB} best arm identification framework to quickly align and adapts to the \ac{ME} motion over time.  In Section \ref{sec:analysis}, we develop closed-form expressions of the sample complexity that inform the choice of parameters in the \ac{DBZ} algorithm.  In Section \ref{sec:sims}, we  present extensive numerical simulations demonstrating the performance \ac{DBZ} performance across a wide range of scenarios.

\section{System Model}\label{sec:problemstatement}

\subsection{\mrb{System Model}}
\label{sec:channel-model}


At each discrete time step $n \in \mathbb{N}$, a \ac{BS} transmits a \ac{SS}/\ac{RS}, $\mathbf{s} \in \mathbb{C}^{Q\times 1}$, consisting of $Q$ samples at finer time or frequency granularity. Each signal consists of time-frequency \ac{RE}s across multiple \ac{OFDM} symbols similar to that of the \ac{SSB} or \ac{CSI}-\ac{RS} used in the current 5G standard \cite[7.4]{3GPP_TS_38211_R17}. Each transmission, $\mathbf{s}$, has a cell identification number that is unique to one \ac{BS}, where $\mathbf{s}^H\mathbf{s} = 1$.  We use a \ac{ULA} of $M$ elements to transmit signals over which we apply a \ac{HAD} beamforming vector to electronically steer each transmission of $\mathbf{s}$.  The  \ac{HAD} beamforming vectors belong to a beamforming codebook, $\mathcal{F}$, that consists of analog phase shifts for $M$ antenna elements with $N_{\textnormal{RF}}$ \ac{RF} chains, $\mathbf{F}_{\textnormal{RF}}\in \mathbb{C}^{M\times N_{\textnormal{RF}}}$, and a digitally applied baseband precoder, $\mathbf{F}_{\textnormal{BB}}\in\mathbb{C}^{N_{\textnormal{RF}}\times N_S}$,  for each \ac{RF} chain to feed $N_S$ datastreams.  We denote the combined beamforming vector for a single datastream, $u$, as $\mathbf{f} = \mathbf{F}_{\textnormal{RF}}[\mathbf{F}_{\textnormal{BB}}]_u$ ($\mathbf{f}^H\mathbf{f} = 1$), where $[\mathbf{F}_{\textnormal{BB}}]_u$ is the $u^{\textnormal{th}}$ column of  $\mathbf{F}_{\textnormal{BB}}$.  The  beamforming pattern for each of the vectors, $\mathbf{f}\in \mathcal{F}$, has a unique pointing angle, $\bar{\phi}\in \Phi \eqdef [\phi_{\textnormal{min}},\phi_{\textnormal{max}}]$, with $\bar{\phi}$ evenly spaced across a predefined range $\Phi$ and with each pattern having an equivalent \emph{beamwidth}, $\phi_{\textnormal{bw}}$, and \emph{gain}, $g$.  We require steering the beamforming pattern in the angular direction of the receiving \ac{ME}, $\theta_k(n)$.

\subsection{\mrb{Channel and Kinematic Models}}\label{ssec:channel_model}
We assume that the receiving \ac{ME} forms a single beam, such that it is always directed at the transmitting \ac{BS} or omni-directional.\footnote{While this model abstracts away the joint alignment process at the \ac{ME} and \ac{BS}, it still captures the essence of the problem and has been widely adopted~\cite{chiu2019active,wei2022fast,wu2019fast,noh2017multi}.}  We consider only a single subcarrier to allow for the narrowband channel representation~\cite[Eq. (7)]{heath2016overview},
\begin{align}\label{eq:channelresponse}
    \mathbf{h}(n) = \sum_{k = 1}^K\alpha_k(n) \mathbf{a}^H(\theta_k(n)),
\end{align}
where
\begin{align}\label{eq:arrayresponse}
    \mathbf{a}(\theta) \eqdef \begin{bmatrix} e^{-j\frac{M-1}{2}\frac{2\pi d \cos(\theta)}{\lambda}} & \cdots & e^{j\frac{M-1}{2}\frac{2\pi d \cos(\theta)}{\lambda}}\end{bmatrix}^T
\end{align}
is the array response for a \ac{ULA}, $j^2 \eqdef -1$, and $d$ and $\lambda$ are the array element spacing and wavelength, respectively. For each path, $\alpha_k(n) \in \mathbb{C}$ represents the complex gain caused by large and small scale fading.\footnote{For each path, $\alpha_{k,I}(n) + j\alpha_{k,Q}(n) = \alpha_k(n) \in \mathbb{C}$, we specify later that each component amplitude changes over time according to $\alpha_{k,I}(n+1) = \rho\alpha_{k,I}(n) + \omega_I(n)$, where $\omega_I(n) + j\omega_Q(n) = \omega(n)\sim\mathcal{CN}(0,(1-\rho^2))$ \cite{va2016beam} or by the Rician AR-1 channel model in \cite{chiu2019active}.} We assume $\theta_1(n)$ is the dominant path in a \ac{LOS} scenario with the receiving \ac{ME}.  The received signal takes the form
\begin{align}\label{eq:complexmeasurement}
    z(n) &=\mathbf{h}(n)\mathbf{f}\mathbf{s}^T\mathbf{s}^* + \mathbf{v}^T(n)\mathbf{s}^*(n)= \mathbf{h}(n)\mathbf{f} + v(n)
\end{align}
where $\mathbf{v}(n) \sim \mathcal{CN}(0,\sigma_v^2\mathbf{I})$.  The \ac{RSRP} measurement is
\begin{align}\label{eq:rsrp}
y(n) = \abs{z(n)}^2.
\end{align}
The receiving \ac{ME} regularly communicates control data or measurements to the \ac{BS} advising beamforming vector selection, similar to the 5G/\ac{NR} standard \cite[Section 6]{3GPP_TS_38213_R17}\cite[Section 5.6.1]{3GPP_TS_38214_R17}\cite[Section 5]{3GPP_TS_38331_R17}. 

Between discrete time steps $n$ and $n+1$, spaced $\tau$ seconds apart, the \ac{BS} and \ac{ME} experience relative motion according to a \ac{DWNA} motion model \cite[Chapter 6.3.2]{bar2004estimation},
\begin{align}\label{eq:kinematicmodel}
\theta(n) &=  \theta(n-1) + \tau\dot{\theta}(n-1)+ \frac{\tau^2}{2}u(n-1),\\
\dot{\theta}(n) &= \dot{\theta}(n-1) + \tau u(n-1),
\end{align}
where $\dot{\theta}(n)$ is the angular velocity and $u(n) \sim \mathcal{N}(0,\sigma_u^2)$. Standard deviation of the acceleration, $\sigma_u$, governs the severity of the motion between time steps.  We simulate the operation of \ac{DBZ} with the \ac{DWNA} model in Section~\ref{sec:sims} for various values of $\sigma_u$, governing the severity of the motion. 

\subsection{\mrb{Alignment Problem}}\label{ssec:problem_statement}
The relative motion requires adjustments to the beamforming vector to maintain \emph{alignment}.  We define alignment as the state in which we choose the beamforming vector, $\mathbf{f}^*(n)$ at time step $n$,  such that\footnote{To ensure that our codebook contains a beamforming vector that points in the direction $\theta(n)$ in our simulations, we wrap the angle $\theta(n)$ to constrain it to $\Phi$.} 
\begin{align}\label{eq:isacgoal}
    \mathbf{f}^*(n) = \argmin_{\mathbf{f}\in\mathcal{F}}\norm{\mathbf{f}-\mathbf{a}(\theta(n))}.
\end{align}
For a beamforming vector $\mathbf{f}^*(n)$ pointed towards the angle $\bar{\phi}$, we define $N^{\textnormal{a}}$ as the maximum number of consecutive time steps during which $\theta(n) \in \mathcal{R} = [\bar{\phi} - \phi_{\textnormal{bw}}/2, \bar{\phi} + \phi_{\textnormal{bw}}/2]$, where $\mathcal{R}$ represents the coverage region of the beamforming vector.  To achieve \eqref{eq:isacgoal} with a probability of at least $1-\delta$, we employ a \ac{MAB} best-arm identification strategy, based on \cite{gabillon2012best}, to select a beamforming vector. Following the beamforming vector selection, we monitor the \ac{RSRP} measurements over time, utilizing the same signals used for communication, where abrupt changes in power serve as indicators of misalignment.


\section{\mrb{Hierarchical Codebook and Structure}}\label{sec:hierarchical}
\begin{figure}[b]
\includegraphics[width = .9\linewidth]{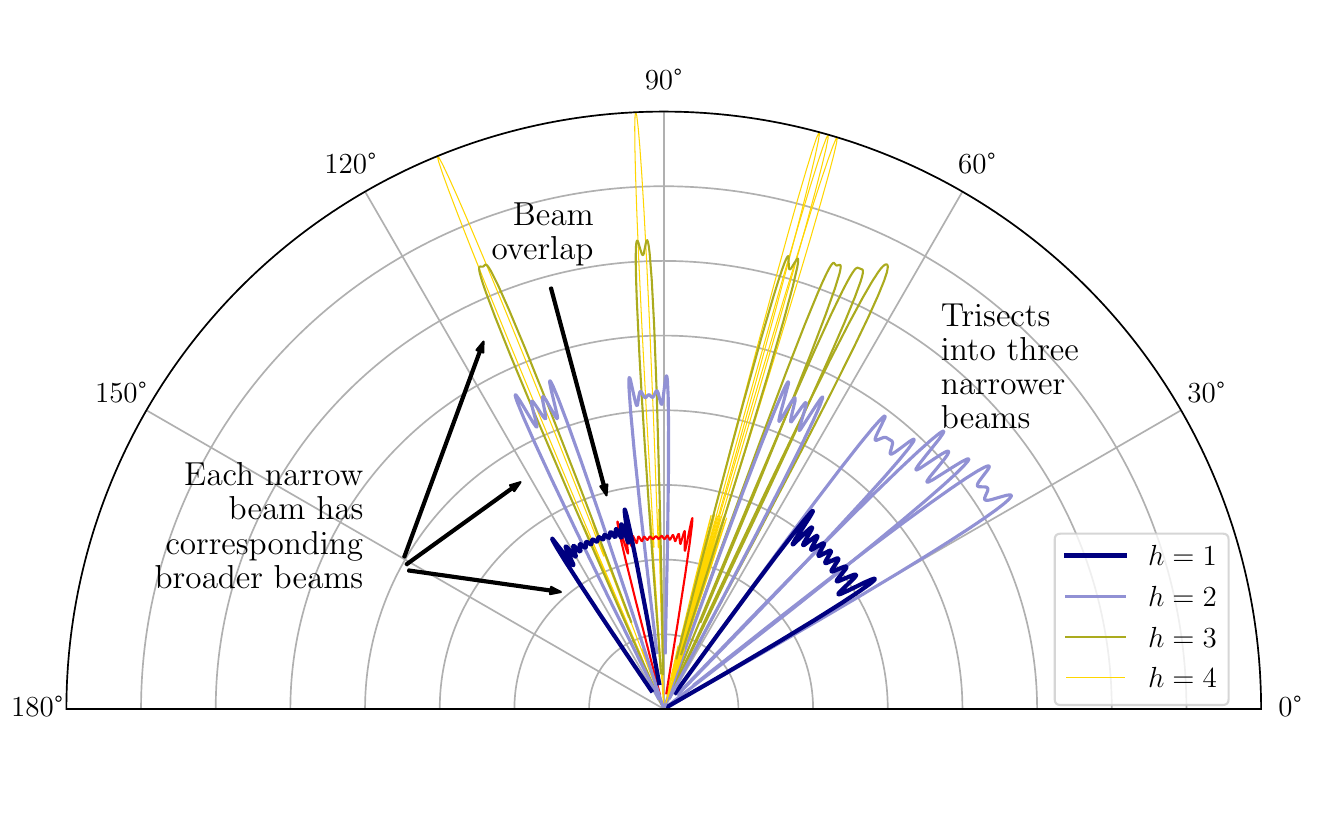}
\centering
\caption{Example beamforming patterns for the hierarchical codebook.}
\label{fig:beamzooming}
\end{figure}
\mrb{Our work exploits a {HAD} \emph{hierarchical} codebook $\mathcal{F}^H$ adapted from \cite{alkhateeb2014channel}. As illustrated in Fig.~\ref{fig:beamzooming}, our construction is as follows.
  \begin{itemize}
  \item The codebook consists of $H$ levels, each level $h\in \set{1,\dots,H}$ corresponding to beams with beamwidth $\phi_{\textnormal{bw},h}$;
  \item Each beam at level $h$ is split into three non-overlapping narrower beams at level $h+1$ so that $\phi_{\textnormal{bw},h} = 3\phi_{\textnormal{bw},h+1}$;
  \item The gain at each level $h$ is $g_h = g^{h-H+1}$;
  \item Each beam at level $h$ is also associated to a broader beam pointed in the same angle for all $h <H$.
  \end{itemize}
  Mathematically, this means there are exactly $I=3^{H-1}\abs{\mathcal{I}_1}$ beamforming vectors at each level $h$ with pointing angles
  \begin{align}\label{eq:pointingangle}
    \bar{\phi}_{h,i} = \phi_{\textnormal{min}} + \frac{\phi_{\textnormal{bw},H}}{2} + (i-1)\phi_{\textnormal{bw},H},\quad i\in\left\{1,\dots,I\right\}.
  \end{align}
  Each beam identified by $(h,i)$ aggregates a unique set of three non-overlapping patterns with indices $(h+1,j)$ with $j\in\{i-3^{H-h},i,i+3^{H-h}\}$. Each beam $(h,i)$ has a corresponding beam $(h-1,i)$ with the same pointing angle.}
  \mrb{This construction allows one to quickly ``zoom in'' from a beam $(h,i)$, narrowing the beamwidth\footnote{In our simulations showing initial alignment performance comparison, we adopt the binary codebook from \cite{chiu2019active,alkhateeb2014channel}.} by aggregating the beams in the set
  \begin{align}\label{eq:zoominindices}
    \mathcal{Z}_{h,i} \eqdef \begin{cases}\{(h+1,i),(h+1,i \pm 3^{H-h})\} &\textnormal{if }h < H,\\
      \set{(H,i)} &\textnormal{if }h = H.
    \end{cases}
  \end{align}
  In case of misalignment, the codebook allows one to zoom out from $(h,i)$, to $(h-1,i)$ without changing the pointing angle.  
}

\subsection{\mrb{Codebook Characteristics}}{\label{ssec:codebook_construction}

\mrb{We briefly discuss codebook depth, branching factors, and design methodology in how they impact \ac{DBZ} alignment accuracy, computational complexity, and robustness in dynamic environments. \ac{DBZ} is agnostic to codebook depth (determined by $H$). We show in Section \ref{ssec:configuration} that the algorithm parameters may be calibrated to support codebook designs with different choices of beamwidths at various depth levels to most efficiently ensure alignment accuracy with an \ac{ME}.  The branching factor is the main restricting codebook design characteristic for \ac{DBZ}.  To ensure alignment accuracy when broadening a beam and zooming out, each broad beam must be divisible into an odd-numbered quantity of beams to preserve the pointing angle of the previous narrow beam.  We exclusively use a branching factor of $3$ for this work, but \ac{DBZ} easily adopts any codebook with an odd-numbered branching factor.  Increased depth and larger branching factors contribute to higher sample complexity due to the increased total number of beamforming vectors.  However, the increase in depth or branching factor provides more precise alignment with the finer resolution of the search space, $\Phi$.  On the other hand, \ac{DBZ} benefits from shallower codebooks in the case of highly sporadic motion to more quickly adapt to the \ac{ME} position and maintain alignment accuracy.  \ac{DBZ} may adapt other \ac{HAD} codebook construction methodologies outside of \cite{alkhateeb2014channel} without compromising performance.  Additional logic for \ac{DBZ} allows easy extension to use adaptive constructed codebooks, as in \cite{qi2020hierarchical}, to further reduce training overhead for multiple \ac{ME}.  
}

\subsection{\mrb{Induced Mean Reward Structure}}\label{ssec:induced_mean_reward_structure}
For any beamforming vector $\mathbf{f}_{h,i}\in \mathcal{F}^H$, $\abs{z_{h,i}(n)}^2$ is a $\sigma_v^2/2$-scaled non-central chi-squared random variable with two degrees of freedom, and has non-centrality parameter, $2\zeta_{h,i}(n)/\sigma_v^2$, where $\zeta_{h,i}(n) = \abs{\mathbf{h}(n)\mathbf{f}_{h,i}}^2$. We define the mean-reward function generated by the \ac{RSRP} measurements~\eqref{eq:rsrp} of the channel as
\begin{align}\label{eq:freward}
    \mu_{h,i}(n) \eqdef \E{\abs{\mathbf{h}(n)\mathbf{f}_{h,i} + v(n)}^2}.
\end{align}
Using the hierarchical codebook results in an induced structure of the mean rewards, we make two assumptions.
\begin{assumption}\label{asp:uniquemax}
For each $h$, at any time step, $n$, there exists a unique beamforming vector $\mathbf{f}_{h,i^*}$ such that
\begin{align}\label{eq:levelmaxreward}
\mu^*_h(n) = \mu_{h,i^*_h}(n)  = \max_{i \in \set{1,\dots,I}}\mu_{h,i}(n).
\end{align}
\end{assumption}
\noindent We define an $\epsilon$-optimal arm as $(H,i^\epsilon) \in \set{(H,i) : \mu_{H,i}(n) + \epsilon \geq \mu_H^*(n)}$.  By definition, $(H,i^*)$ is $\epsilon$-optimal. 
\begin{assumption}{\textnormal{(Unimodality)}}\label{asp:unimodality}
For all $n$, if $\mu_{H,i^\epsilon}(n) + \epsilon \geq \mu_H^*(n)$ then there exist paths $((1,i_1),(2,i_2), \dots, (H-1,i_{H-1}),(H,i^\epsilon))$ through the tree graph defining the codebook where 
\begin{align}\label{eq:rewardmaxperlayer}
\mu_{H,i^\epsilon}(n) > \mu_{H-1,i_{H-1}}(n) > \dots >\mu_{2,i_2}(n)  > \mu_{1,i_1}(n)
\end{align} 

\end{assumption}
\noindent The sparsity and high path loss attenuation associated with \ac{mmWave} propagation \cite{rappaport2013millimeter} suggest that Assumptions \ref{asp:uniquemax} and \ref{asp:unimodality} hold in most situations.  We denote the difference between mean rewards at a particular level $h$ as
\begin{align}\label{eq:meandifferences}
    \Delta_{h,i}(n) \eqdef\begin{cases}
    \mu^*_h(n) - \mu_{h,i}(n) &\textnormal{if }i \neq i^*,\\
    \mu^*_h(n) - \max_{i\neq i^*}\mu_{h,i}(n) & \textnormal{if }i = i^*.
    \end{cases}
\end{align}
\noindent \mrb{Our analysis and discussion in Section IV-E emphasizes that the spacing between mean rewards, $\Delta_{h,i}(n)$, significantly contributes to overall sample complexity.  In particular, broader beams will have smaller values of $\Delta_{h,i}(n)$, and therefore higher sample complexity. Section IV-F shows how to configure \ac{DBZ} such that we play certain levels and optimize the trade off of sample complexity and number of beamforming vectors played.} From our codebook construction, for any $\epsilon$-optimal arm, there exists a path $\set{(h,i_h)}|_{h=1}^H$ such that
\begin{align}\label{eq:gaineps}
    \frac{\mu_{H,i^\epsilon}(n)}{\mu_{H-1,i_{H-1}}(n)} = \frac{\mu_{H-1,i_{H-1}}(n)}{\mu_{H-2,i_{H-2}}(n)} = \dots = \frac{\mu_{2,i_{2}}(n)}{\mu_{1,i_{1}}(n)} = g.
\end{align}
From $\mu_{H,i^\epsilon}(n) + \epsilon \geq \mu_H^*(n)$, \eqref{eq:gaineps} ensures that $\mu_{H,i^\epsilon}(n)  + \epsilon \geq g\mu_{H-1}^*(n)$, from which we obtain $\mu_{h,i_h}(n) + \epsilon_h \geq \mu_{h}^*(n)$, where $\epsilon_h \eqdef g^{-(H-h)}\epsilon$.  If the average reward corresponding to beamforming vector $\mathbf{f}_{h,i}$ meets the criteria of $\mu_{h,i}(n) \geq \mu_h^*(n) + \epsilon_h$ then it is $\epsilon_h$-optimal.  We relate the relative cost to spectral efficiency to $\epsilon$ in Section \ref{sec:analysis}.   In our model \eqref{eq:kinematicmodel}, the mean rewards are non-stationary, causing the unique maximum mean-reward, $\mu_H^*(n)$, to change over time.   The next section introduces our algorithm, \ac{DBZ}, that dynamically adjusts the beamwidth used for communication by selecting beamforming vectors under certain \emph{zoom-in} and \emph{zoom-out} criteria, based on \ac{MAB} best arm identification and power threshold, respectively, to maintain alignment with the \ac{ME}.

\section{Algorithm: Dynamic Beam Zooming}\label{sec:algs}
\label{sec:algs}
\ac{DBZ} uses the hierarchical codebook described in Section~\ref{sec:hierarchical} and efficiently exploits the induced dynamic reward structure.  \ac{DBZ} exploits the representation of each beamforming vector $\mathbf{f}_{h,i}$ as a vertex $(h,i)$ in \mrb{a tree} and uses a best arm identification \ac{MAB} framework \cite{gabillon2012best} and power threshold to dynamically \mrb{navigate the tree and maintain alignment with the \ac{ME}.  Informally, the algorithm operates as illustrated in Fig.~\ref{fig:dbz_visual} to show example of traversing the graph vertices for beam refinement. Vertices with an asterisk indicate the beam used to communicate and the triangle moving along the bottom represents an \ac{ME}.  The leafs at the bottom of each tree represent the narrowest beams.  }
\mrb{
  \begin{itemize}
  \item Steps~\nbbox{1} and~\nbbox{2}: to initially align, we identify with probability $1-\delta$ the beamforming vector $\mathbf{f}_{H,i^*_H}$ that most closely matches the \ac{ULA} response to $\theta(n)$  according to \eqref{eq:isacgoal} within $N_h^\textnormal{a}$ time steps.  This is achieved with \ac{MAB} algorithms that, at levels $h$, play beamforming vectors $\mathbf{f}_{h,i}$ viewed as arms in a \ac{MAB} best-arm identification fixed confidence setting. The chosen arm, corresponding to a narrower beam, is used for increasing the rate at which we communicate data. We then put the chosen arm's zoom-in indices $\mathcal{Z}_{h,i}$ in \eqref{eq:zoominindices} in contention to play a subsequent \ac{MAB} game, to continue to refine the communication beamwidth. 
  \item  Steps~\nbbox{2} to~\nbbox{3}: \ac{DBZ} detects beam misalignment by the \ac{RSRP} failing to meet a power threshold, and ``zooms out'', adjusting the set of active vertices.  
  \item Step~\nbbox{4}: the broader beam is adjusted to realign.
  \item Step~\nbbox{5}: the beam is correctly re-adjusted to the narrowest width.  
  \end{itemize}
}
\begin{figure}[h]
\includegraphics[scale = .35]{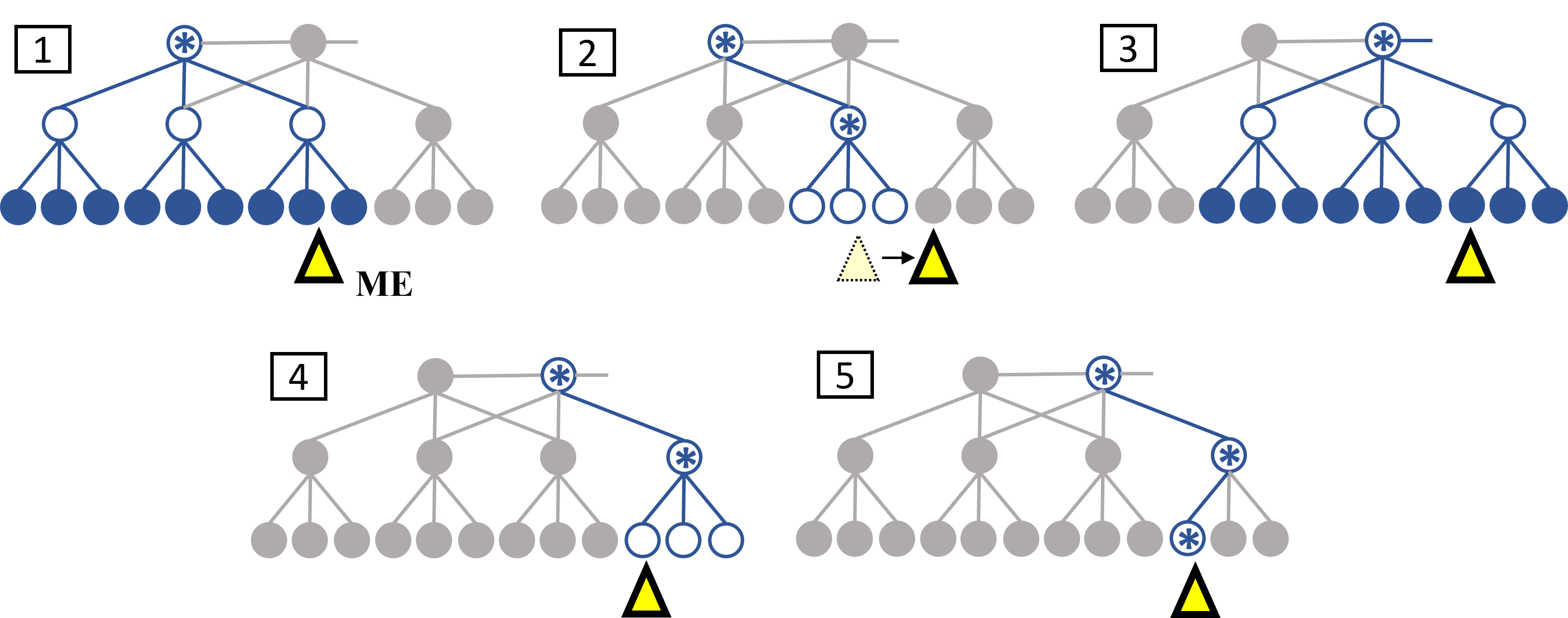}
\centering
\caption{Illustration of beamforming vector selection in \ac{DBZ} over time.}
\label{fig:dbz_visual}
\end{figure}
\subsection{Baseline Framework}\label{ssec:baselineframework}
\ac{DBZ} proceeds mathematically using the \ac{LUCB} best-arm identification framework \cite{gabillon2012best,kalyanakrishnan2012pac} for each \ac{MAB} game.  \ac{LUCB} uses empirical statistics derived from the sample reward values (in this case \ac{RSRP}) that represent the estimation and uncertainty on the mean rewards.  Due to the non-stationary rewards from the \ac{ME} motion, we only consider a finite set of size $\eta$ consisting of the most recent samples to compute the \ac{LUCB} statistics.  We refer to the finite set as the \emph{sample window} \cite{besbes2014stochastic}.  At each time step $n$, we select a specific beamforming vector $(h,i)$ whose indices are stored as $S(n)$ and observe the corresponding reward, $y(n)$. The mean rewards, $\mu_{h,i}(n)$ \eqref{eq:freward}, at time step $n$ are estimated by the empirical mean, using the $\eta$ most recent samples according to
\begin{align}
    \hat{\mu}_{h,i}(\eta,n) = \frac{1}{N_{h,i}(\eta,n)}\sum_{p = \max\set{1,n-\eta}}^n y(p)\indic{S(p) = (h,i)},\label{eq:empmean1}
\end{align}
where\footnote{For the indicator function, $\indic{S(n) = (h,i)} = 1$ when $S(n) = (h,i)$ and $0$ otherwise.}
\begin{align}\label{eq:numberofplays}
    N_{h,i}(\eta,n) = \sum_{p = \max\set{1,n-\eta}}^n \indic{S(p) = (h,i)}.
\end{align}
To allow further generalization later on, we let $\mathcal{I}_h$ denote the set of arms in contention at level $h$, noting that $\abs{\mathcal{I}_h} = 3$ for all levels except $h=1$, and let $\mathcal{I}^H = \sum_h\abs{\mathcal{I}_h}$.   Pictorially, each level's active vertices in Fig. \ref{fig:dbz_visual} represent arms in $\mathcal{I}_h$.   For constants $B,C \geq 1$ to be chosen later, we use a confidence term, empirical observation variance estimate, and exploration rate
\begin{align}\label{eq:confidenceterm}
    D_{h,i}(\eta,n)\eqdef \sqrt{\frac{4B\hat{\nu}_{h,i}^2(\eta,n)\beta(\eta,n,\delta)}{N_{h,i}(\eta,n)}} + \frac{2\sqrt{2BC}\beta(\eta,n,\delta)}{N_{h,i}(\eta,n)-1},
\end{align}
\begin{align}\label{eq:empvariance}
    \hat{\nu}_{h,i}^2&(\eta,n)\notag\\ &= \sum_{p = \max\set{1,n-\eta}}^n\frac{(y(p) - \hat{\mu}_{h,i}(\eta,p))^2\indic{S(n) = (h,i)}}{N_{h,i}(\eta,n)},
\end{align}
\begin{align}\label{eq:explorationrate}
\beta(\eta,n,\delta) \eqdef \log\left(15\mathcal{I}^H(\min\set{n,\eta})^4/(2\delta)\right),
\end{align}
respectively, in the \ac{UCB} and \ac{LCB} terms
\begin{align}
    U_{h,i}(\eta,n) = \hat{\mu}_{h,i}(\eta,n) + D_{h,i}(\eta,n)\label{eq:ucbterm},
\end{align}
\begin{align}
    L_{h,i}(\eta,n) = \hat{\mu}_{h,i}(\eta,n) - D_{h,i}(\eta,n)\label{eq:lcbterm},
\end{align}
respectively.  The terms, \eqref{eq:ucbterm} and \eqref{eq:lcbterm}, capture the best and worst performance, respectively, of a beamforming vector, that we use to define the \emph{gap} for each arm,
\begin{align}\label{eq:gapterm}
    G_{h,i}(\eta,n) = \max_{j \neq i} U_{h,j}(\eta,n-1) - L_{h,i}(\eta,n-1),
\end{align}
and the indices 
\begin{align}
\gamma(n) &= \argmin_{i: (h,i) \in \mathcal{I}_h}G_{h,i}(\eta,n)\label{eq:empiricalbest},\\
u(n) &= \argmax_{i: (h,i) \in \mathcal{I}_h, i\neq \gamma(n)}U_{h,i}(\eta,n-1)\label{eq:bestofworst}.
\end{align}
\mrb{We sample a beamforming vector $\mathbf{f}_{S(n)}$ with index tuple
\begin{align}\label{eq:sample_selection}
S(n) \eqdef \argmax_{(h,i):i\in\set{\gamma(n),u(n)}}D_{h,i}(\eta,n-1), 
\end{align}
or all $(h,i) \in \mathcal{I}_h$ in round-robin fashion to first initialize a new level.}  The individual \ac{MAB} games are independent across levels, where ``zooming in" to the next level is governed by \emph{termination} at the current level.  Termination and zooming in at a particular level $h$ occurs when the gap term for $\gamma(n)$ first satisfies:
\begin{align}\label{eq:stoppingcriteria}
G_{h,\gamma(n)}(n) = U_{h,u(n)}(\eta,n-1) - L_{h,\gamma(n)}(\eta,n-1) < \epsilon_h,
\end{align}
at which point, we choose $(h,\gamma(n))$ for communication\mrb{, and store it as $(h,i^c)$}.  Intuitively, $L_{h,\gamma(n)}(\eta,n-1)$ is the worst performance of the estimated best beamforming vector and $U_{h,u(n)}(\eta,n-1)$ is the best performance of the runner-up beamforming vector.  We show in our analysis in Section \ref{sec:analysis} that with \eqref{eq:stoppingcriteria} we make a correct selection beam, i.e., $(h,\gamma(n)) = (h,i^*)$, of a beamforming vector at level $h$ with probability at least $1-\delta$.  \mrb{We next show how this \ac{LUCB} mathematical framework is used to facilitate \ac{DBZ}.}

\begin{figure}[h]
\includegraphics[width = .5\textwidth]{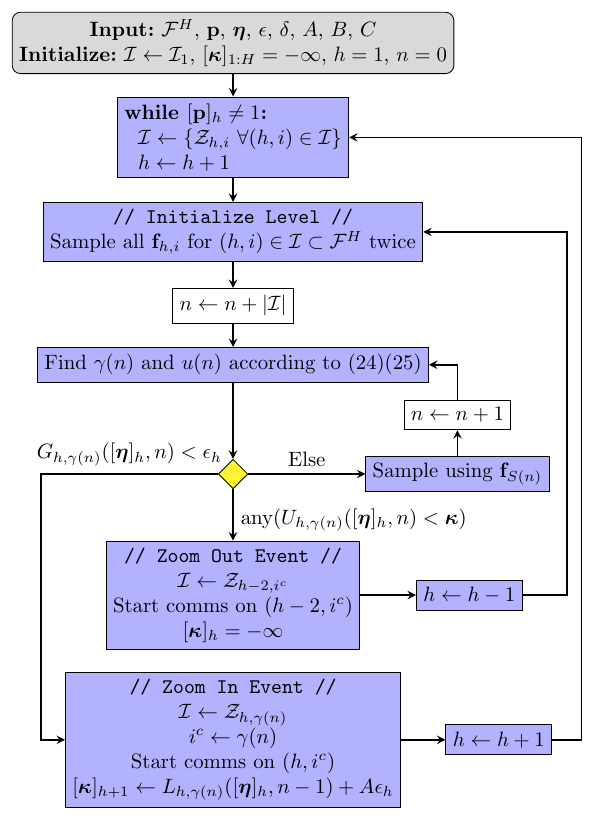}
\centering
\caption{\mrb{Flowchart of DBZ algorithm.}}
\label{fig:dbz_flowchart}
\end{figure}

\subsection{DBZ Algorithm}\label{ssec:dbz}

\mrb{Fig. \ref{fig:dbz_flowchart} shows how the \ac{DBZ} algorithm proceeds with zooming in or zooming out using the baseline \ac{LUCB} \ac{MAB} framework.}  Each level has a different beamwidth, $\phi_{\textnormal{bw},h}$, which implies a different alignment time.  We use a vector of hyperparameters, $\boldsymbol\eta\in \mathbb{N}^H$, whose $h^{\textnormal{th}}$ element is the sampling window length used at level $h$.  \mrb{For now, assume $\mathbf{p} = [1,\dots,1]$.}  \ac{DBZ} initially samples from a fixed set of beamforming vectors, $\mathcal{I}_1$.\footnote{\mrb{Note that in Fig. \ref{fig:dbz_flowchart} we store $\mathcal{I}_1$, or $\mathcal{I}_h$, as $\mathcal{I}$.}}  $\mathcal{I}_1$ possesses only the broadest beam patterns that are non-overlapping and perfectly cover $\Phi$.  \mrb{\ac{DBZ}} checks the termination criteria in \eqref{eq:stoppingcriteria} to determine zooming in at \mrb{the western path of the decision diamond in Fig. \ref{fig:dbz_flowchart}}.  Once the algorithm terminates at the initial level, $h=1$, we begin to communicate using the chosen beamforming vector $(1,\gamma(n))$\mrb{, stored as $(h,i^c)$}.  \ac{DBZ} continues to play \ac{MAB} games at subsequent levels, choosing $(h,\gamma(n))$ upon termination at level $h$, and refines the communication beam with the subsequent \ac{MAB} games at $h>1$.  As shown in Fig. \ref{fig:dbz_visual}, this operation continues until \ac{DBZ} terminates with a narrowest beam at level $H$.  \mrb{Fig. \ref{fig:dbz_flowchart} shows \ac{DBZ} loops back to initialize the next level after zooming in.} Conversely\mrb{, following the southern path of the decision diamond in Fig. \ref{fig:dbz_flowchart}}, we control ``zooming out'' to a previous level with a wider beam by establishing a vector of hyperparameters, $\boldsymbol\kappa$, whose elements are the threshold \ac{RSRP} after termination and zooming in at each level, 
\begin{align}\label{eq:rsrpthresh}
	[\boldsymbol\kappa]_{h+1} \eqdef L_{h,\gamma(n)}(\eta,n-1) + A\epsilon_h,
\end{align}
for $A\leq 1$ to be chosen later.  If at any time step
\begin{align}\label{eq:zoomoutcriteria}
 U_{h,\gamma(n)}(\eta,n-1) < [\boldsymbol\kappa]_{h'},
\end{align} 
for any $h' \leq h$, \ac{DBZ} zooms out choosing $(h-1,\mrb{i^c})$ as the new communication beam and \mrb{loops back on the flowchart} re-initializ\mrb{ing} all arms in $\mathcal{Z}_{h-2,i}$ that are now in contention.\footnote{The logical test $\texttt{any}(\cdot)$ (\mrb{southern path of the decision diamond in Fig. \ref{fig:dbz_flowchart}}) returns Boolean $\texttt{True}$ if any element in a logic vector returns $\texttt{True}$.}  \mrb{In the case of zooming out at $h=2$, we reset and store $\mathcal{I}_1$ as $\mathcal{I}$. } The intuition for our choice of of adaptive threshold in \eqref{eq:rsrpthresh} is that we base it on the worst performance of a previous level's ($h'<h$) best-performing beamforming vector, $L_{h',\gamma(n)}(\eta,n-1)$.  If the best performance of the current level's best-performing beamforming vector, $U_{h,\gamma(n)}(\eta,n-1)$, does not exceed the threshold, one concludes misalignment and zooms out.  These discrete decision points for beam transitions allow the transmission of control information between the \ac{BS} and \ac{ME} to adjust the corresponding rate and beamforming vector \cite{3GPP_TS_38213_R17,3GPP_TS_38214_R17}.  \mrb{In the case of neither zooming in or out, we take the eastern path of the decision diamond and sample beamforming vector corresponding to $S(n)$.} 

Playing each level reduces the number of arms considered overall, but \emph{naively} exploits the hierarchical codebook.  Certain levels of the codebook are more beneficial to play than others based on the number of beams eliminated per number of samples required.  Consequently, one might benefit from skipping some levels at the expense of contending more arms in a best-arm identification \ac{MAB} game.  We characterize the strategy used to navigate the codebook levels by a pruning vector of hyperparameters $\mathbf{p}$ with elements $[\mathbf{p}]_h\in \set{0,1}$. Specifically, assume that at level $h-1$, the beamforming vectors corresponding to the vertices in $\mathcal{I}_{h-1}$ are used in the best-arm identification \ac{MAB} game to take samples as \ac{RSRP} measurements.  Upon termination, if $[\mathbf{p}]_h =1$, the children vertices' beamforming vectors of the arm chosen at level $h-1$ are played in the next level $h$.  If $[\mathbf{p}]_h = 0$, we bypass level $h$ and, pending $[\mathbf{p}]_{h+1} = 1$, put all descendant arms in contention (\mrb{See block prior to level initialization in Fig. \ref{fig:dbz_flowchart}}). If $[\mathbf{p}]_{h+1} = 0$, we bypass this level, and so on.  As an example, the size of a set of beamforming vectors after skipping one level is $\abs{\mathcal{I}}=9$ for a ternary tree, or $\abs{\mathcal{I}}=4$ for a binary tree. Note that $[\mathbf{p}]_H$ must be set to $1$ because we require a choice of one of the narrowest beamforming patterns.  Our simulations in Section \ref{sec:sims} show that many of the hyperparameters may be generically set over a broad range of channel conditions and maintain performance.


\section{Analysis}\label{sec:analysis}

For initial alignment, \ac{DBZ} adapts the fixed-confidence best arm identification framework in \cite{gabillon2012best}, in which the algorithm requires choosing the correct beam with high probability. \ac{DBZ} requires accurate estimation of the mean reward values in  \eqref{eq:freward} for $(h,i) \in \mathcal{I}_h$ while maintaining alignment.  For these two requirements, we define the events,
\begin{align}\label{eq:confidenceintervalevent}
\mathcal{B}_h \eqdef \bigl\{\forall &(h,i) \in \mathcal{I}_h, \forall n > 2\abs{\mathcal{I}_h},\\ &\abs{\hat{\mu}_{h,i}(\eta,n) - \mu_{h,i}(n)}<D_{h,i}(\eta,n)\bigr\}, 
\end{align}
and
\begin{align}\label{eq:coherenceevent}
\mathcal{A}_h \eqdef \set{\exists (h,i) \in \mathcal{I}_h:  \theta(n) \in \calR_{h,i}},
\end{align}
for a beamforming vector $\mathbf{f}_{h,i}$ whose beampattern is pointed toward angle $\bar{\phi}_i$ and has beamwidth $\phi_{\text{bw},h}$.  As a reminder,   $\calR_{h,i}= [\bar{\phi}_{h,i} -\phi_{\textnormal{bw,h}}/2,\bar{\phi}_{h,i} +\phi_{\textnormal{bw,h}}/2]$ is the coverage region of the beam pattern corresponding to $(h,i)$.  Under events \eqref{eq:confidenceintervalevent} and \eqref{eq:coherenceevent} $\forall h\in\set{1,\dots,H}$, we show that \ac{DBZ} zooms in to choose an $\epsilon$-optimal beamforming vector with probability at least $1-\delta$.  In the case the \ac{ME} is out of alignment, $\mathcal{A}_h^c$, we zoom out to mitigate severe outages, which keeps the \ac{ME} aligned in the wider beam.   We confirm a zoom-out action from $(h,i)$ to $(h',i)$, where $h' < h$, as being \emph{correct} when the mean reward of the broader beam $\mu_{h',i}(n) > \mu_{h,i}(n)$.  We first prove in Section \ref{ssec:confidenceanalysis} why $\mathcal{B}_h$ holds with high probability, ensuring $\mu_{h,i}(n)$ is well estimated.  In Section \ref{ssec:coherencetime}, based on our kinematic motion model \eqref{eq:kinematicmodel}, we analyze the maximum sample window lengths that may be chosen at each level, $[\boldsymbol\eta]_h$, such that $\mathcal{A}_h$ holds (aligned) with high probability. We then show the correctness of our decision criteria for zooming in \eqref{eq:stoppingcriteria} and zooming out \eqref{eq:zoomoutcriteria} to adjust to the dynamically changing mean reward values in Section \ref{ssec:zoominzoomout}.  Finally, in Section \ref{ssec:complexity} we develop a means to calculate the sample complexity required to zoom in with respect to the spacings of mean rewards \eqref{eq:meandifferences} and choice of $\epsilon$.  We use the sample complexity to select the hyperparameters corresponding to sample window lengths, $\boldsymbol\eta$, pruning vector, $\mathbf{p}$.

\subsection{Confidence}\label{ssec:confidenceanalysis}

We first provide two supporting lemmas that lay the foundation for ensuring that \ac{DBZ} correctly zooms in and zooms out with high confidence \mrb{using the mathematical components in Section \ref{ssec:baselineframework}}. We use the lemmas to show $\mu_{h,i}$ is well estimated if the event $\mathcal{B}_h$ occurs with high probability during \ac{DBZ} for all levels.  We conclude that with probability at least $1-\delta$, the true mean reward satisfies $\mu_{h,i}(\eta,n)\in [L_{h,i}(\eta,n),U_{h,i}(\eta,n)]$ during execution of \ac{DBZ}.  
We let $n_{h,i} \in \mathbb{N}$ denote the time steps at which arm $(h,i)$ is sampled.
\begin{lemma}\label{lem:subexponentiallemma}{For the sequence of observations, $\set{y(n_{h,i}) : n_{h,i} \geq 2}$}, which follow a $\sigma_v^2/2$-scaled non-central chi-squared distribution,
\begin{align}\label{eq:probsubexponential}
\P{\abs{\hat{\mu}_{h,i}(\eta,n) - \mu_{h,i}(n) > \delta}} \leq 2\exp\left(-\frac{N_{h,i}(\eta,n)\delta^2}{4\nu_{h,i}^2(n)}\right),
\end{align}
where $\nu_{h,i}^2(n) = \sigma^4+ 2\sigma^2\zeta_{h,i}(n)$ is the variance of $y(n_{h,i})$, and $\zeta_{h,i}(n) = \abs{\mathbf{h}(n)\mathbf{f}_{h,i}}^2$.
\end{lemma}

\begin{IEEEproof}
    Our proof follows the steps from \cite[Appendix E]{ghosh2022learning} and \cite[Section 2.1.3]{wainwright2019high}. Note that $\mu_{h,i} = \zeta_{h,i} + \sigma^2$, with the moment generating function of $y(n_{h,i})$ and dropping the time dependence temporarily, we write,
\begin{align}
    \E{\exp(\lambda(y(n_{h,i})-\mu_{h,i}))} &= \frac{\exp(-\lambda \mu_{h,i})}{1 - \sigma^2\lambda}\exp\left(\frac{\lambda \zeta_{h,i}}{1-\sigma^2\lambda}\right)\\
    &\leq \exp(\sigma^4\lambda^2) \exp(2\zeta_{h,i}\sigma^2\lambda^2)\label{eq:in_alphaconditionsubexponential}\\
    &= \exp\left(\frac{2\nu^2_{h,i}\lambda^2}{2}\right),\label{eq:sncx2mgf}
\end{align}
where \eqref{eq:in_alphaconditionsubexponential} holds when $\abs{\lambda}< 1/(2\sigma^2)$.  We use \eqref{eq:sncx2mgf} with the Cramer-Chernoff method to derive our concentration bound.  
The full steps are available in Appendix \ref{ssec:pf_lem_subexponentiallemma} of the supplementary material.
In particular, we are interested in the empirical mean of $y(n_{h,i})$ over time \eqref{eq:empmean1}, hence,
\begin{align}
    \P{\hat{\mu}_{h,i}(\eta,n) - \mu_{h,i}(n) \geq \delta}\leq\exp\left(-\frac{N_{h,i}(\eta,n)\delta^2}{4\nu_{h,i}^2(n)}\right).
\end{align}
A union bound completes our proof.
\end{IEEEproof}

\noindent Lemma \ref{lem:subexponentiallemma} enables us to write the concentration expression using our choice of confidence term \eqref{eq:confidenceterm} and exploration rate \eqref{eq:explorationrate} for the next lemma.

\begin{lemma}\label{lem:confidenceempvar}{Let $\set{y(n_{h,i}) : n_{h,i} \geq 2}$ be the sequence of \ac{iid} random variables in Lemma \ref{lem:subexponentiallemma}, then for any $B,C \geq 1$,  $0<\delta \leq \nu_{h,i}^2(n)/\sigma^2$, exploration rate $\beta(\eta,n,\delta)$ in \eqref{eq:explorationrate}, and confidence term $D_{h,i}(\eta,n)$ in \eqref{eq:confidenceterm},}
\begin{align}\label{eq:probempvar3}
    \P{\abs{\hat{\mu}_{h,i}(\eta,n) - \mu_{h,i}(n)} \geq D_{h,i}(\eta,n)} \leq 3\exp(-\beta(\eta,n,\delta)).
\end{align}
\end{lemma}

\begin{IEEEproof}
 We use the one-sided version of \eqref{eq:probsubexponential} from our result in Lemma \ref{lem:subexponentiallemma} with a constant $B \geq1$,
     \begin{align}\label{eq:probsubexponentialwb}
         \P{\hat{\mu}_{h,i}(\eta,n) - \mu_{h,i}(n) \geq\sqrt{\frac{4B\nu_{h,i}^2(\eta,n)\beta(\eta,n,\delta)}{N_{h,i}(\eta,n)}}}&\notag \\  \leq \exp(- \beta(\eta,n,\delta))&,
     \end{align}
     and the result in \cite[Theorem 10]{maurer2009empirical} with $C \geq 1$ to bound the difference between standard deviation $\nu_{h,i}$ and its empirical estimate, $\hat{\nu}_{h,i}(\eta,n)$, as
     \begin{align}\label{eq:empvarbound}
         \P{\nu_{h,i}(n) > \hat{\nu}_{h,i}(\eta,n) + \sqrt{\frac{2C\beta(\eta,n,\delta)}{N_{h,i}(\eta,n)-1}}}&\notag\\ \leq \exp(-\beta(\eta,n,\delta)).&
     \end{align}
By replacing $\nu_{h,i}^2(n)$ in \eqref{eq:probsubexponentialwb} with 
\begin{align}
\hat{\nu}_{h,i}(\eta,n) + \sqrt{\frac{2C\beta(\eta,n,\delta)}{N_{h,i}(\eta,n)-1}},
\end{align}
simplifying, and using union bounds, we obtain our result.  The full steps are available in Appendix \ref{ssec:pf_lem_confidenceempvar} of the supplementary material.
\end{IEEEproof}
 
 The concentration expressions in \eqref{eq:probsubexponential} and \eqref{eq:probempvar3} do not explicitly account for the changing mean reward, $\mu_{h,i}(n)$, over time. However, our choice of confidence term \eqref{eq:confidenceterm} incorporates the empirical variance, which past works have shown can suffice to adjust for the dynamic rewards \cite{wei2016tracking,maurer2009empirical}.

\subsection{\mrb{Confidence with }Alignment Time}\label{ssec:coherencetime}
Determining the likelihood of event $\mathcal{A}_h$ \eqref{eq:coherenceevent} requires a probabilistic description of the angle, $\theta(n)$, over time.   With the random variable model and distribution in hand (full derivation in Appendix \ref{sec:deriveprobcoherence} of the supplementary material), we determine the likelihood of $\theta(n)$ remaining in the region, $\mathcal{R}_{h,i} = [\bar{\phi}_{h,i}-\phi_{\text{bw},h}/2,\bar{\phi}_{h,i} + \phi_{\text{bw},h}/2]$, under the kinematic motion described in Section \ref{ssec:channel_model}. We express the probability of alignment after $n$ timesteps as
\begin{multline}
\P{\abs{\bar{\phi}_{h,i} - \theta(n)}\leq \frac{\phi_{\text{bw},h}}{2}} = \\
\frac{\sqrt{2}\sigma_n}{\phi_{\text{bw},h}\sqrt{\pi}} \left( \exp\left( -\frac{\phi_{\text{bw},h}^2}{2\sigma_n^2}\right)  - 1\right) \\
+ \erf\left( \frac{\phi_{\text{bw},h}}{\sqrt{2}\sigma_n} \right) ,
\label{eq:probcoherence}
\end{multline}
with
\begin{align}\label{eq:sigman}
\sigma_n^2 \eqdef\frac{\tau^4}{4}\left(\frac{4n^3}{3} -4n^2 + \frac{11n}{3} -1 \right)\sigma^2_u + \tau^2(n-1)\sigma_u^2,
\end{align}
where $\tau$ is the time difference, in seconds, between $n-1$ and $n$.\footnote{We use the error function defined as $\erf(z) = \frac{2}{\sqrt{\pi}}\int_0^z\exp(-t^2)dt$.} We use \eqref{eq:probcoherence} with the bounds on complexity of \ac{DBZ}, which we determine in Section \ref{ssec:complexity}, to characterize the limits of kinematic motion that \ac{DBZ} is capable of performing.  We must choose sample window lengths, $\boldsymbol\eta$, at each level, $h$, such that  

%
%
%


\noindent 
\begin{align}\label{eq:etamaxlimit}
[\boldsymbol\eta]_h < N^\textnormal{a}_h.
\end{align}
Offline numerical methods provide a means to select elements of $\boldsymbol\eta$ that meet the criteria of \eqref{eq:etamaxlimit}.  Our following lemma establishes guarantees on correctness when we choose $[\boldsymbol\eta]_h$ properly.  \mrb{In the following lemma, we combine Lemma \ref{lem:confidenceempvar} with our new insights on event $\mathcal{A}_h$ to show confidence of correct beamforming vector selection with a \ac{ME}.}

\begin{lemma}{With the choice of $[\boldsymbol\eta]_h < N^\textnormal{a}_h$ such that $\P{\mathcal{A}_h} \leq \delta/(2H)$, under Assumptions 1 and 2, $\mathcal{B}_h$ and $\mathcal{A}_h$ for all $1\leq h\leq H$ hold with probability $1-\delta$.}\label{lem:confidence}
\end{lemma}
Our proof shows that $P(\mathcal{B}_h \cap \mathcal{A}_h, \forall 1\leq h\leq H) > 1-\delta$ over all time steps, $n$, and all arms, $i$.   
\begin{IEEEproof}
    We apply \eqref{eq:probempvar3} from Lemma \ref{lem:confidenceempvar} for one level, $h$, where each $(h,i)$ has  $N_{h,i}(\eta,n) = u \geq 2$ samples taken.  
Then, using a union bound over all levels,
    \begin{align}
    \P{\mathcal{B}_1^c\cup\cdots\cup\mathcal{B}_H^c}
 	&\leq \P{\mathcal{B}_1^c} + \cdots + \P{\mathcal{B}_H^c}\\   
    &\leq  \sum_{n = 1}^\infty\sum_{h= 1}^H\sum_{i:(h,i) \in \mathcal{I}_h}\sum_{u=1}^{n} 3\exp(-\beta(\eta,n,\delta))\\
    &\leq \sum_{h=1}^H\frac{\abs{\mathcal{I}_h}\delta}{2\mathcal{I}^H} = \frac{\delta}{2}\label{eq:deltaconfB}.
    \end{align}
With the appropriate choice of $\eta$, we combine \eqref{eq:deltaconfB} with 
\begin{align}\label{eq:deltaconfA}
	\P{\mathcal{A}_1^c\cup\cdots\cup\mathcal{A}_H^c}
 	&\leq \P{\mathcal{A}_1^c} + \cdots + \P{\mathcal{A}_H^c}\\
 	&\leq \sum_{h=1}^H \frac{\delta}{2H} = \frac{\delta}{2},
\end{align}
 from which we conclude that $P(\mathcal{B}_h \cap \mathcal{A}_h, \forall 1\leq h\leq H) > 1-\delta$.
\end{IEEEproof}

\subsection{Sampling Strategy Performance}\label{ssec:samplingstrategyanalysis}
\ac{DBZ} adapts the sampling and termination policy of~\cite{gabillon2012best} in order to zoom in. We adapt~\cite[Lemma 4, Lemma 2 and Corollary 1]{gabillon2012best} to show that at each level $h$, $(h,u(n))$ and $(h,\gamma(n))$ are good choices for sampling, where the policy is \emph{greedy} toward the termination criteria \eqref{eq:stoppingcriteria}. \ac{DBZ} differs from~\cite{gabillon2012best} in the confidence term \eqref{eq:confidenceterm} and exploration rate \eqref{eq:explorationrate}, which include the empirical variance, $\nu^2_{h,i}(\eta,n)$ \eqref{eq:empvariance}, and the total number of arms, $\calI^H$.\footnote{$\calI^H = \sum_h \abs{\mathcal{I}_h}$ is the total quantity of beamforming vectors participating in \ac{MAB} games at all levels, and is fixed for any codebook.}  The operation of \ac{DBZ} consists in playing \emph{independent} \ac{MAB} games at each level $h$ dictated by the pruning vector $\mathbf{p}$, hence each lemma extends to all choices of $\mathbf{p}$.  

\begin{lemma}\label{lem:samplingchoices}
Let $S(n)\in\set{u(n),\gamma(n)}$ denote the arm pulled at time step $n$.  At each time step $n \geq 2$,
\begin{align}
    S(n) &= u(n) \implies L_{h,u(n)}(\eta,n) \leq L_{h,\gamma(n)}(\eta,n),\label{eq:samplingchoicecase1}\\
    S(n) &= \gamma(n) \implies U_{h,u(n)}(\eta,n) \leq U_{h,\gamma(n)}(\eta,n),\label{eq:samplingchoicecase2}
\end{align}
and if $S(n) = (h,i)$ then 
\begin{align}\label{eq:samplingchoicecase3}
G_{h,\gamma(n)}(n) \leq 2D_{h,i}(\eta,n-1).
\end{align}
\end{lemma}
\begin{IEEEproof}
The proof requires basic handling of each case, as outlined in~\cite{gabillon2012best}, applied to a single level $h$.  We provide the detailed proof in Appendix \ref{ssec:pf_lem_samplingchoices} of the supplementary material.
\end{IEEEproof}

From Lemma \ref{lem:samplingchoices}, we provide an upper bound on $G_{h,\gamma(n)}(\eta,n)$ adapted from \cite[Lemma 2]{gabillon2012best}.  The upper bound allows us to derive an expression in the Section \ref{ssec:complexity} to describe the complexity of the \ac{DBZ} algorithm.
\begin{lemma}\label{lem:boundonG}
   On event $\mathcal{B}_h$, if $(h,i) \in \set{(h,u(n)),(h,\gamma(n))}$ at time step $n \geq 2$, then
   \begin{align}\label{eq:boundonG}
       G_{h,\gamma(n)}(\eta,n) &\leq \min\set{0,2D_{h,i}(\eta,n-1)-\Delta_{h,i}(n)}\notag\\ &+ 2D_{h,i}(\eta,n-1).
   \end{align}
\end{lemma}
\begin{IEEEproof}
Similar to Lemma \ref{lem:samplingchoices}, this proof requires bookkeeping to analyze each statement, as outlined in~\cite{gabillon2012best}, at level $h$.  We provide the detailed steps in Appendix \ref{ssec:pf_lem_boundonG} of the supplementary material.
\end{IEEEproof}


\subsection{Zooming In and Zooming Out}\label{ssec:zoominzoomout}
Because of the \ac{ME} motion \eqref{eq:kinematicmodel}, the unique maximum mean reward, $\mu_h^*(n)$, and paths within the reward structure change over time (see Assumptions \ref{asp:uniquemax} and \ref{asp:unimodality}). \mrb{Section \ref{ssec:dbz} describes the mechanics uses to adapt the beamwidth to compensate for \ac{ME} motion, but we must ensure correct decisions to zoom in or out.}
The following lemma, adapted from \cite[Lemma 1]{gabillon2012best} ensures an arm, $(h,i)$, will not be mistakenly chosen as an $\epsilon_h$-optimal arm and zoomed in on, under event $\mathcal{B}_h$.

\begin{lemma}\label{lem:cannotzoomin}
	If $\mathcal{B}_h$ holds, for any $(h,i) \notin \set{(h,i) : \mu_{h,i}(n) + \epsilon_h \geq \mu_h^*(n)}$, $G_{h,i}(\eta,n) \geq \epsilon_h$ for all $n\geq 2$.
\end{lemma}
\begin{IEEEproof}
\begin{align}
	G_{h,i}(\eta,n) &= \max_{j\neq i}U_{h,j}(\eta,n-1) - L_{h,i}(\eta,n-1)\\
	&\geq \max_{j\neq i} \mu_{h,j}(n) - \mu_{h,i}(n) \label{eq:in_pfconnotzoominln3}\\
	&= \mu_h^*(n) - \mu_{h,i}(n) > \epsilon_h, \label{eq:in_pfconnotzoominln4}
\end{align}
where \eqref{eq:in_pfconnotzoominln3} is from $\mathcal{B}_h$. \eqref{eq:in_pfconnotzoominln4} comes from the fact that $i \neq i^*$ and the definition of an $\epsilon_h$-optimal arm.
\end{IEEEproof}

\noindent  Say \ac{DBZ} terminates with $(h',i')$ from a previous level $h'<h$, and produces threshold $[\boldsymbol\kappa]_{h'+1} = L_{h',i'}(\eta,n-1) + A\epsilon_{h'}$.  Let $n$ and $n'$ represent time steps at levels $h$ and $h'$, respectively. Complementing Lemma \ref{lem:cannotzoomin}, our next lemma ensures that \ac{DBZ} zooms out according to changes in reward structure due to motion.

\begin{lemma}\label{lem:wontzoomout}
If $\mathcal{B}_h$ and $\mathcal{B}_{h'}$ hold, with threshold $[\boldsymbol\kappa]_{h'+1}$ in \eqref{eq:rsrpthresh} based on termination at level $h'$ with $(h',i')$, and $A \leq 1$, \ac{DBZ} zooms out correctly with probability greater than $1-\delta$ if $\nexists(h,i) \in \mathcal{I}_h$ such that $\mu_{h,i}(n) > \mu_{h',i'}(n') + \epsilon_{h'}$.
\end{lemma}
\begin{IEEEproof}
If \ac{DBZ} zooms out from level $h$, we have that $U_{h,\gamma(n)}(\eta,n-1) <[\boldsymbol\kappa]_{h'+1}$, and therefore
\begin{align}
	\mu_{h,i}(n) &\leq \max_{i:(h,i) \in \mathcal{I}_h} \mu_{h,i}(n)\label{eq:pfwontzoomout1}\\
	 &\leq U_{h,\gamma(n)}(\eta,n-1)\label{eq:pfwontzoomout2}\\
	  &< [\boldsymbol\kappa]_{h'+1}\label{eq:pfwontzoomout3}\\
	   &= L_{h',i'}(\eta,n'-1) + A\epsilon_{h'}\label{eq:pfwontzoomout4}\\ 
	   &\leq \mu_{h',I}(n') + \epsilon_{h'}\label{eq:pfwontzoomout5}.
\end{align}
The relationship of \eqref{eq:pfwontzoomout1} to \eqref{eq:pfwontzoomout2} and \eqref{eq:pfwontzoomout4} to \eqref{eq:pfwontzoomout5} come from event $\mathcal{B}_h$ and $\mathcal{B}_{h'}$, respectively, where  Lemma \ref{lem:confidenceempvar} 
 shows both events hold with probability greater than $1-\delta$.
\end{IEEEproof}
Together, Lemmas \ref{lem:cannotzoomin} and \ref{lem:wontzoomout} show that despite the time-varying mean rewards, \ac{DBZ} will correctly zoom in and out with at least probability $1-\delta$ under event $\mathcal{B}_h$ for all $h$.

\subsection{Sample Complexity}\label{ssec:complexity}
We now provide an analysis of the sample complexity of \ac{DBZ} to zoom in at each level.  The sample complexity enables setting sample window lengths $\boldsymbol\eta$ large enough to accommodate the number of samples to zoom in \eqref{eq:stoppingcriteria}. For zooming in, \ac{DBZ} exploits the structure induced by the hierarchical codebook by reducing the overall total number of arms considered along a path (see Assumption \ref{asp:unimodality}), $\mathcal{I}^H$, by order of the logarithm of the number of beamforming vectors required in traditional \ac{MAB} strategies using the narrowest beams \cite{wu2019fast,hashemi2018efficient}.  The reduction in beamforming vectors considered directly attributes to an overall reduction in sample complexity. However, the variance and individual spacing between mean rewards of arms also play a significant role in determining overall complexity.  Let $\Delta_{h,i,\epsilon}(n) \eqdef \max\set{(\Delta_{h,i}(n) + \epsilon_h)/4,\epsilon_h/2}$, and 
\begin{align}\label{eq:sampleH}
\begin{split}
\mathlarger{\aleph}_{h,\epsilon}(n) &\eqdef \sum_{i : (h,i) \in \mathcal{I}_h} \frac{2B\nu^2_{h,i}(n) + 2\sqrt{2BC}\Delta_{h,i,\epsilon}(n)}{\Delta_{h,i,\epsilon}^2(n)} \\
&\quad + \frac{\sqrt{4B^2\nu^4_{h,i}(n) + 2\sqrt{2C}B^{3/2}\nu^2_{h,i}(n)\Delta_{h,i,\epsilon}(n)}}{\Delta_{h,i,\epsilon}^2(n)}.
\end{split}
\end{align}
The relevance of $\mathlarger{\aleph}_{h,\epsilon}(n)$ is justified by the following lemma.
\begin{lemma}\label{lem:upperboundsamples}
If $\mathcal{B}_h$ holds, \ac{DBZ} ensures that the number of samples of beamforming vector $(h,i)$ after $n$ total samples at level $h$, satisfies
\begin{align}\label{eq:upperboundsamples}
\begin{split}
    N_{h,i}(\eta,n) \leq &\frac{2B\nu^2_{h,i}(n) + 2\sqrt{2BC}\Delta_{h,i,\epsilon}(n)}{\Delta_{h,i,\epsilon}^2(n)}\beta(\eta,n-1,\delta) \\
    &+ \frac{\sqrt{4B^2\nu^4_{h,i}(n) + 2\sqrt{2C}B^{3/2}\nu^2_{h,i}(n)\Delta_{h,i,\epsilon}(n)}}{\Delta_{h,i,\epsilon}^2(n)} \\
    &\times \beta(\eta,n-1,\delta) + 2,
\end{split}
\end{align}
or rounded to the next largest integer, $N^*_{h,i}(\eta,n) = \ceil{N_{h,i}(\eta,n)}$.
\end{lemma}
\begin{IEEEproof}
The proof involves writing \eqref{eq:boundonG} and replacing with our expression for $D_{h,i}(\eta,n-1)$ in \eqref{eq:confidenceterm}, then solving for $N_{h,i}(\eta,n)$.  The full steps are available in Appendix \ref{ssec:pf_lem_upperboundsamples} of the supplementary material.
\end{IEEEproof}
%
%


\subsection{Configuring DBZ}\label{ssec:configuration}
This section provides \mrb{\ac{DBZ} users with a} practical methodology for selecting a sample window length, $\boldsymbol\eta$, pruning vector, $\mathbf{p}$, and parameter $\epsilon$. \mrb{To set $\boldsymbol\eta$ and $\mathbf{p}$, we use Lemma \ref{lem:upperboundsamples} that describes} the total number of samples required at each level, $N^*_{h,i}(\eta,n)$, \mrb{which} scales directly with noise variance, $\sigma_v^2$.  In order for \ac{DBZ} to take sufficient samples such that it meets either the criteria of \eqref{eq:stoppingcriteria} or \eqref{eq:zoomoutcriteria}, we require elements of $\boldsymbol\eta$ large enough, such that for a single element $\eta$,
\begin{align}\label{eq:etalowerrequirement}
\eta \geq \sum_{i:(h,i)\in\mathcal{I}_h}N^*_{h,i}(\eta,n).
\end{align}
Note that $\sum_{i:(h,i)\in\mathcal{I}_h}N_{h,i}(\eta,n) = \eta$ if $n \geq \eta$ and $n$ otherwise.  We obtain an estimate of how to set $\eta$ by further analyzing \eqref{eq:upperboundsamples}, where the total number of samples required at each level is
\begin{align}
    \eta = \sum_{i:(h,i)\in\mathcal{I}_h}N_{h,i}(\eta,n) \leq \mathlarger{\aleph}_{h,\epsilon}(n)\log\left(\frac{15N_H\eta^4}{2\delta}\right) + 2\abs{\mathcal{I}_h}
\end{align}
and has the closed form solution to suggest the value,
\begin{align}\label{eq:lambertsamples}
\eta_\textnormal{est}= \ceil*{   -4\mathlarger{\aleph}_{h,\epsilon}(n)    W\left(  -\frac{  \exp\left(-\frac{2\abs{\mathcal{I}_h}-1}{4\mathlarger{\aleph}_{h,\epsilon}(n) }\right)}   {  4\mathlarger{\aleph}_{h,\epsilon}(n)  \left(\frac{15\mathcal{I}^H}{2\delta}\right)^{1/4}   }\right)} + 1,
\end{align}
where $W(\cdot)$ is the Lambert-W function.\footnote{The Lambert-W function enables the relation $x_h\eta \geq\log(y_h\eta) \iff \eta \leq -\frac{1}{x_h}W\left(-\frac{x_h}{y_h}\right)$, where $x_h = 1/(4\mathlarger{\aleph}_{h,\epsilon}(n))$ and $y_h = (15\mathcal{I}^H/(2\delta))^{1/4}\exp\left((2\abs{\mathcal{I}_h}-1)/(4\mathlarger{\aleph}_{h,\epsilon}(n) )\right)$.}  The sample window length should be chosen such that 
\begin{align}\label{eq:samplemincomplexity}
	[\boldsymbol\eta]_h \geq \eta_{\textnormal{est}}
\end{align}
In cases of extreme motion with very large $\sigma_u$ and/or especially low \ac{SNR} with large $\sigma_v$, we conclude that \ac{DBZ} delivers poor performance. When $\eta_\textnormal{est} > N_h^\textnormal{a}$, \ac{DBZ} cannot guarantee selection of $\epsilon$-optimal beamforming vectors with at least probability $1-\delta$.  \mrb{For practical implementation, a user should choose $\sigma_u$ in \eqref{eq:sigman} such that it approximates the highest angular acceleration possible by the \ac{ME} intended to track.}

We perform optimization of the pruning vector, $\mathbf{p}$, in an offline manner to optimize utilization of the beamforming codebook.  We show in Section V that the choices of $\mathbf{p}$ generalize over a broad range of \ac{SNR}.   Minimizing \eqref{eq:lambertsamples} over the range of possible path angles provides an assessment of which pruning vector, $\mathbf{p}$, is optimal.  We require the expected number of samples at level $h$, $\E[\theta_1]{\eta_\textnormal{est}(\theta)}$. ``Averaging'' over the range of angles $\Phi$ eliminates dependence on the angle. Furthermore, the sparsity of the \ac{mmWave} channel allows us to focus on the dominant path, $\theta_1$ \cite{rappaport2013millimeter}.  The vector $\mathbf{p}^*$ minimizes the average complexity, such that 
\begin{align}\label{eq:optimizecomplexity}
    \mathbf{p}^* = \argmin_{\mathbf{p}}\E[\theta_1]{\sum_{h:[\mathbf{p}]_h =1}\eta_\textnormal{est,h}(\theta)},
\end{align}
and we estimate the expected number of samples, 
\begin{align}
\E[\theta_1]{\sum_{h:[\mathbf{p}]_h =1}\eta_\textnormal{est,h}(\theta)},
\end{align}
numerically. Our numerical simulations in the next section show the samples required for initial alignment with different choices of $\mathbf{p}$ for comparison. \mrb{We include example code for computing $\boldsymbol\eta$ and $\mathbf{p}$ in our source code \cite{mlcommrepo}.}


\ac{DBZ} uses the parameter $\epsilon$ to compensate for cases with especially small $\Delta_{h,i}(n)$, when two mean rewards are very close in value.  The case of small $\Delta_{h,i}(n)$ occurs when $\theta_1 \approx \bar{\phi}_i \pm \phi_{\textnormal{bw},h}/2$ \mrb{or \ac{NLOS} scenarios where there is no clear dominant path}, causing the \ac{RSRP} \eqref{eq:rsrp} of \mrb{multiple} beamforming vectors to be very similar.
%
As a reminder, the $\epsilon$ parameter in the termination criteria allows \ac{DBZ} to terminate with a sub-optimal arm $(H,i)$, such that $\mu_H^* \leq \mu_{H,i} + \epsilon$. The sub-optimal choice impacts the relative spectral efficiency with respect to $\epsilon$ as
\begin{align}\label{eq:relativespectralefficiency}
    \tilde{\xi}_{h,i} \eqdef\frac{\log_2\left(1 +(\zeta_{h,i^*}(n)-\epsilon)/\sigma_v^2\right)}{\log_2\left(1 +\zeta_{h,i^*}(n)/\sigma_v^2\right)}
\end{align}
for $h = H$ and for all $\epsilon > 0$.  We set $\epsilon$ such that $\tilde{\xi}_{h,i} > .95$ 
 for all $h$. \mrb{In practice, $\zeta_{h,i^*}(n)$ corresponds to some maximum \ac{RSRP}, while $\epsilon$ denotes the penalty allowed with communication persisting.}  We note that choosing an $\epsilon$-optimal arm is unique to \ac{DBZ} compared to existing algorithms \cite{wei2022fast,liu2022adaptive} that fall victim to high complexity with small $\Delta_{h,i}(n)$.  With $\epsilon_h$, we use our scaling of $\epsilon$ with respect to the gain at level $h$, $g^{-(H-h)}$, for each subsequent level of the hierarchical beamforming codebook $\mathcal{F}^H$. We expect $\Delta_{h,i}(n)$ to be smaller at lower levels\mrb{, or overall in \ac{NLOS} scenarios}.  By scaling $\epsilon$ to $\epsilon_h$  for the corresponding level, $h$, we ensure that there is no unnecessarily high penalty to relative spectral efficiency incurred for our beamforming vector selection at termination.

\section{Numerical Simulations}\label{sec:sims}
Our numerical simulations assess the \ac{ISAC} performance of \ac{DBZ} to quickly align, i.e., choose a beamforming vector at level $H$, and adjust the beamforming pattern width over time to compensate for motion while communicating.  Our simulation source code is available at \cite{mlcommrepo}.

\subsection{Methodology for Initial Alignment Simulations}\label{ssec:methodologyia}
We execute each simulation by first making $K$ uniformly random selections $\theta_k \in \Phi$, each representing the $k^\textnormal{th}$ path.  We use a unique random number generator seed for each individual simulation that we denote with index $\ell$. The $K$-length vector of angles chosen for simulation $\ell$ is denoted $\boldsymbol\theta_\ell$ with a corresponding vertex $(H,i^*_\ell)$.  We use $\boldsymbol\theta_\ell$ to then construct the array response \eqref{eq:arrayresponse}.  We take samples by applying beamforming vectors to the channel model observations, as in \eqref{eq:rsrp}, that are chosen based on the algorithm policy.  Each simulation terminates after the stopping criteria \eqref{eq:stoppingcriteria} is met.  We compare the performance of \ac{DBZ} across several \ac{SNR} values with various pruning vectors, $\mathbf{p}$ (which we denote by their decimal values), and directly with \ac{HPM} from \cite{chiu2019active} and \ac{2PHTS} from \cite{wei2022fast}.\footnote{Note that there is some degradation at high \ac{SNR} for HPM \cite{chiu2019active} due to not perfectly compensating for the multi-path effects.  Additionally, we could only simulate the behavior of \ac{2PHTS} in the high-\ac{SNR} regime because of numerical issues intrinsic to the algorithm.}  The \ac{HPM} algorithm acts a baseline of performance in utilizing perfect channel knowledge in the posterior computations to deploy the hierarchical codebook. Another potential comparison candidate algorithm, \ac{HBA}, aggressively searches the range of $\Phi$, sacrificing performance under lower-\ac{SNR} conditions to terminate quickly. \ac{2PHTS} adapts the state-of-the-art \ac{TAS} \ac{MAB} framework using an approximation of the stochastic channel model that works for high \ac{SNR}.  We dynamically determine the number of total simulations required, $L$, by utilizing the Wilson score \cite{wilson1927probable} interval width.  Further details of the confidence intervals are available in Appendix \ref{sec:additionalsimulation} of the supplementary material.  Let $T_h(\ell)$ denote the samples required for level $h$ in simulation $\ell$, the average sample complexity, or number of beamforming vectors required, is
\begin{align}\label{eq:samplecomplexityoversims}
\hat{T}(L) = \frac{1}{L}\sum_{\ell=1}^L\sum_{h:[\mathbf{p}]_h =1}T_h(\ell).
\end{align}
For the initial alignment performance, algorithms utilize a common beamforming codebook with $H = 7$ levels ($128$ pointing angles at the finest resolution) organized by a binary tree graph with $M = 128$ antenna elements in a \ac{ULA}. 
We design the beamforming architecture to support as few as a single \ac{RF} chain in a \ac{HAD} configuration based on the design in \cite{alkhateeb2014channel}.  The gain parameter is set as $g = 10^{.2}$ which corresponds to $2$~dB of gain per level with the increasingly narrow beams.  We fix $P_k = 1$ and assume no knowledge of the channel \ac{SNR}. We also do not use any knowledge of the channel fading factors, $\alpha_k(n)$ \eqref{eq:channelresponse}, in \ac{DBZ}.  Our results show that \ac{DBZ} is robust to the time-varying $\alpha_k(n)$.  The sequence $\mathbf{p}$ is chosen as decimal values $p_{\textnormal{dec}} \in \set{0,3,4,7,8}$, identified with the methodology in  Section \ref{ssec:configuration} to be a good set of $\mathbf{p}$ to compare.  We summarize the details of each selection of $\mathbf{p}$ in Table \ref{tab:pvals}.  We use the convention ``\texttt{<algorithm>}$p_\textnormal{dec}$", i.e., \ac{DBZ}$7$, to indicate the algorithm and selection of pruning vector.

\begin{table}
\begin{center}
\caption{Details on pruning vector values, $\mathbf{p}$.}
\label{tab:pvals}
\centering
\begin{tabularx}{\columnwidth}{  >{\hsize=0.8\hsize\linewidth=\hsize\centering\arraybackslash}X X X}
$p_{\textnormal{dec}}$ & $\mathbf{p}$ & $h$ \textbf{Traversed} \\ 
\hline\hline
$0$ & \texttt{0000001} & $7$ \\
$3$ & \texttt{0000111} & $5,6,7$ \\
$4$ & \texttt{0001001} & $4,7$ \\
$7$ & \texttt{0001111} & $4,5,6,7$ \\
$8$ & \texttt{0010001} & $3,7$ \\
$63$ & \texttt{1111111} & All $h$ \\
\hline
\end{tabularx}
\end{center}
\end{table}

\subsection{Results and Discussion}\label{ssec:resultsdiscussion}

Our initial alignment experiments investigate the overall complexity \eqref{eq:samplecomplexityoversims}, which is the key metric for the fixed confidence best arm identification setting. We emphasize that the \ac{ME} is NOT mobile during these initial alignment simulations, as to have a fair comparison with other algorithms.  We also verify that the expected relative spectral efficiency \mrb{after $n$ samples and the algorithm chooses a beamforming vector},
\begin{align}\label{eq:relativespectralefficiencysim}
\mrb{\xi(n,L) \eqdef \frac{1}{L}\sum_{\ell =1}^L \frac{\log_2\left(1 +\zeta_{h,i^c}(n)/\sigma_v^2\right)}{\log_2\left(1 +\zeta_{H,i^*}(n)/\sigma_v^2\right)},}
\end{align}
is obtained after obtaining samples with chosen $\epsilon$. \mrb{Fig. \ref{fig:samplesvsnr} and \ref{fig:rsevsnr} provide a comparison of the sample complexity and resulting relative spectral efficiency for several \ac{mmWave} beam alignment algorithms:}
\begin{itemize}
\item \mrb{\ac{DBZ} with several configurations of the pruning vector, $\mathbf{p}$, along with two values of $\epsilon$.}
\item \mrb{\ac{HPM} from \cite{chiu2019active} utilizing perfect \ac{CSI} of both the channel fading coefficient, $\alpha_1(n)$, and noise variance, $\sigma_v^2$.}
\item \mrb{\ac{HBA} from \cite{wu2019fast}, which bisects the search space according to the \ac{MAB} policy in \cite{bubeck2011x}.}
\item \mrb{\ac{HOSUB} from \cite{blinn2022mmwave} that has operates as a fixed-budget (or fixed number of samples) algorithm using the \ac{MAB} framework in \cite{combes2014unimodal} to explore the hierarchical codebook graph.  We show the performance with two different budget constraints, $50$ and $100$.}
 \end{itemize}
As anticipated, \ac{HPM} provides a baseline for performance in that it optimally exploits the induced structure by using \ac{CSI} to compute the posteriors at each time step. In general, we anticipate many of the algorithms that compute the explicit distributions \cite{noh2017multi,liu2022adaptive} offer similar performance, but with the price of significant computational overhead to compute the posteriors.  At very low \ac{SNR}, an exhaustive search (\ac{DBZ}0) outperforms any other \ac{DBZ} variation in Fig. \ref{fig:samplesvsnr}.  This is expected, in fact, works such as \cite{wei2022fast} hinge on the assumption of exclusively operating in a high-\ac{SNR} regime.  Our results show that values of \ac{SNR} roughly between $-6$ to $6$ are the target \ac{SNR} regimes in which \ac{DBZ} achieves better complexity than an exhaustive search.  \mrb{Fig. \ref{fig:rsevsnr} shows a significant reduction in relative spectral efficiency at low \ac{SNR} for \ac{HBA} and \ac{HOSUB}, which both sacrifice some performance for lower complexity, shown in Fig \ref{fig:samplesvsnr}.  \ac{DBZ} lowers its complexity by utilizing larger values of $\epsilon$, however, there is a corresponding loss in relative spectral efficiency shown in Fig. \ref{fig:rsevsnr}.}

\begin{figure}[h]
\includegraphics[scale = .43]{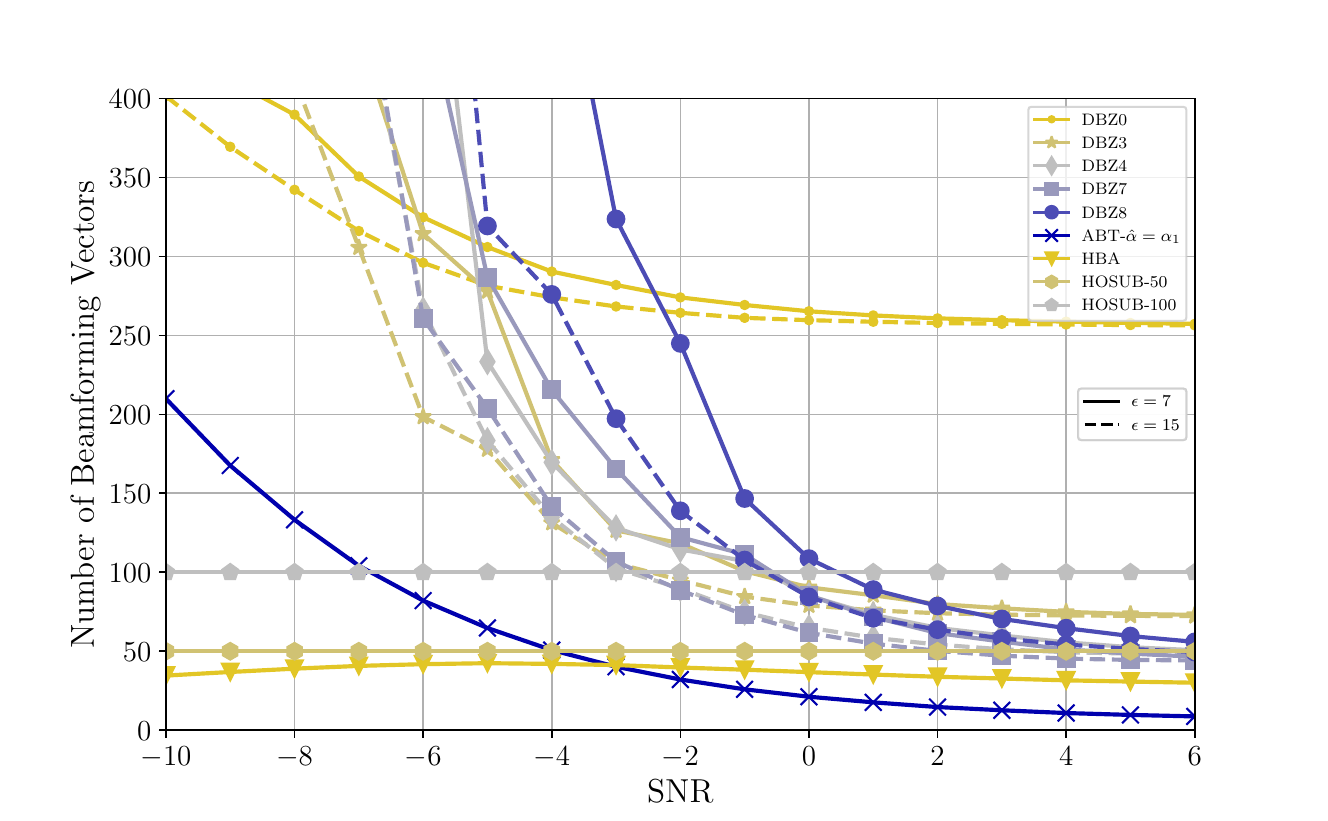}
\centering
\caption{Comparison of complexity at various \ac{SNR}.}
\label{fig:samplesvsnr}
\end{figure}

\begin{figure}[h]
\includegraphics[scale = .43]{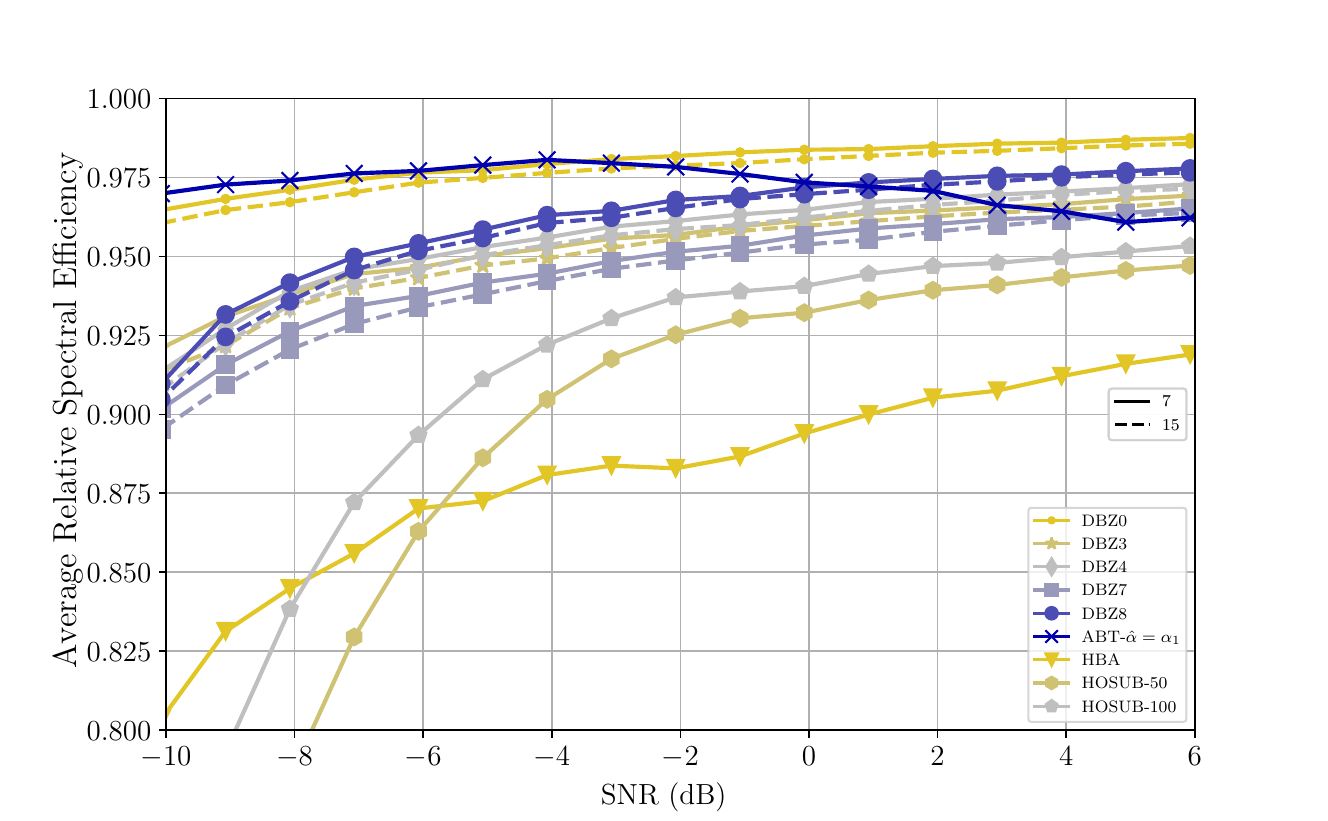}
\centering
\caption{\mrb{Comparison of resulting relative spectral efficiency at various \ac{SNR}.}}
\label{fig:rsevsnr}
\end{figure}

We provide one last numerical result for initial alignment at \ac{SNR} $= 20$~dB (a high \ac{SNR} regime) to compare the performance of \ac{DBZ} to that of \ac{TAS} methods in Table \ref{tab:highsnrcomplexity}.  We adapt the \ac{TAS} framework in \cite{garivier2021nonasymptotic} for identifying an $\epsilon$-optimal arm, and similar to \cite{wei2022fast}, apply \ac{TAS} over subsequent levels, $h$, as shown in Table \ref{tab:pvals}.  We also use the assumption in \cite{wei2022fast}, in which the observation is close to a Heteroscedastic Gaussian to compute the relative entropy in the \ac{TAS} steps, and apply our scaling of $\epsilon$ as $\epsilon_h$ at each level. As expected, \ac{HPM} achieves the best results given the full \ac{CSI}.  A particular point of interest is that strategies considering fewer arms, $\mathcal{I}^H$, perform significantly better at high \ac{SNR}.  One concludes that at high \ac{SNR}, $\mathbf{p}$ should be chosen to minimize $\calI^H$.  While \ac{TAS} methods perform better overall, the number of samples for \ac{DBZ}63 and \ac{DBZ}31 are only marginally worse than \ac{TAS}63 and \ac{TAS}31.  
\mrb{Some of the algorithms in Table \ref{tab:algcomparisonsT} do not have an explicit initial alignment component or guarantees on accuracy in the algorithm, \cite{zhang2020beam,va2016beam,chung2020adaptive}.  For the tracking simulations in the next subsection, we assume \cite{zhang2020beam,va2016beam,chung2020adaptive} incur an $O(I)$ sample complexity with an exhaustive search of all narrow beams to initially align. There may be space in future work to combine initial alignment approaches like \ac{DBZ}, \ac{HPM}, or \ac{2PHTS} with \cite{va2016beam,chung2020adaptive} to enhance algorithm performance.}

\begin{table}
\begin{center}
\caption{Complexity in high \ac{SNR} regime.}
\label{tab:highsnrcomplexity}
\centering
\begin{tabularx}{\columnwidth}{  >{\hsize=0.8\hsize\linewidth=\hsize\centering\arraybackslash}X X X X X X X X X} 
\textbf{DBZ4} & \textbf{DBZ7} & \textbf{DBZ31} & \textbf{DBZ64} & \textbf{TAS4} & \textbf{TAS7} & \textbf{TAS31} & \textbf{TAS64} & \textbf{HPM}\\ 
\hline\hline
57.0 & 53.7 & 30.4 & 30.8 & 26.0 & 26.0 & 20.0 & 21.0 & 8.2\\
\hline
\end{tabularx}
\end{center}
\end{table}

\subsection{Methodology for Tracking Simulations}\label{ssec:methodologytrack}
We provide a series of numerical simulations to demonstrate the performance of \ac{DBZ} under different channel \ac{SNR}, $\sigma_v$, and magnitude of motion, $\sigma_u$ (and more extensively in Appendix \ref{sec:additionalsims} of the supplementary material).  We fix the interval in which samples are taken, $\tau = 1$, and execute each simulation by first choosing $\theta_k(1)$ as in our initial alignment simulations, according to a random number generator seed, $\ell$.  We take a single sample \eqref{eq:complexmeasurement} at each time step by applying beamforming vectors to the channel model observations, as in \eqref{eq:rsrp}, \mrb{that are chosen based on the flowchart in Fig. \ref{fig:dbz_flowchart}}.  After each sample the \ac{ME} undergoes the kinematic motion transition in \eqref{eq:kinematicmodel}.  We execute the main algorithm loop for \ac{DBZ} (\mrb{After input and parameter initialization in Fig. \ref{fig:dbz_flowchart}}) until a specified number of time steps occur, $N$. We perform $L$ simulations of $N$ time steps, and calculate the average relative spectral efficiency at each time step, $n$, \eqref{eq:relativespectralefficiencysim}.
The indices \mrb{$(h,i^c)$} in the numerator of \eqref{eq:relativespectralefficiencysim} corresponds to the beam currently being used for communication.  \ac{DBZ} adapts to the changing $\theta(n)$ by broadening (zooming out) and narrowing (zooming in) the beam used to communicate on the events in lines 21 and 5, respectively.  
We use a ternary hierarchical codebook with $\mathcal{I}_1 = 5$, with depth $H = 4$, and each beam splits into $3$ narrower beams, creating $135$ narrow beams at $h = H$.  We use $\mathbf{p} = [1,1,1,1]$ for all tracking simulations.  The degradation of each algorithm's performance at later time steps comes from the \ac{ME} possibly accelerating to reach faster speeds \eqref{eq:kinematicmodel}, making that tracking task more difficult over time.

\subsection{\mrb{Comparison of Algorithms}}\label{ssec:comparison_algs}
\mrb{We provide a performance comparison across several algorithms for \ac{mmWave} beam tracking by assessing the time-average relative spectral efficiency, of (72),} 
\begin{align}
\mrb{\frac{1}{N}\sum_{n=1}^N\xi(n,L).}
\end{align}
\mrb{In particular, we use the following algorithms for comparison in simulation:}
\begin{itemize}
\item \mrb{The \ac{ABT} algorithm from \cite{ronquillo2021active,ronquillo2023integrated}, which acts as our baseline algorithm by exploiting full \ac{CSI} and knowledge of the \ac{ME} motion to compute the Bayesian posteriors and select beamforming vectors.  We assess two variations, one in which the fading coefficient, $\alpha_1(n)$, is known and one variation that uses a noisy estimate of $\alpha_1(n)$.}
\item \mrb{\ac{PF} approach from \cite{chung2020adaptive}, which uses the covariance of the particles to broaden or narrow the beam by activating a specified number of antenna elements. To offer a more fair comparison, we assess the effective beamwidth produced by the number of elements, and use a level in our ternary codebook $\calF^H$ that most closely matches the effective beamwidth.}
\item \mrb{\ac{MAB} approach in \cite{zhang2020beam}, which periodically sweeps neighboring ``offset'' narrow beams in a different type of \ac{MAB} application.}
\item \mrb{\ac{EKF} approach in \cite{va2016beam}, where we use the angle estimations to select the narrow beamforming vectors.}
\end{itemize}
\mrb{Our implementation of each algorithm is in the source code \cite{mlcommrepo}. Each algorithm has different trade-offs with respect to the characteristics listed in Table \ref{tab:algcomparisonsT}.  Fig. \ref{fig:compare_algs_los} and \ref{fig:compare_algs_nlos} show the performance of each algorithm in \ac{LOS} and \ac{NLOS} scenarios with different severity of motion, $\sigma_u$.  In our \ac{LOS} scenario, the dominant path is $10$~dB above the others, where the \ac{NLOS} has no clear dominant path.  We see that \ac{DBZ} outperforms all other algorithms except \ac{ABT} \cite{ronquillo2021active}, as expected.  The adaptive beamwidth control for the \ac{PF} approach, \cite{chung2020adaptive}, allows for better performance than the offset sweep in \cite{zhang2020beam} or the \ac{KF} in \cite{va2016beam}.  However, the adaptive beamwidth control for \ac{DBZ} exceeds that of the \ac{PF}.  Combination of the \ac{PF} or \ac{KF} with \ac{DBZ} could yield a potent algorithm for \ac{mmWave} tracking.}
     \begin{figure}[h]
		\includegraphics[width = .5\textwidth]{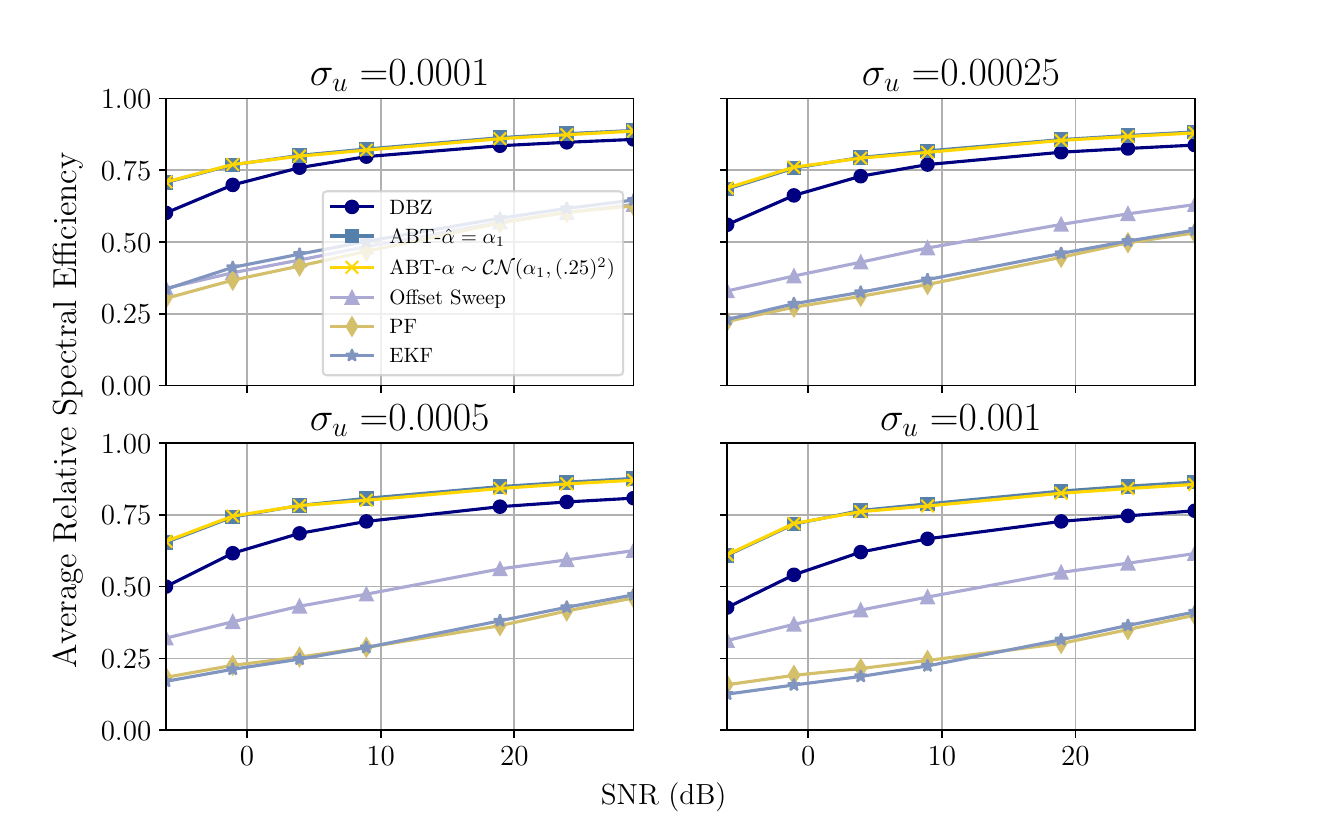}
		\centering
		\caption{\mrb{Comparison of performance between algorithms in \ac{LOS} scenario.}}
		\label{fig:compare_algs_los}
	\end{figure}
     \begin{figure}[h]
		\includegraphics[width = .5\textwidth]{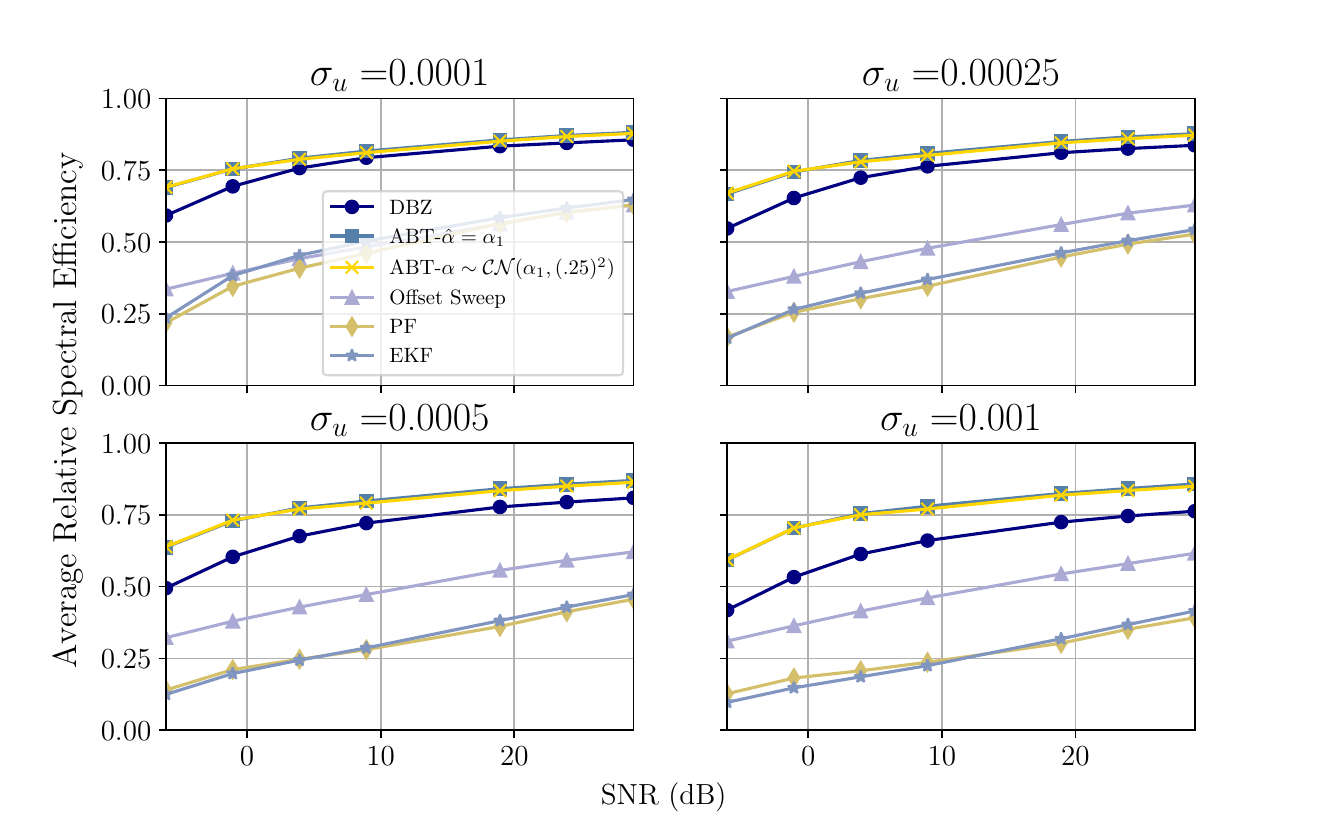}
		\centering
		\caption{\mrb{Comparison of performance between algorithms in \ac{NLOS} scenario.}}
		\label{fig:compare_algs_nlos}
	\end{figure}
\mrb{In the next section, we take a closer look at the results of comparing \ac{DBZ} with the Bayesian algorithm, \ac{ABT} from \cite{ronquillo2021active}.  We show there are instances where \ac{DBZ} indeed performs better if \ac{ABT} does not have access to exquisite channel information, failing to be \ac{CSI} adaptive.}

\subsection{Comparison to Bayesian Method}\label{ssec:comparisonbayesiannumerical}

Our first experiment compares \ac{DBZ} performance with an extension of \ac{HPM} to compensate for motion, \ac{ABT} \cite{ronquillo2021active,ronquillo2023integrated}\mrb{. We use \ac{ABT} as a baseline of performance, the Bayesian framework} leverages full channel information to compute the posteriors on beamforming vectors $\mathbf{f}_{H,i}$ after each observation.  The broader beams posteriors are the sum of the posteriors for narrower ones.  The framework in \cite{ronquillo2023integrated} shows a way to optimize \ac{RS} (pilot signals) or data sent based on optimizing spectral efficiency. However, for this comparison, we fix the interval in which \ac{RS} are sent.  We calculate the posterior for the measurement \eqref{eq:complexmeasurement}, which is corrupted by \ac{AWGN} under multiplicative fading coefficient $\alpha_1(n)$.  We also apply the density for the entity motion, (see \eqref{eq:pdfcoherence} in Appendix \ref{sec:deriveprobcoherence} of the supplementary material), to the posterior.  Fig. \ref{fig:performancecompareabt} shows our results for \ac{DBZ} using sample windows set by the estimated complexity, $\eta_\textnormal{est}$, with a subset of \ac{SNR}s compared with the performance of \ac{ABT}.  \ac{ABT} performs extremely well when the fading coefficient, $\alpha_1(n)$, is used in the computation of the posterior.  In practice, this comes from a method to make a precise estimate of the coefficient. To assess performance when the estimation is in slight error, we choose the fading coefficient as a random variable distributed as $\mathcal{CN}(\alpha_1(n),(.25)^2)$.  We see a slight degradation in performance with the error in estimation of the fading coefficient.  What may be more interesting however, is the performance disparity between high and low \ac{SNR} (\ac{SNR}$= 14$ versus \ac{SNR}$=-6$), in that one would expect better performance at higher \ac{SNR}, but the opposite is shown in Fig. \ref{fig:performancecompareabt}.  This is due to the imperfect posterior computed at high \ac{SNR} creating ``overconfidence'' in the selection of narrower beamforming vectors.  At lower \ac{SNR}, \ac{ABT} is more discerning (broader distributions) in its choices to narrow or broaden the beam, hence there is less emphasis on accurate estimations of $\alpha_1(n)$.  We see \ac{DBZ} is competitive with \ac{ABT} given no channel knowledge other than the \ac{SNR} to compute the sampling window lengths, $\boldsymbol\eta$. \ac{DBZ} only requires $O(\abs{\mathcal{I}_h}=3)$ \ac{FLOP}s ($O(\abs{\mathcal{I}_1} = 5)$ in our case) versus \ac{ABT} with $O(128)$ \ac{FLOP}s (with the codebook used), for each of the algorithm's computational cost at each sample.  The $O(128)$ in \ac{ABT} comes from the need to update each posterior for each beamforming vector at each sampling iteration with the binary codebook used therein.\footnote{The relative spectral efficiency metric normalizes any differences in codebook selection between the two algorithms} 

\begin{figure}[h]
\includegraphics[scale = .43]{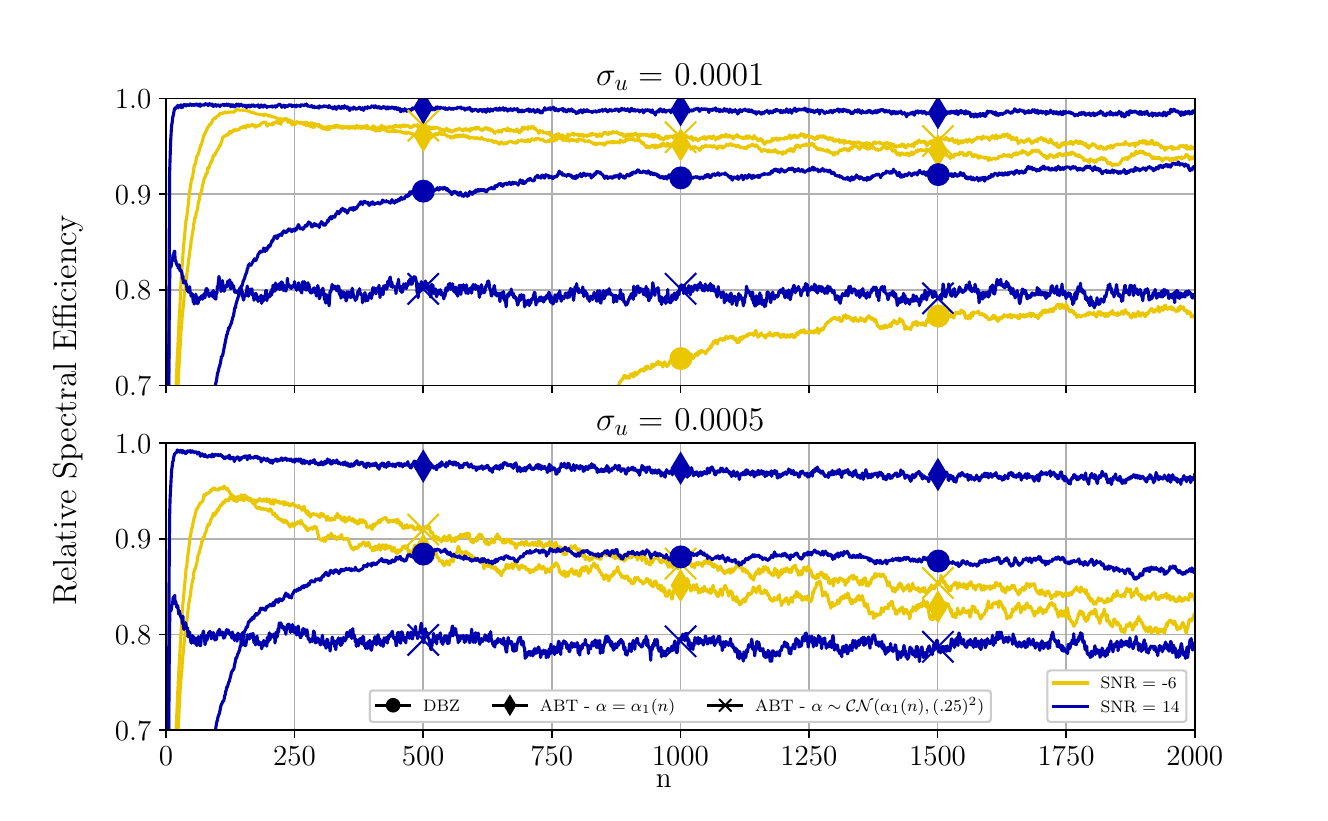}
\centering
\caption{Comparing performance between \ac{ABT} with full channel knowledge and \ac{DBZ} with $\eta$ set by estimated complexity.}
\label{fig:performancecompareabt}
\end{figure}

\subsection{Performance in NYU Sim Model}\label{ssec:nyusimruns}

The question of \ac{DBZ} performance in the presence of realistic multi-path environments remains.  We use NYU Sim \cite{ju2018nyusimmotion,sun2017nyusim} to generate $L = 100$ unique spatially consistent trajectories of a moving entity within $\Phi$.  
In using NYU Sim, we compute the estimated relative spectral efficiency \eqref{eq:relativespectralefficiencysim} over each trajectory consisting of $N = 600$\footnote{This number of timesteps worked out to be an integer number with the actual time, in seconds, between time steps $n$ and the length of the track. See Appendix \ref{sec:nyusimparameters} in the supplementary material.} time steps and create an average result for each scenario and track, i.e. \texttt{UMa: Linear}.  We provide the full list of parameters to configure NYU Sim in Appendix \ref{sec:nyusimparameters} of the supplementary material.  \mrb{\ac{DBZ} (and other algorithms alike) struggle to handle drastic large-scale fading and outage models where there may be variations of up to $\sim 50$~dB of power between each time step.}  This is especially  true in urban cases (UMa and UMi).  We assume an analog front end that applies an \ac{AGC} mechanism, which we model here as normalization of the channel vector \eqref{eq:channelresponse}, $\sqrt{M}\mathbf{h}(n)/\abs{\mathbf{h}(n)}$.  The urban cases still see swings in receive power that would be indicative that severe multi-path is present, despite normalization.  We see \ac{DBZ} performs relatively well in all scenarios.  In particular, the rural scenarios, \texttt{RMa}, \ac{DBZ} matches or exceeds it's performance against the \ac{DWNA} motion model.  The severe multi-path elements in the urban scenarios cause edge cases of the induced structure described in Section \ref{ssec:induced_mean_reward_structure}, where perturbations induced by noise, even small, cause significant degradation in performance.  The consistent spikes and valleys in spectral efficiency at specific time steps come from using the same track, which is especially true in the hexagonal track case.

\begin{figure}[h]
\includegraphics[scale = .43]{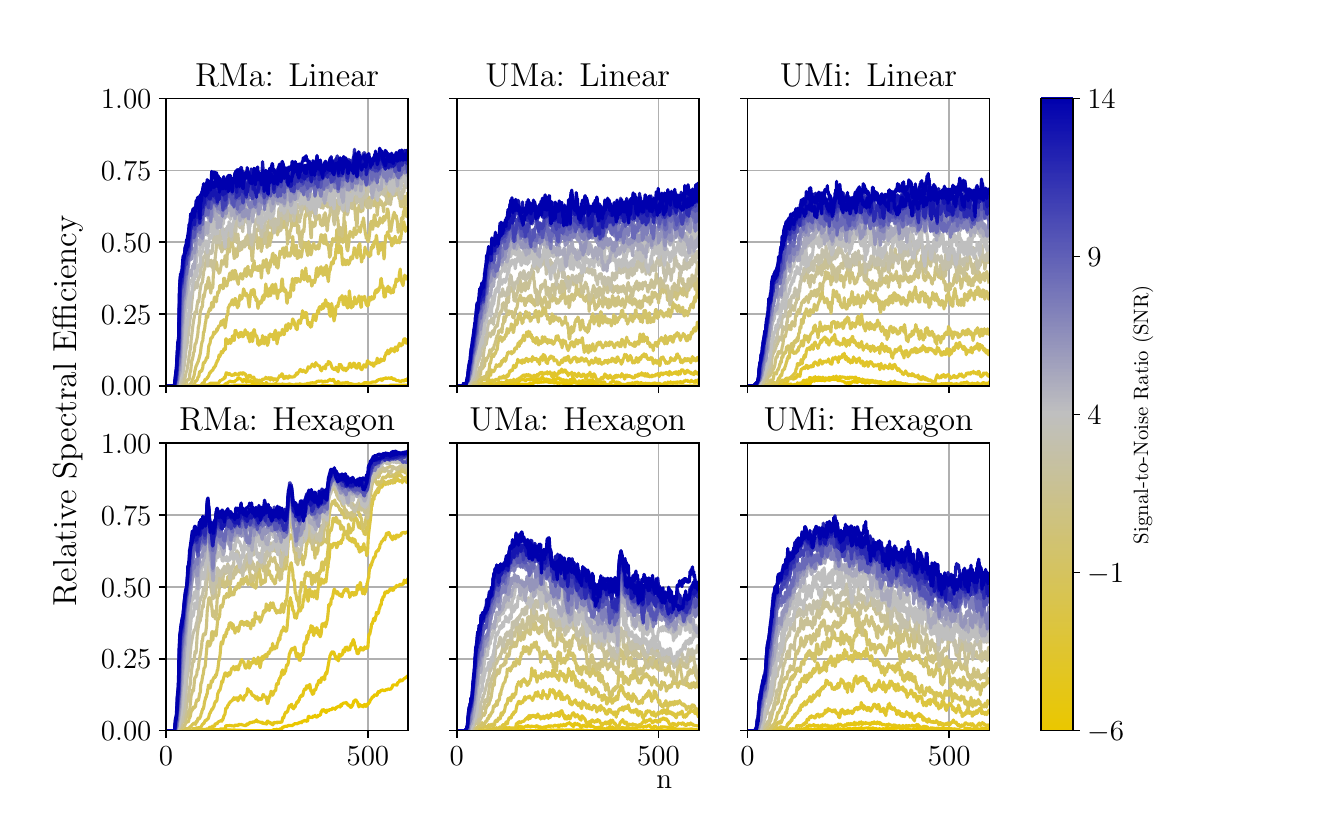}
\centering
\caption{Comparing performance with NYUsim.}
\label{fig:performanceNYU}
\end{figure}

\section{Conclusion}\label{sec:conclusion}

We presented \ac{DBZ}, an algorithm with low computational overhead, that encompasses all defining features of \ac{ISAC}. Exploiting the structure induced by the hierarchical codebook, we adapted the \ac{MAB} best arm identification framework from \cite{gabillon2012best} to handle a \ac{ME}.  Our analysis shows the correctness guarantees on beamforming vector selection.  Additionally, we have characterized how to set the sample window lengths based on a \ac{DWNA} channel model and the complexity expected of the \ac{DBZ} algorithm.  The beamwidth adjustments over time prevent severe outages typically associated with \ac{mmWave} systems. We show \ac{DBZ} strikes competitive performance against Bayesian methods exploiting full channel knowledge and \ac{ME} motion \cite{ronquillo2021active}. Finally, our simulations with NYU Sim show \ac{DBZ}'s efficacy in realistic fading environments over several scenarios.

\bibliographystyle{IEEEtran}
\bibliography{ref}

\pagebreak
\newcommand{\appendixfigscaling}{0.8}
\onecolumn
\appendices

\section{Full Proofs}\label{sec:fullproofs}
\subsection{Proof of Lemma \ref{lem:subexponentiallemma}}\label{ssec:pf_lem_subexponentiallemma}

\begin{IEEEproof}
    Our proof follows the steps from \cite[Appendix E]{ghosh2022learning} and \cite[Section 2.1.3]{wainwright2019high}. Note that $\mu_{h,i} = \zeta_{h,i} + \sigma^2$, the variance of $y(n_{h,i})$, $\nu_{h,i}^2  = \sigma^4 + 2\zeta_{h,i}\sigma^2$, and with the moment generating function of $y(n_{h,i})$, dropping the time dependence temporarily, we write,
\begin{align}
    \E{\exp(\lambda(y(n_{h,i})-\mu_{h,i}))} &= \frac{\exp(-\lambda \mu_{h,i})}{1 - \sigma^2\lambda}\exp\left(\frac{\lambda \zeta_{h,i}}{1-\sigma^2\lambda}\right)\\
    &= \frac{\exp(-\lambda(\zeta_{h,i} + \sigma^2))}{1 - \sigma^2\lambda}\exp\left(\frac{\lambda \zeta_{h,i}}{1-\sigma^2\lambda}\right)\\
    &=\frac{\exp(-\sigma^2\lambda)}{1-\sigma^2\lambda}\exp\left(\frac{\zeta_{h,i}\lambda^2\sigma^2}{1-\sigma^2\lambda}\right)\\
    &\leq \exp(\sigma^4\lambda^2) \exp(2\zeta_{h,i}\sigma^2\lambda^2)\label{eq:alphaconditionsubexponential}\\
    &= \exp\left(\lambda^2(\sigma^4 + 2\zeta_{h,i}\sigma^2)\right)\\
    &= \exp\left(\frac{2\nu^2_{h,i}\lambda^2}{2}\right),\\
\end{align}
where \eqref{eq:alphaconditionsubexponential} holds when $\abs{\lambda}< 1/(2\sigma^2)$.  We next prove \eqref{eq:probsubexponential} using the Cramer-Chernoff method.  We start with a "boosted" form of Markov's Inequality, 
\begin{align}
    \P{y(n_{h,i})-\mu_{h,i} \geq \delta}&\leq \frac{\E{\exp(\lambda(y(n_{h,i})-\mu_{h,i}))}}{\exp(\lambda\delta)}\\
    & \leq \exp(-\lambda\delta)\exp\left(\frac{2\nu_{h,i}^2\lambda^2}{2}\right)\\
    & = \exp(-g(\lambda,\delta))\label{eq:probwithg}.
\end{align}
We find $\lambda^*$, such that $g(\lambda^*,\delta) \eqdef \inf_{\lambda \in (0,1/a)} g(\lambda,\delta)$, by solving of $g'(\lambda,\delta) = 0$.  The solution yields $\lambda^* = \delta/(2\nu_{h,i}^2)$ and $g(\lambda^*,\delta) = \delta^2/(2\nu_{h,i}^2)$, which we plug into \eqref{eq:probwithg}.  In particular, we are interested in the empirical mean of $y(n_{h,i})$ over time, hence,
\begin{align}
    \P{\hat{\mu}_{h,i}(n_{h,i}) - \mu_{h,i} \geq \delta}&\leq\exp\left(-N_{h,i}(\eta,n_{h,i})g(\lambda^*,\delta)\right)\\
    &=\exp\left(-\frac{N_{h,i}(\eta,n_{h,i})\delta^2}{4\nu_{h,i}^2}\right).
\end{align}
A union bound completes our proof.
\end{IEEEproof}

\subsection{Proof of Lemma \ref{lem:confidenceempvar}}\label{ssec:pf_lem_confidenceempvar}

\begin{IEEEproof}
 We use the one-sided version of \eqref{eq:probsubexponential} from our result in Lemma \ref{lem:subexponentiallemma} with a constant $B \geq1$,
     \begin{align}\label{eq:probsubexponentialwb}
         \P{\mathcal{C}_1} = \P{\hat{\mu}_{h,i}(\eta,n) - \mu_{h,i}(n) \geq\sqrt{\frac{4B\nu_{h,i}^2(n)\beta(\eta,n,\delta)}{N_{h,i}(\eta,n)}}}\leq \exp(- \beta(\eta,n,\delta)),
     \end{align}
     and the result in \cite[Theorem 10]{maurer2009empirical} with $C \geq 1$ to bound the difference between standard deviation $\nu_{h,i}(n)$ and its empirical estimate, $\hat{\nu}_{h,i}(\eta,n)$, as
     \begin{align}\label{eq:empvarbound}
         \P{\mathcal{C}_{2}} = \P{\nu_{h,i}(n) > \hat{\nu}_{h,i}(\eta,n) + \sqrt{\frac{2C\beta(\eta,n,\delta)}{N_{h,i}(\eta,n)-1}}}\leq \exp(-\beta(\eta,n,\delta))
     \end{align}
     We have
     \begin{align}
         &2\exp(-\beta(\eta,n,\delta))\\
         &\geq \P{\mathcal{C}_1} +  \P{\mathcal{C}_2}\\ 
         &\geq\P{\mathcal{C}_1 \cup \mathcal{C}_2} \\
         &\geq\P{\hat{\mu}_{h,i}(\eta,n) - \mu_{h,i} \geq \sqrt{\frac{4B\beta(\eta,n,\delta)}{N_{h,i}(\eta,n)}}\left(\hat{\nu}_{h,i}(\eta,n) + \sqrt{\frac{2C\beta(\eta,n,\delta)}{N_{h,i}(\eta,n)-1}}\right)}\label{eq:combinedbounds}\\
         &= \P{\hat{\mu}_{h,i}(\eta,n) - \mu_{h,i} \geq \sqrt{\frac{4B\hat{\nu}_{h,i}^2(\eta,n)\beta(\eta,n,\delta)}{N_{h,i}(\eta,n)}} + \frac{4BC\beta(\eta,n,\delta)}{\sqrt{N_{h,i}(\eta,n)(N_{h,i}(\eta,n)-1)}}}\\
         &\geq \P{\hat{\mu}_{h,i}(\eta,n) - \mu_{h,i} \geq \sqrt{\frac{4B\hat{\nu}_{h,i}^2(\eta,n)\beta(\eta,n,\delta)}{N_{h,i}(\eta,n)}} + \frac{4BC\beta(\eta,n,\delta)}{N_{h,i}(\eta,n)-1}},
     \end{align}
     where we combine \eqref{eq:probsubexponentialwb} with \eqref{eq:empvarbound} in \eqref{eq:combinedbounds}.  Equivalent steps provide the proof for the other case, which we denote as,
     \begin{align}\label{eq:empvarbound}
         \P{\mathcal{C}_{3}} = \P{\mu_{h,i}(n)-\hat{\mu}_{h,i}(\eta,n) \geq \sqrt{\frac{4B\hat{\nu}_{h,i}^2(\eta,n)\beta(\eta,n,\delta)}{N_{h,i}(\eta,n)}} + \frac{4BC\beta(\eta,n,\delta)}{N_{h,i}(\eta,n)-1}} \leq 2\exp(-\beta(\eta,n,\delta)).
     \end{align}
     The two-sided version follows,
     \begin{align}
         &3\exp(-\beta(\eta,n,\delta))\\
         &\geq \P{\mathcal{C}_1} +  \P{\mathcal{C}_2} + \P{\mathcal{C}_3}\\ 
         &\geq\P{\mathcal{C}_1 \cup \mathcal{C}_2 \cup \mathcal{C}_3} \\
         &\geq\P{\abs{\hat{\mu}_{h,i}(\eta,n) - \mu_{h,i}(n)} \geq \sqrt{\frac{4B\beta(\eta,n,\delta)}{N_{h,i}(\eta,n)}}\left(\hat{\nu}_{h,i}(\eta,n) + \sqrt{\frac{2C\beta(\eta,n,\delta)}{N_{h,i}(\eta,n)-1}}\right)}\label{eq:combinedbounds}\\
         &= \P{\abs{\hat{\mu}_{h,i}(\eta,n) - \mu_{h,i}(n)} \geq \sqrt{\frac{4B\hat{\nu}_{h,i}^2(n_{h,i})\beta(\eta,n,\delta)}{N_{h,i}(\eta,n)}} + \frac{4BC\beta(\eta,n,\delta)}{\sqrt{N_{h,i}(\eta,n)(N_{h,i}(\eta,n)-1)}}}\\
         &\geq \P{\abs{\hat{\mu}_{h,i}(\eta,n) - \mu_{h,i}(n)} \geq \sqrt{\frac{4B\hat{\nu}_{h,i}^2(n_{h,i})\beta(\eta,n,\delta)}{N_{h,i}(\eta,n)}} + \frac{4BC\beta(\eta,n,\delta)}{N_{h,i}(\eta,n)-1}},
     \end{align}
 \end{IEEEproof}

\subsection{Proof of Lemma \ref{lem:confidence}}\label{ssec:pf_lem_confidence}

\begin{IEEEproof}
    We apply \eqref{eq:probempvar3} from Lemma \ref{lem:confidenceempvar} for one level, $h$,
    \begin{align}
    &\P{\abs{\hat{\mu}_{h,i}(\eta,n) - \mu_{h,i}(n)} \geq D_{h,i}(\eta,n), N_{h,i}(\eta,n) = u}\\
    &=\P{\abs{\hat{\mu}_{h,i}(\eta,n) - \mu_{h,i}(n)} \geq \sqrt{\frac{4B\hat{\nu}_{h,i}^2(\eta,n)\beta(\eta,n,\delta)}{u}} + \frac{2\sqrt{2BC}\beta(\eta,n,\delta)}{u-1}, N_{h,i}(\eta,n) = u}\\ 
    &\leq 3\exp(-\beta(\eta,n,\delta))
\end{align}
and use a union bound with the exploration rate,  $\beta(t,\delta)$ in \eqref{eq:explorationrate}, over all levels
    \begin{align}
    \P{\mathcal{B}_1^c\cup\cdots\cup\mathcal{B}_H^c}
 	&\leq \P{\mathcal{B}_1^c} + \cdots + \P{\mathcal{B}_H^c}\\   
     &\leq  \sum_{n = 1}^\infty\sum_{h= 1}^H\sum_{i:(h,i) \in \mathcal{I}_h}\sum_{u=1}^{n} \P{\abs{\hat{\mu}_{h,i}(\eta,n) - \mu_{h,i}(n)} \geq D_{h,i}(\eta,n), N_{h,i}(\eta,n) = u}\\
    &\leq  \sum_{n = 1}^\infty\sum_{h= 1}^H\sum_{i:(h,i) \in \mathcal{I}_h}\sum_{u=1}^{n} 3\exp(-\beta(\eta,n,\delta))\\
    &=\sum_{n = 1}^\infty\sum_{h= 1}^H\sum_{i:(h,i) \in \mathcal{I}_h}\sum_{u=1}^{n}\frac{2\delta}{\mathcal{I}^H5\eta^4}\\
    &<\sum_{h = 1}^H\int_1^\infty\frac{2\abs{\mathcal{I}_h}\delta}{\mathcal{I}^H5n^3}dn\\
    &\leq \sum_{h=1}^H\frac{\abs{\mathcal{I}_h}\delta}{2\mathcal{I}^H} = \frac{\delta}{2}\label{eq:deltaconfB}.
    \end{align}
With the appropriate choice of $\eta$, we combine \eqref{eq:deltaconfB} with 
\begin{align}\label{eq:deltaconfA}
	\P{\mathcal{A}_1^c\cup\cdots\cup\mathcal{A}_H^c}
 	\leq \P{\mathcal{A}_1^c} + \cdots + \P{\mathcal{A}_H^c}
 	\leq \sum_{h=1}^H \frac{\delta}{2H} = \frac{\delta}{2},
\end{align}
 from which we conclude that $P(\mathcal{B}_h \cap \mathcal{A}_h, \forall 1\leq h\leq H) > 1-\delta$.
\end{IEEEproof}

\subsection{Proof of Lemma \ref{lem:samplingchoices}}\label{ssec:pf_lem_samplingchoices}

\begin{IEEEproof}
Each \ac{MAB} game plays independently of other levels, so without loss of generality, we drop the index indicating the specific level, i.e., $t$ versus $n$.  In each scenario, where either $\hat{\mu}_{u(n)}(\eta,n-1) \leq \hat{\mu}_{\gamma(n)}(\eta,n-1)$ or $\hat{\mu}_{u(n)}(\eta,n-1) > \hat{\mu}_{\gamma(n)}(\eta,n-1)$, we first show that if $S(n) \in \set{(h,u(n)),(h,\gamma(n))}$ then neither \eqref{eq:samplingchoicecase1} or \eqref{eq:samplingchoicecase2} is violated.  As a result of \eqref{eq:samplingchoicecase1} and \eqref{eq:samplingchoicecase2}, we then show that for any $(h,i) \in \mathcal{I}_h$ that \eqref{eq:samplingchoicecase3} holds.
\begin{enumerate}
    \item  If $\hat{\mu}_{u(n)}(\eta,n-1) \leq \hat{\mu}_{\gamma(n)}(\eta,n-1)$, we have the following two sub-cases
    \begin{enumerate}
        \item If $X(n) = u(n)$, then $D_{u(n)}(\eta,n-1) \geq D_{\gamma(n)}(\eta,n-1)$.  \eqref{eq:samplingchoicecase1} holds because
        \begin{align}
            L_{u(n)}(n) = \hat{\mu}_{u(n)}(\eta,n-1) - D_{u(n)}(\eta,n-1) \leq \hat{\mu}_{\gamma(n)} - D_{u(n)}(\eta,n-1) \leq \hat{\mu}_{\gamma(n)}(\eta,n-1) = L_{\gamma(n)}(\eta,n-1)
        \end{align}
        \item If $X(n) = \gamma(n)$, then $D_{\gamma(n)}(\eta,n-1) \geq D_{u(n)}(\eta,n-1)$.  \eqref{eq:samplingchoicecase2} holds because
        \begin{align}
            U_{\gamma(n)}(n) = \hat{\mu}_{\gamma(n)}(\eta,n-1) + D_{\gamma(n)}(\eta,n-1) &\geq \hat{\mu}_{\gamma(n)}(\eta,n-1) + D_{u(n)}(\eta,n-1)\notag\\ &\geq \hat{\mu}_{u(n)}(\eta,n-1) + D_{u(n)}(\eta,n-1) = U_{u(n)}(n)
        \end{align}
    \end{enumerate}
    \item If $\hat{\mu}_{u(n)}(\eta,n-1) > \hat{\mu}_{\gamma(n)}(\eta,n-1)$, we have the following two sub-cases
    \begin{enumerate}
        \item If $X(n) = u(n)$, then $D_{u(n)}(\eta,n-1) \geq D_{\gamma(n)}(\eta,n-1)$, so $U_{u(n)}(n) > U_{\gamma(n)}(n)$. We show by contradiction that \eqref{eq:samplingchoicecase1} holds.  Assume that $L_{u(n)}(n) > L_{\gamma(n)}(n)$, we may write
        \begin{align}
            G_{u(n)}(n) = \max_{j \neq u(n)}U_j(n) - L_{u(n)}(n) < U_{u(n)}(n) - L_{\gamma(n)}(n) = G(n),
        \end{align}
        but by definition, $u(n) \neq \gamma(n)$, $\gamma(n) = \argmin_{i : i \in \mathcal{S}}G_i(n)$, and $G_{u(n)}(n) \geq G(n)$.  Therefore $G_{u(n)}(n) \leq G(n)$ is a contradiction and \eqref{eq:samplingchoicecase1} holds.
        \item If $X(n) = \gamma(n)$, then $D_{\gamma(n)}(\eta,n-1) \geq D_{u(n)}(\eta,n-1)$, so $L_{u(n)}(n) > L_{\gamma(n)}(n)$. We show by contradiction that \eqref{eq:samplingchoicecase2} holds.  Assume that $U_{u(n)}(n) > U_{\gamma(n)}(n)$, we may write
        \begin{align}
            G(n) = U_{u(n)}(n) - L_{\gamma(n)}(n) > \max_{j\neq u(n)}U_j(n) - L_{u(n)}(n) = G_{u(n)}(n),
        \end{align}
        which, again, is a contradiction due to the sample reasoning in case 2a.
    \end{enumerate}
\end{enumerate}    
Finally,  $i \in \set{u(n),\gamma(n)}$,  \eqref{eq:samplingchoicecase1} allows us to write
\begin{align}
    G(n) = U_{u(n)}(n) - L_{\gamma(n)}(n) \leq U_{u(n)}(n) - L_{u(n)}(n) = 2D_{u(n)}(\eta,n-1) = 2D_{i}(\eta,n-1),
\end{align}
and \eqref{eq:samplingchoicecase2} allow us to write
\begin{align}
    G(n) = U_{u(n)}(n) - L_{\gamma(n)}(n) \leq U_{\gamma(n)}(n) - L_{\gamma(n)}(n) = 2D_{\gamma(n)}(\eta,n-1) = 2D_{i}(\eta,n-1).
\end{align}
\end{IEEEproof}

\subsection{Proof of Lemma \ref{lem:boundonG}}\label{ssec:pf_lem_boundonG}

\begin{IEEEproof}
Without loss of generality, we again drop the index indicating the specific level.  Recall that the beamforming vector with the highest average rewards, value denoted $\mu_h^*$, has indices $(h,i^*)$.
    \begin{enumerate}
        \item $S(n) = u(n)$ 
        \begin{enumerate}
            \item $u(n) = i^*$ and by definition $u(n) \neq \gamma(n)$, then $\exists j \neq i^*$ such that $j = \gamma(n)$.  We write
            \begin{align}
                \mu^* \geq \mu_j \geq L_j(\eta,n) &= L_{\gamma(n)}(\eta,n) \label{eq:boundonGproofstep1a1}\\
                &\geq L_{u(n)}(\eta,n) = \hat{f}_{i}(\eta,n-1) - D_i(\eta,n-1)\label{eq:boundonGproofstep1a2}\\
                &\geq \mu_i) - 2D_{i}(\eta,n-1)\label{eq:boundonGproofstep1a3},
            \end{align}
            which allows us to conclude that $-2D_i(\eta,n-1) - \Delta_i \geq 0$.  \eqref{eq:boundonGproofstep1a1} and \eqref{eq:boundonGproofstep1a3} are true under event $\mathcal{B}$, and \eqref{eq:boundonGproofstep1a2} is from \eqref{eq:samplingchoicecase1} in Lemma \ref{lem:samplingchoices}.
            \item $u(n) \neq i^*$
            \begin{enumerate}
                \item $\gamma(n) = i^*$, 
                \begin{align}
                    G(\eta,n) = U_{u(n)}(\eta,n) - L_{\gamma(n)}(\eta,n) &\leq \mu_{u(n)} + 2D_{u(n)}(\eta,n-1) - \mu_{\gamma(n)} + 2D_{\gamma(n)}(\eta,n-1)\label{eq:boundonGproofstep1bi1}\\
                    & \leq \mu_{u(n)} + 2D_{u(n)}(\eta,n-1) - \mu^* + 2D_{u(n)}(\eta,n-1)\label{eq:boundonGproofstep1bi2}\\
                    & \leq -\Delta_{u(n)} + 4D_{u(n)}(\eta,n-1)\label{eq:boundonGproofstep1bi3}.\\
                \end{align}
                Under event $\mathcal{B}$, \eqref{eq:boundonGproofstep1bi1} holds, and \eqref{eq:boundonGproofstep1bi3} is from our sampling strategy, $S(n) = u(n)$ when  $D_{u(n)}(\eta,n-1) \geq D_{\gamma(n)}(\eta,n-1)$. 
                \item $\gamma(n) \neq i^*$, then under event $\mathcal{B}$, $\mu_{u(n)} + 2D_{u(n)}(\eta,n-1) \geq U_{u(n)}(\eta,n) \geq U_{i^*}(\eta,n) \geq \mu^*$.
            \end{enumerate}
        \end{enumerate}
        \item $S(n) = \gamma(n)$
        \begin{enumerate}
            \item $\gamma(n) = i^*$, then under event $\mathcal{B}$, $\mu^* = \mu_{\gamma(n)} \leq \hat{\mu}_{\gamma(n)}(\eta,n-1) + D_{\gamma(n)}(\eta,n-1) \leq \mu_{\gamma(n)} + 2D_{\gamma(n)}(\eta,n-1)$.

            \item $\gamma(n) \neq i^*$
            \begin{enumerate}
                \item $u(n) = i^*$ and $\gamma(n) \neq u(n)$, then $\exists i \neq i^*$ such that $\gamma(n) = i$.  We write
                    \begin{align}
                        \mu^* \leq U_{i^*}(\eta,n) &= U_{u(n)}(\eta,n)\label{eq:boundGproofstep2bi1}\\
                        &\leq U_{\gamma(n)}(\eta,n)\label{eq:boundGproofstep2bi2}\\
                        &= \hat{f}_{\gamma(n)}(\eta,n-1) + D_{\gamma(n)}(\eta,n-1)\label{eq:boundGproofstep2bi2}\\
                        &\leq \mu_{\gamma(n)} + 2D_{\gamma(n)}(\eta,n-1)\label{eq:boundGproofstep2bi3}\\
                        &= \mu_{i} + 2D_{i}(\eta,n-1),\\
                    \end{align}
                    where \eqref{eq:boundGproofstep2bi2} is from \eqref{eq:samplingchoicecase2} in Lemma \eqref{lem:samplingchoices}. Under event $\mathcal{B}$, \eqref{eq:boundGproofstep2bi1} and \eqref{eq:boundGproofstep2bi3} hold.
                \item $u(n) \neq i^*$, then
                    \begin{align}
                        \mu^* \leq U_{i^*}(\eta,n) &\leq U_{u(n)}(\eta,n)\label{eq:boundGproofstep2bii1}\\
                        &\leq U_{\gamma(n)}(\eta,n)\label{eq:boundGproofstep2bii2} \\
                        &={f}_{\gamma(n)}(\eta,n) + D_{\gamma(n)}(\eta,n-1)\label{eq:boundGproofstep2bii3} \\
                        &\leq \mu_{\gamma(n)} + 2D_{\gamma(n)}(\eta,n-1).\label{eq:boundGproofstep2bii4},
                    \end{align}
                    where \eqref{eq:boundGproofstep2bii1} follows from the definition of $u(n)$. \eqref{eq:boundGproofstep2bii2} is from \eqref{eq:samplingchoicecase2} in Lemma \eqref{lem:samplingchoices} and \eqref{eq:boundGproofstep2bii4} holds under event $\mathcal{B}$.
            \end{enumerate}
        \end{enumerate}
    \end{enumerate}
\end{IEEEproof}

\subsection{Proof of Lemma \ref{lem:cannotzoomin}}\label{ssec:pf_lem_cannotzoomin}
The beamforming vector $\mathbf{f}_{h,i}$ with mean reward $\mu_{h,i}(n)$ is not $\epsilon_h$-optimal arm, then
\begin{IEEEproof}
\begin{align}
	G_{h,i}(\eta,n) &= \max_{j\neq i}U_{h,j}(\eta,n-1) - L_{h,i}(\eta,n-1)\\
	&= \max_{j \neq i}(\hat{\mu}_{h,j}(\eta,n-1) +  D_{h,j}(\eta,n-1)) - (\hat{\mu}_{h,i}(\eta,n-1) - L_{h,i}(\eta,n-1))\label{eq:pfconnotzoominln2}\\
	&\geq \max_{j\neq i} \mu_{h,j}(n) - \mu_{h,i}(n) \label{eq:pfconnotzoominln3}\\
	&= \mu_h^*(n) - \mu_{h,i}(n) > \epsilon_h, \label{eq:pfconnotzoominln4}
\end{align}
where \eqref{eq:pfconnotzoominln3} is from $\mathcal{B}_h$. \eqref{eq:pfconnotzoominln4} comes from the fact that $i \neq i^*$ and the definition of an $\epsilon_h$-optimal arm.
\end{IEEEproof}

\subsection{Proof of Lemma \ref{lem:wontzoomout}}\label{ssec:pf_lem_wontzoomout}

\begin{IEEEproof}
Suppose that $U_{h,\gamma}(\eta,n-1) <[\boldsymbol\kappa]_{h'+1}$, we write
\begin{align}
	\mu_{h,i}(n) &\leq \max_{i:(h,i) \in \mathcal{I}_h} \mu_{h,i}(n)\label{eq:app_pfwontzoomout1}\\
	 &\leq U_{h,\gamma}(\eta,n-1)\label{eq:app_pfwontzoomout2}\\
	  &< [\boldsymbol\kappa]_{h'+1}\label{eq:app_pfwontzoomout3}\\
	   &= L_{h',i'}(\eta,n'-1) + A\epsilon_{h'}\label{eq:app_pfwontzoomout4}\\ 
	   &\leq \mu_{h',i'}(n') + \epsilon_{h'}\label{eq:app_pfwontzoomout5}.
\end{align}
We see $\mu_{h,i}(n) < \mu_{h',I}(n') + \epsilon_{h'}$ and therefore Assumption 2 no longer holds for the current path of mean rewards.  Remember, \ac{DBZ} correctness guarantees hinge on taking one of the paths where Assumption 2 holds to choose the correct beam.  The relationship of \eqref{eq:app_pfwontzoomout1} to \eqref{eq:app_pfwontzoomout2} and \eqref{eq:app_pfwontzoomout4} to \eqref{eq:app_pfwontzoomout5} come from event $\mathcal{B}_h$ and $\mathcal{B}_{h'}$, respectively, where  Lemma \ref{lem:confidenceempvar} 
 shows both events hold with probability greater than $1-\delta$.
\end{IEEEproof}

\subsection{Proof of Lemma \ref{lem:upperboundsamples}}\label{ssec:pf_lem_upperboundsamples}

 \begin{IEEEproof}
 If \ac{DBZ} has not terminated at a particular level, $h$, then $G_{h,\gamma(n)}(\eta,n) \geq \epsilon_h$.  From Lemma \ref{lem:boundonG}, $\epsilon_h \leq G_{h,\gamma(n)}(\eta,n) \leq \min\set{0,2D_{h,i}(\eta,n-1)-\Delta_{h,i}(n)} + 2D_{h,i}(\eta,n-1)$ implies that $\Delta_{h,i,\epsilon}(n) \leq D_{h,i}(\eta,n-1)$.  We first substitute our expression for $D_{h,i}(\eta,n-1)$ from \eqref{eq:confidenceterm}, we write
 \begin{align}
     \Delta_{h,i,\epsilon}(n) &\leq D_{h,i}(\eta,n-1)\\
      &\leq \sqrt{\frac{4B\nu_{h,i}^2\beta(\eta,n,\delta)}{N_{h,i}(\eta,n)}} + \frac{2\sqrt{2BC}\beta(\eta,n,\delta)}{N_{h,i}(\eta,n-1)}\\
      &\leq \sqrt{\frac{4B\nu_{h,i}^2\beta(\eta,n,\delta)}{N_{h,i}(\eta,n-1}} + \frac{2\sqrt{2BC}\beta(\eta,n,\delta)}{N_{h,i}(\eta,n-1)}\label{eq:upperboundsamplesproofstep2}.
 \end{align}
 The step in \eqref{eq:upperboundsamplesproofstep2} holds because $\beta(n,\delta)$ is monotonically increasing with respect to $n$.  After some algebra and solving a quadratic, we obtain the expression,
 \begin{align}
     N_{h,i}(\eta,n) - 1 \leq \frac{2B\nu^2_{h,i} + 2\sqrt{2BC}\Delta_{h,i,\epsilon}(n) + \sqrt{4B^2\nu^4_{h,i} + 2\sqrt{2C}B^{3/2}\nu^2_{h,i}\Delta_{h,i,\epsilon}(n)}}{\Delta_{h,i,\epsilon}^2(n)} \beta((\eta,n),\delta)\label{eq:upperboundsamplesproofstep3}.
 \end{align}
 At some future time step we sample $(h,i)$, such that $N_{h,i}(n) = N_{h,i}((\eta,n)) + 1$, and by substituting into \eqref{eq:upperboundsamplesproofstep3} we obtain the required result.
 \end{IEEEproof}

\section{Derivation of Alignment Time Probability}\label{sec:deriveprobcoherence}
We require the current angle state $\theta(n)$ to be described as a random variable with respect to time step $n$ to determine probability of alignment over time.  We first derive the random variable model for the angular position $\theta(n)$, then derive the distribution describing the alignment time within an angular region, $[-\phi_\textnormal{max},\phi_\textnormal{max}]$.  The region is a generalization of the beamwidth region covered by a beamforming vector's beam pattern. Assume $\tau = \tau(n)$ for all $n$, we rewrite the expressions for $\theta(n)$ and $\dot{\theta}(n)$ as 
\begin{align}
\theta(n) =  \theta(n-1) + \tau\dot{\theta}(n-1)+ \frac{\tau^2}{2}u(n-1)\label{eq:angleupdateapp}\\
\dot{\theta}(n) = \dot{\theta}(n-1) + \tau u(n-1).\label{eq:velocityupdateapp}
\end{align}
We modify the expressions to purely be functions of the state at $n=1$, starting with \eqref{eq:velocityupdateapp},
\begin{align}
\dot{\theta}(n) = \dot{\theta}(1) + \tau \sum_{p = 1}^{n-1} u(p),\label{eq:velocityupdatefrominitial}
\end{align}
which we plug into \eqref{eq:angleupdateapp}, simplify, and obtain the induced expression for $\theta(n-1), \theta(n-2),\dots,\theta(1)$, 
\begin{align}
\theta(n) &=  \theta(n-1) + \tau(\dot{\theta}(1) + \tau \sum_{p = 1}^{n-1} u(p))+ \frac{\tau^2}{2}u(n-1)\\
&= \theta(1) + (n-1)\tau\dot{\theta}(1) + \tau^2\sum_{p = 1}^{n-1}\frac{2p-1}{2}u(n-p).\label{eq:angleupdatefrominitial}
\end{align}


We model the initial states as $\theta(1)\sim\mathcal{U}(-\theta_\text{max},\theta_\text{max})$ and $\dot{\theta}(1)\sim\mathcal{N}(0,\sigma^2_u)$.  The uniform distribution of $\theta(1)$ comes from the fact that $\theta(1)$ could appear anywhere within the angular coverage, $[\bar{\phi}_i-\phi_{\textnormal{bw}}/2,\bar{\phi}_i+\phi_{\textnormal{bw}}/2]$ of the beamforming vector $\mathbf{f}_{h,i}$.  We represent the angular position state at time $n$ by the random variable
\begin{align}\label{eq:gaussianinitvelocity}
\theta(n) \sim \mathcal{N}\left(\theta(1), \frac{\tau^4}{4}(\frac{4n^3}{3} -4n^2 + \frac{11n}{3} -1 )\sigma^2_u + \tau^2(n-1)\sigma_u^2\right),
\end{align}
where we denote
\begin{align}\label{eq:sigmanapp}
\sigma_n^2 \eqdef\frac{\tau^4}{4}(\frac{4n^3}{3} -4n^2 + \frac{11n}{3} -1 )\sigma^2_u + \tau^2(n-1)\sigma_u^2.
\end{align}
The variance expression in \eqref{eq:gaussianinitvelocity} comes from the sum term in \eqref{eq:angleupdatefrominitial}. The random variable $\theta(n)$ is distributed according to the compound density 
\begin{align}\label{eq:pdfcoherence}
f(\theta(n); \theta_\text{max})&=\int_{-\theta_\text{max}}^{\theta_\text{max}} \frac{1}{2\theta_\text{max}} \frac{1}{\sqrt{2\pi}\sigma_n}\exp\left(-\frac{(\theta(n)-\theta(1))^2}{2\sigma^2_n}\right)d\theta(1)\\
&= \frac{1}{4\theta_\text{max}} \left(  \erf\left(\frac{\theta(n)+\theta_\text{max}}{\sqrt{2}\sigma_n}\right) - \erf\left(\frac{\theta(n)-\theta_\text{max}}{\sqrt{2}\sigma_n}\right)\right).
\end{align}
For beamforming vector $\mathbf{f}_{h,i}$, we assume $\bar{\phi}_{h,i} = 0$, set $\phi_{\textnormal{max}} = \phi_{\textnormal{bw},h}/2$, and calculate the probability of alignment after $n$ timesteps as 
\begin{align}
\P{\abs{\bar{\phi}_{h,i} - \theta(n)}\leq \frac{\phi_{\text{bw},h}}{2}} &= \int_{-\phi_{\text{bw},h}/2}^{\phi_{\text{bw},h}/2} f(\theta(n);\theta_\text{max}) d\theta(n)\\
&=\frac{1}{2\phi_{\text{bw},h}} \left(  \frac{2\sqrt{2}\sigma_n}{\sqrt{\pi}} \left( \exp\left( -\frac{\phi_{\text{bw},h}^2}{2\sigma_n^2}\right)  - 1\right) +  2\phi_{\text{bw},h} \erf\left( \frac{\phi_{\text{bw},h}}{\sqrt{2}\sigma_n} \right) \right),\label{eq:probcoherenceapp}
\end{align}
using the expression $\int_{0}^z \erf{x}dx = z\erf{z} + \exp(-z^2)/\sqrt{\pi}$. Fig. \ref{fig:coherenceprob} shows \eqref{eq:probcoherence} over several values of $n$ for two values of $\tau(n) = \tau$.  For larger values of $\sigma_u$ in the top half of Fig. \ref{fig:coherenceprob}, alignment time at narrower beams consists of very few or even less than one time step to ensure alignment within $90\%$.  In cases of significant motion, we must decrease the sampling interval, $\tau$, in which we take measurements.  The bottom half of Fig. \ref{fig:coherenceprob} shows the changes in the distribution when reducing the sampling interval by a factor of $4$, $\tau = .25$.  
\begin{figure}[H]
\includegraphics[scale =\appendixfigscaling]{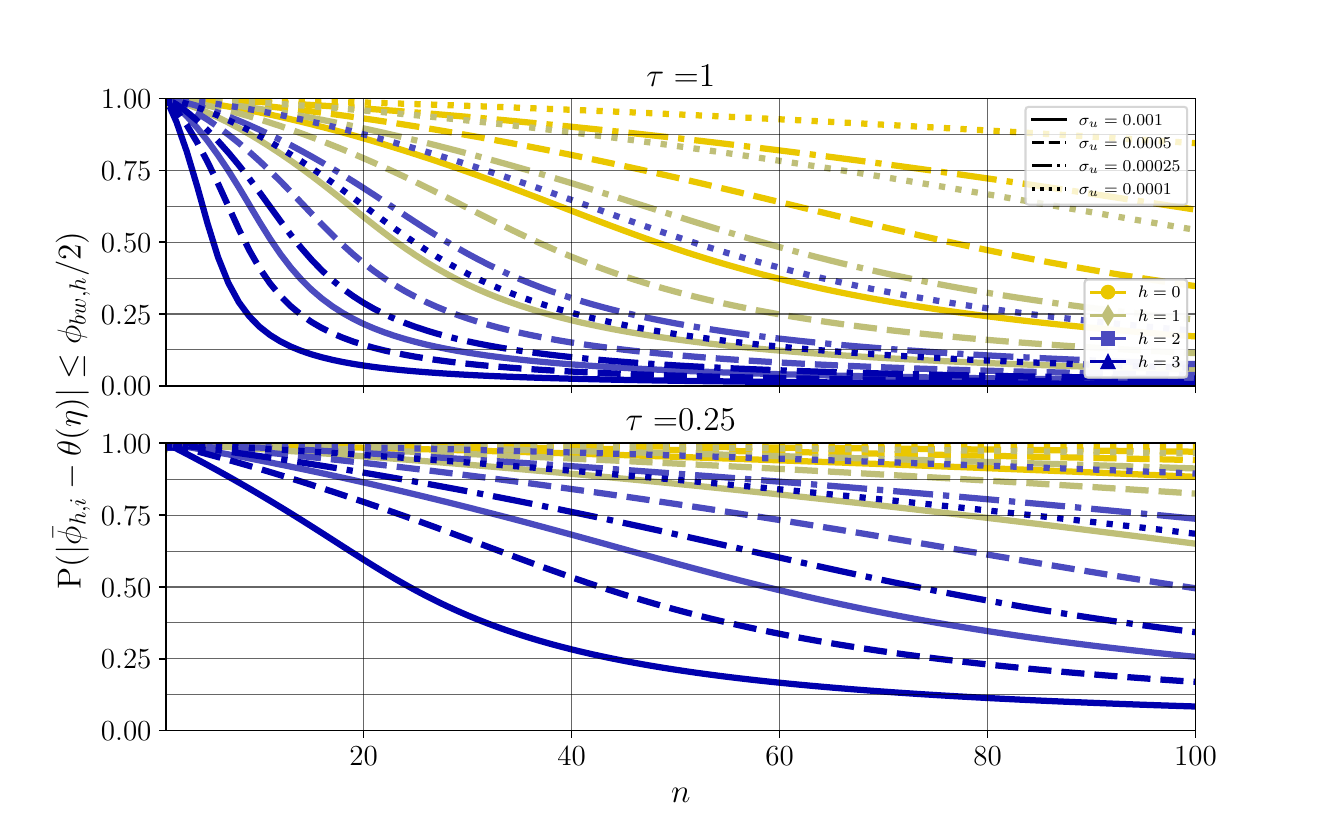}
\centering
\caption{Probability of beam alignment with respect to kinematic motion over time.}
\label{fig:coherenceprob}
\end{figure} 
We use \eqref{eq:probcoherenceapp} with the bounds on complexity of \ac{DBZ} to determine limits on performance.

\section{Additional Content on Simulations}\label{sec:additionalsimulation}

\subsection{Details on Monte-Carlo Simulations}

Specifically, we estimate the probability that an $\epsilon$-optimal arm is chosen over time, $P_c(t)$, as 
\begin{align}\label{eq:estimateprobcorrect}
    \hat{P}_c(n,L) = \frac{1}{L}\sum_{\ell = 1}^L \indic{\theta_{H,\gamma(n)} \in \Theta_\epsilon, \abs{i^*_\ell - \gamma(n)} \leq 1}
\end{align}
at sample $n$.  3GPP recognizes that beam prediction accuracy is a key performance metric for \ac{ML} applications in 5G/\ac{NR} \cite{3gppranaiml}. Let
\begin{align}\label{wpminterval}
w^{\pm}= \frac{1}{1 +z_b/L}\left(\left(\hat{P}_c(n,L) + \frac{z_b^2}{2L}\right) \pm z_b\sqrt{\frac{\hat{P}_c(n,L)(1-\hat{P}_c(n,L))}{L} + \frac{z_b^2}{4L^2}}\right),
\end{align}
we run simulations until the width of the confidence interval, $\abs{w^+ - w^-} < 0.02$, where $z_b = Q^{-1}(b/2)$ and $Q(\cdot)$ is the tail probability function of the standard normal distribution.  For $95\%$ confidence, $b = .95$, $z_b = 1.96$ and for our simulations we set $\delta$ within the set $\set{0.01,0.05,0.1,0.15}$. We extend this Monte-Carlo framework to also track the average relative spectral efficiency,
\begin{align}\label{eq:relativespectralefficiencysim_app}
\xi(t,L) \eqdef \frac{1}{L}\sum_{\ell =1}^L \frac{\log_2\left(1 +\zeta_{h,i'}/\sigma^2\right)}{\log_2\left(1 +\zeta_{H,i^*}/\sigma^2\right)},
\end{align}
and sample complexity.  Recall from \eqref{eq:estimateprobcorrect} that correctness also hinges on the condition that $\abs{i^* -\gamma(n)} \leq 1$, which requires the chosen beamforming vector to be a neighbor of $\mathbf{f}_{H,i^*}$, $\mathbf{f}_{H,i \pm 1}$. 

\section{Additional Simulations}\label{sec:additionalsims}
This appendix further characterizes \ac{DBZ} under different channel conditions.  We show a single ($L = 1$) simulation for $N = 2000$ demonstration in the bottom portion of Fig. \ref{fig:simdemoduo} to illustrate the behavior of \ac{DBZ} over time.  We also plot the corresponding relative spectral efficiency on the top plot of each figure.  The thin gold line indicates the receiving entity's relative angle to the transmitting entity.  The thicker dashed lines represent the limits of the beam currently being used for communication.  The large value of $\sigma_u$ on the left side of Fig. \ref{fig:simdemoduo} causes severe fluctuations at each time step.  In contrast, smaller values of $\sigma_u$ used on the right side of Fig. \ref{fig:simdemoduo}, cause significantly less fluctuations between time steps.  \ac{DBZ} correspondingly maintains tight alignment with a narrow beam for prolonged periods of time with smaller values of $\sigma_u$.

\begin{figure}[h]
\includegraphics[scale = \appendixfigscaling]{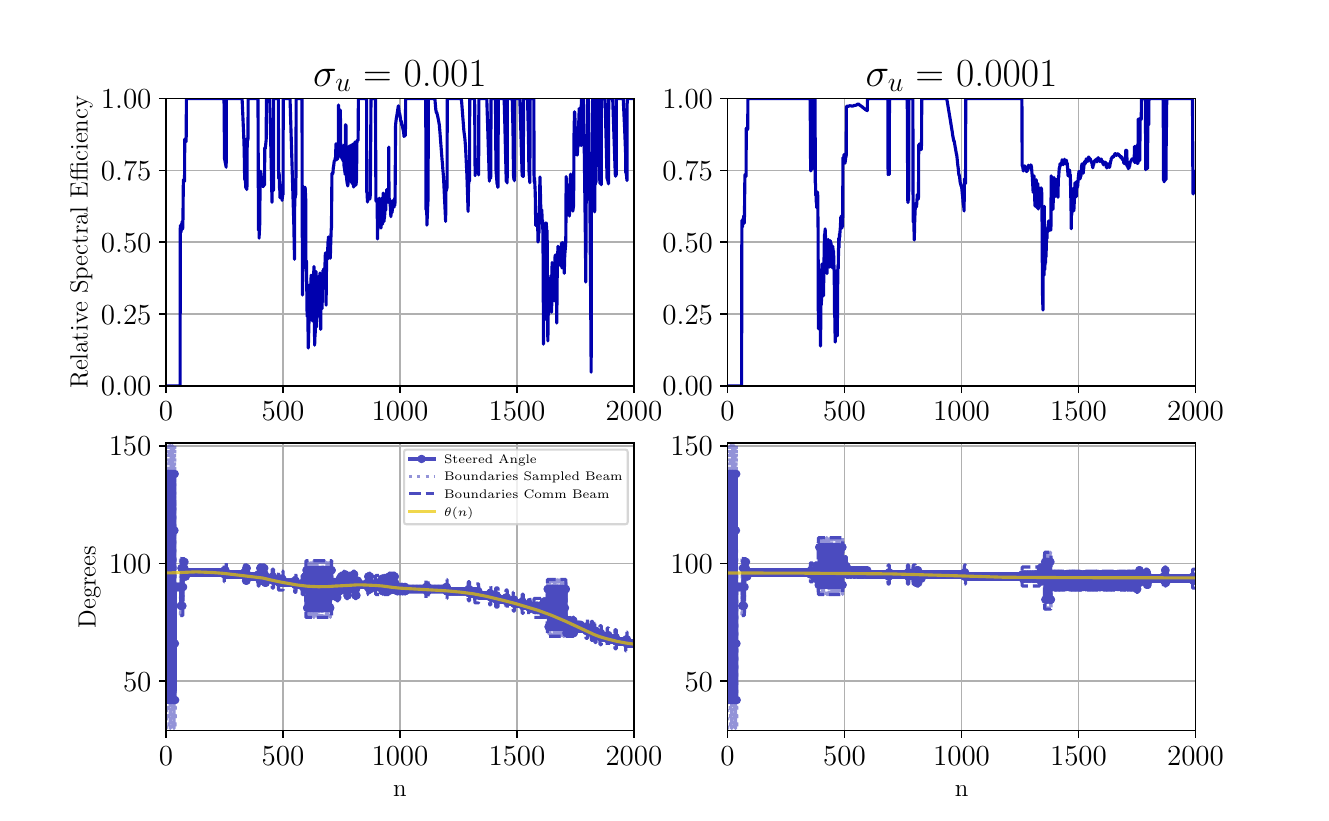}
\centering
\caption{Comparison of example \ac{DBZ} employment with different motion $\sigma_u$, \ac{SNR}$ = 10$, $\tau = 1$.}
\label{fig:simdemoduo}
\end{figure}

We demonstrate \ac{DBZ} with different choices of sample window lengths, in which we choose elements of $\boldsymbol\eta$ based on either $N_h^\textnormal{a}$, number of time steps aligned,  or $\eta_\textnormal{est}$, complexity \eqref{eq:lambertsamples}.  We then show that choosing intermediate values for $[\boldsymbol\eta]_h$ at each level to support a broad range of \ac{SNR} does not greatly impact performance, especially at high \ac{SNR}.  We provide a comparison of the perfomance of \ac{DBZ} with a baseline Bayesian algorithm, \ac{ABT} in \cite{ronquillo2023integrated,ronquillo2021active}. \ac{ABT} utilizes full channel knowledge to compute the exact priors which optimally exploits the reward structure induced by the hierarchical codebook described in Section \ref{sec:hierarchical}, and show performance of \ac{DBZ} is comparable to that of \ac{ABT} with no prior channel fading knowledge.  Finally, we provide simulations of \ac{DBZ} performance in a realistic spatially consistent fading channel model, NYU Sim~\cite{ju2018nyusimmotion,sun2017nyusim}.

\subsection{Sample Window Length Based on Alignment Time Versus Estimated Complexity}\label{ssec:windowcohonumerical}
Our first set of experiments choose elements of $\boldsymbol\eta$ by determining $N_h^\textnormal{a}$ such the probability of alignment \eqref{eq:probcoherence} is greater than $1-\delta/(2H)$ with $\delta = 0.1$, in an offline manner prior to the algorithm execution.  Throughout all our experiments, we choose $\epsilon = 7$, which maintains $95\%$ spectral efficiency upon selection under our zoom-in termination criteria \eqref{eq:stoppingcriteria}.  On average, we expect a degradation in performance over time due to possibility of significantly increased speed.  Hence the almost immediate drop in relative spectral efficiency of $\sigma_u = .001$ in Fig. \ref{fig:performancenco}.   We initialize the \ac{DWNA} model in \eqref{eq:kinematicmodel} with a radial velocity of $\dot{\theta}(n) = 0$.  Successive positive or negative increments results in increased speed, making the beam alignment problem more difficult for the \ac{DWNA} motion model on average as time passes.  The increased speed explains the degradation in performance at later time steps, especially in the bottom right of Fig. \ref{fig:performancenco}.  Results are the average of $L = 2000$ simulations, unless otherwise stated.
\begin{figure}[h]
\includegraphics[scale = \appendixfigscaling]{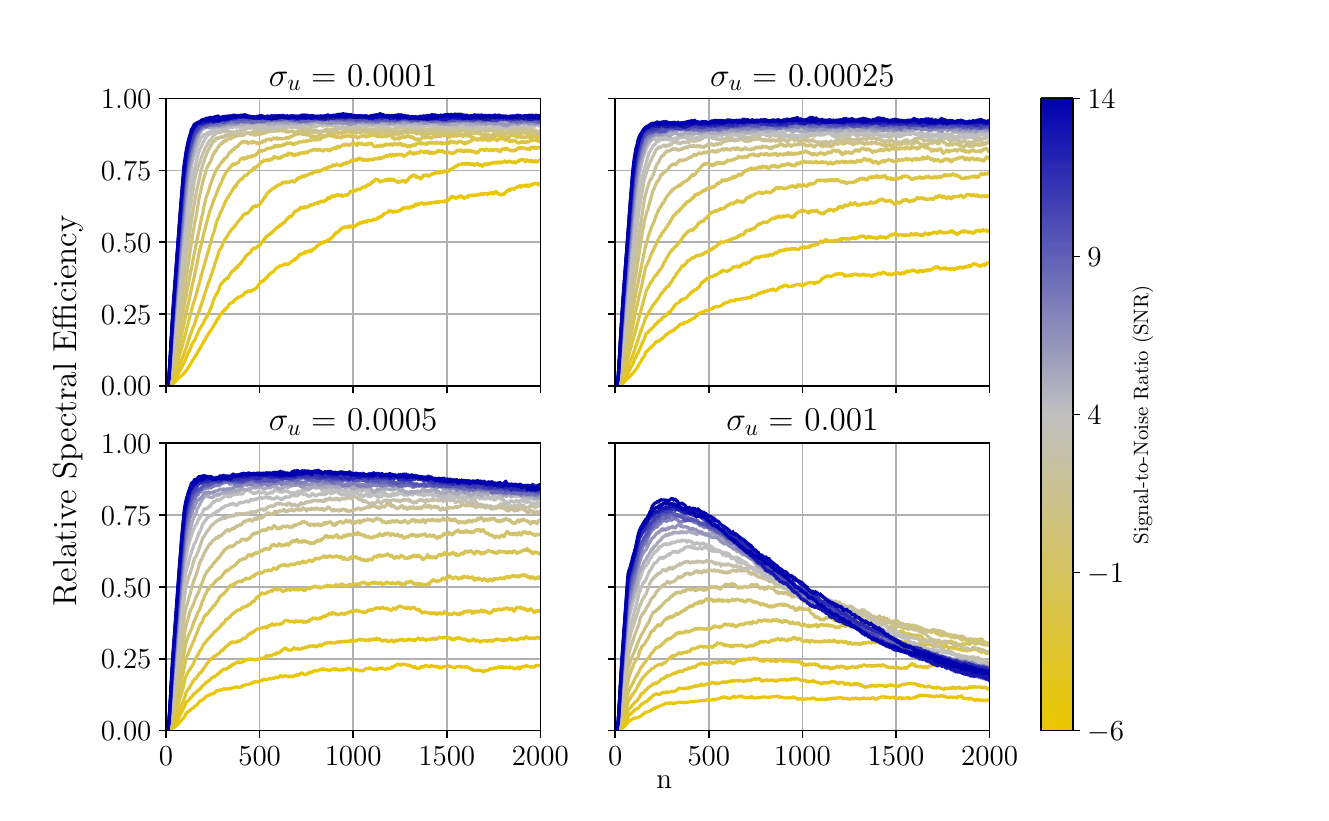}
\centering
\caption{Comparing performance with $\eta$ set by number of time steps in coherence, $N_h^\textnormal{a}$.}
\label{fig:performancenco}
\end{figure}
Our next result shows the performance of \ac{DBZ} when the elements of $\boldsymbol\eta$ are chosen by calculation of $\eta_\textnormal{est}$ \eqref{eq:lambertsamples}.  This requires knowledge of the channel \ac{SNR}, or $\sigma_v$, since we hold the signal power constant and vary $\sigma_v$ exclusively.  We also show that choosing elements of $\boldsymbol\eta$ with respect to an intermediate \ac{SNR} value has little impact on the overall performance.  We see an overall high relative spectral efficiency in simulations using the complexity term, $\eta_{\textnormal{est}}$ \eqref{eq:lambertsamples} , to set the sample window length.  The comparison of results in Fig. \ref{fig:performancenco} versus Fig. \ref{fig:performancewindowcomplexity} suggest that choosing shorter window lengths, i.e.,  $[\boldsymbol\eta]_h < N_h^{\textnormal{a}}$, enables better performance of \ac{DBZ} than longer.  In fact, Fig. \ref{fig:performancewindowintermediate} shows choosing a generic intermediate value enables better performance than choosing sample windows that are too long.
%
\begin{figure}[h]
\includegraphics[scale = \appendixfigscaling]{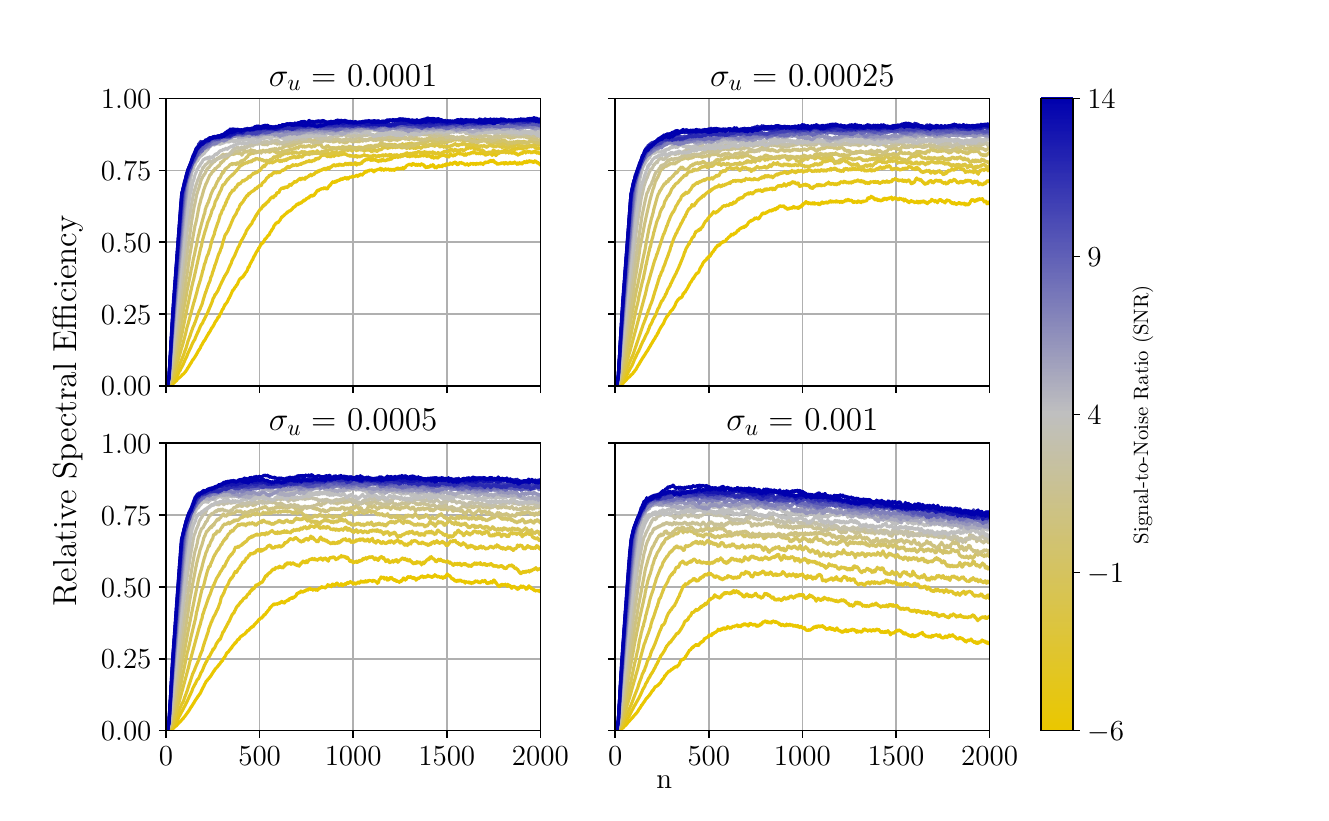}
\centering
\caption{Comparing performance with $\eta$ set by estimated complexity.}
\label{fig:performancewindowcomplexity}
\end{figure}
For one last result in our study of tuning sample window lengths, we use an intermediate value, \ac{SNR}$=4$dB, to set elements of $\boldsymbol\eta$ across all channel values of \ac{SNR}, or proportionally, $\sigma_u$.  Our results in Fig. \ref{fig:performancewindowintermediate} show little impact at high \ac{SNR} and some impact on performance at low \ac{SNR}, indicating that choosing a longer $[\boldsymbol\eta]_h$ is beneficialas a rule of thumb.
\begin{figure}[h]
\includegraphics[scale = \appendixfigscaling]{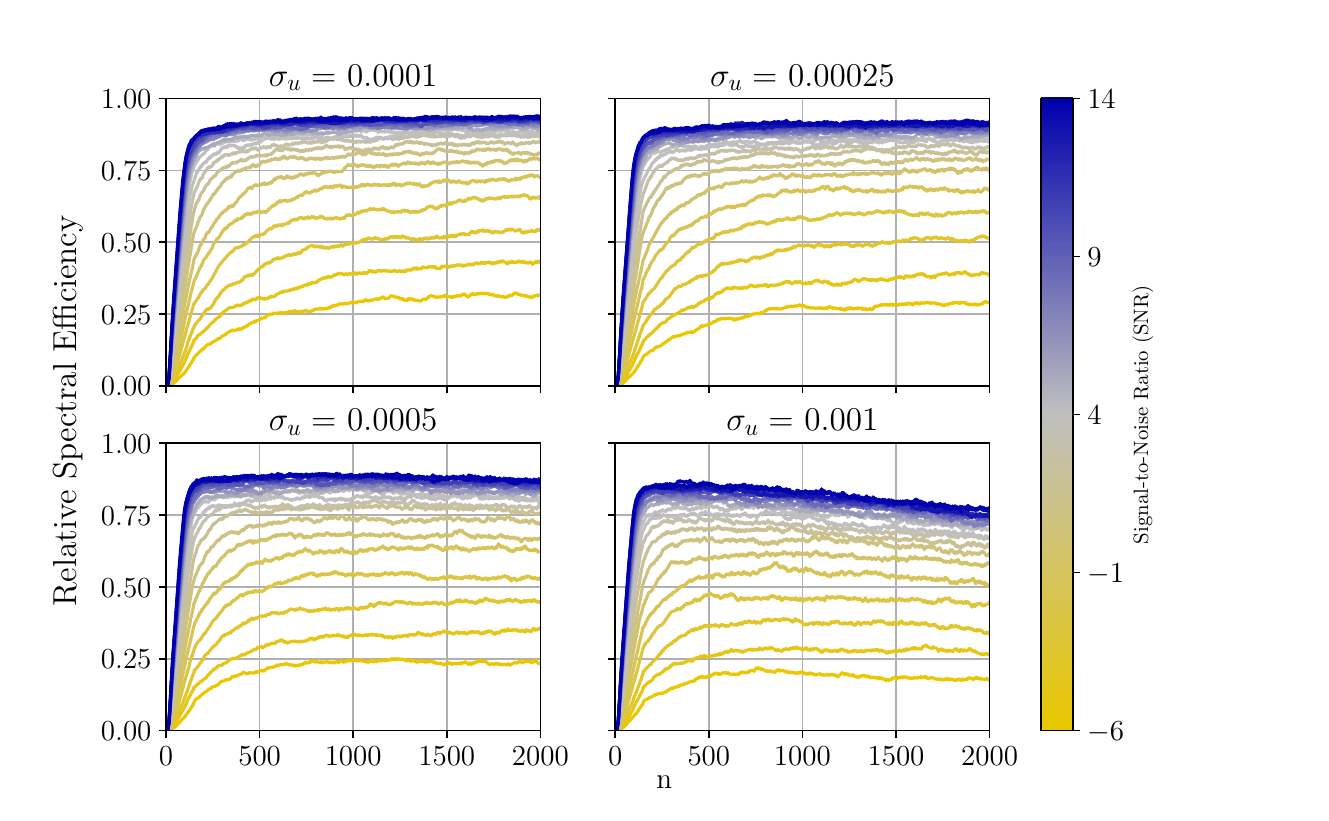}
\centering
\caption{Comparing performance with $\eta$ set by estimated complexity for intermediate value \ac{SNR} $= 4$~dB.}
\label{fig:performancewindowintermediate}
\end{figure}

In addition to the \ac{DWNA} model, we present additional simulations of \ac{DBZ} and \ac{ABT} using two of the motion models used in \cite{ronquillo2021active,ronquillo2023integrated}. These additional simulations provide a more direct comparison to the \ac{ABT} results in \cite{ronquillo2021active,ronquillo2023integrated} with \ac{DBZ}.

\subsubsection{Fixed Velocity}

The fixed velocity motion model updates the angle, $\theta(n)$, at each time step according to 
\begin{align}\label{eq:fixedvmodel}
	\theta(n) = \theta(n-1) + V\tau(n-1),
\end{align}
where $V = 2v\pi/(3M)$, and $v$ is the parameter set in the simulations.  \cite{ronquillo2021active,ronquillo2023integrated} uses the quantity, $v\pi/M$ (since they cover an angular swath of width $\pi$ instead of $2\pi/3$ with $\Phi$), in order to set the velocity to be a multiple of the beamwidth when $M$ antenna elements are in the \ac{ULA}.  The performance of \ac{ABT} hinges on exact knowledge of $v$, such that the posterior adjusts at each time step to accommodate for the motion.  With perfect knowledge of $v$, regardless of the value of $v$, Fig. \ref{fig:fixedvabtperformance} shows \ac{ABT} suffers almost no performance degradation.  Note that we continue to fix $\tau(n) = \tau = 1$ in this Section's simulations as well.
\begin{figure}[H]
\includegraphics[scale = \appendixfigscaling]{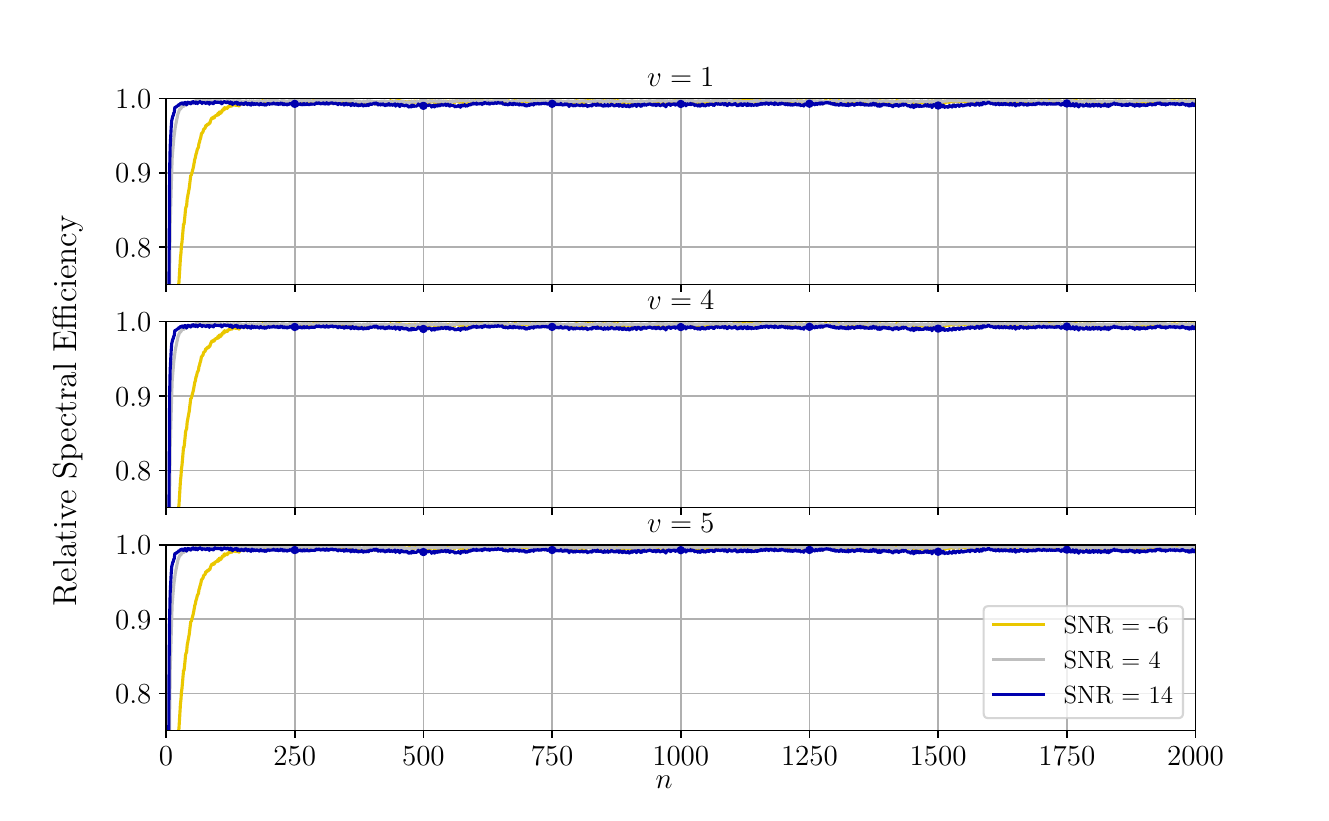}
\centering
\caption{Performance of \ac{ABT} with fixed velocity..}
\label{fig:fixedvabtperformance}
\end{figure}

\ac{DBZ} on the other hand struggles when the $v$ is set such that $\theta(n) - \theta(n-1) > \phi_{\textnormal{bw},h}$.  This makes sense, our analysis in Section \ref{sec:analysis} shows that there are would not enough samples within one alignment period to make a correct decision on beam zooming.  In particular, with integer multiples of $2\pi/3/M$, there is but a single sample at each time step for \ac{DBZ} to use.  We use intermediate values chosen for the sample window lengths in $\boldsymbol\eta$ for \ac{SNR} $= 4$~dB throughout this section.  In general, the fixed velocity model, \eqref{eq:fixedvmodel}, does not adequately describe the anticipated \ac{ME} motion in 5G, 6G, and \ac{IoT} devices \cite{liu2022integrated}.

\begin{figure}[H]
\includegraphics[scale = \appendixfigscaling]{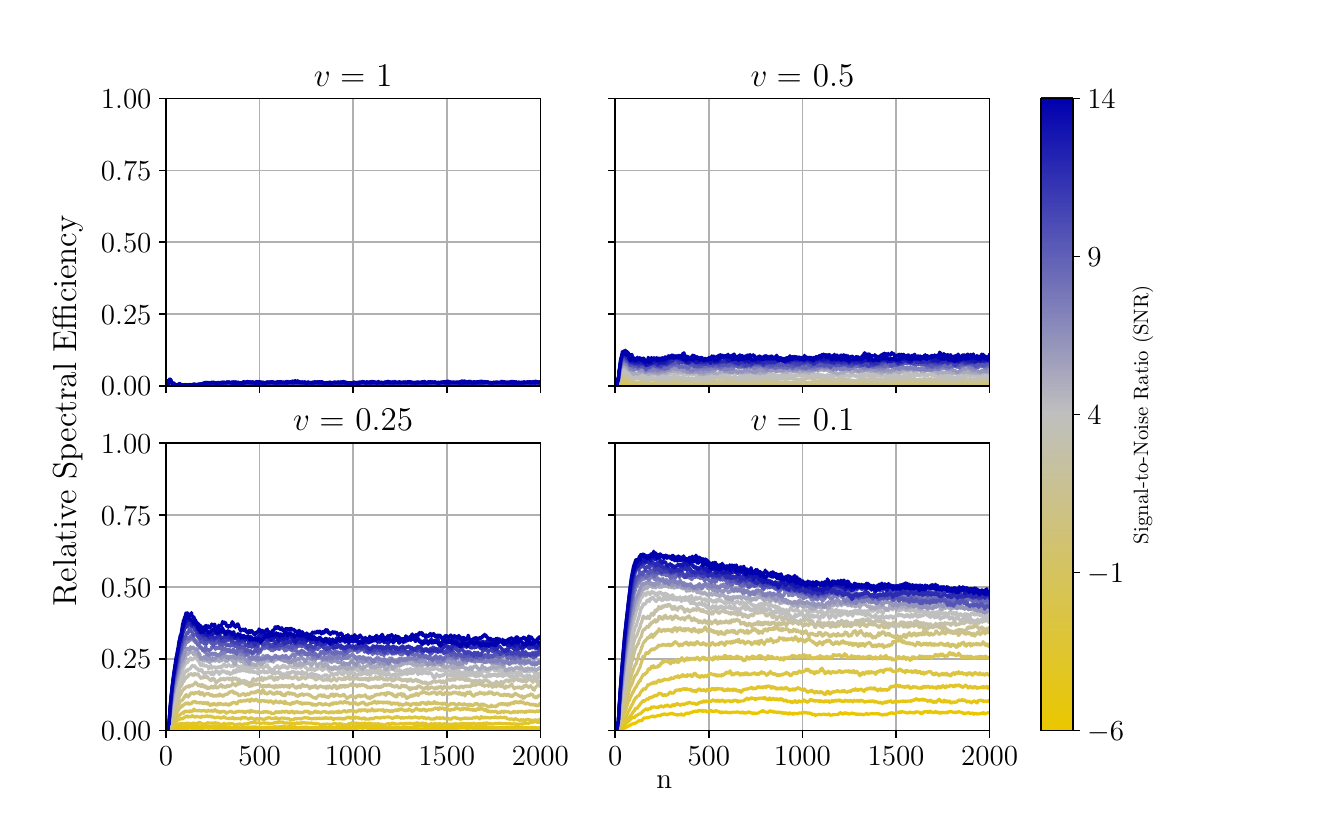}
\centering
\caption{Performance of \ac{DBZ} with fixed velocity.}
\label{fig:fixedvdbzperformance}
\end{figure}

\subsubsection{Discrete Gaussian Jumps}

Popularly used in past works, \cite{va2016beam}\cite{ronquillo2021active}, is the motion model with discrete Gaussian jumps of the angle, $\theta(n)$, at each timestep,
\begin{align}\label{eq:gaussianjumpsmodel}
\theta(n) = \theta(n-1) + \tilde{\theta}(n-1)\tau(n-1),
\end{align}
where $\tilde{\theta}(n) \sim \mathcal{N}(0,\sigma_\theta^2)$.  We use narrower beams than in \cite{ronquillo2021active}, hence we choose smaller values of $\sigma_\theta$ to be proportional to the new beamwidth.  With $M = 128$ elements here, versus $32$ in \cite{ronquillo2021active}, we choose $\sigma_\theta$ to be roughly $25\%$ of $\sigma_\theta = .75^\circ$ chosen in \cite{ronquillo2021active} as the largest value of $\sigma_\theta$.  Fig. \ref{fig:gaussianjumpsperformance} shows competitive performance of \ac{DBZ} at higher \ac{SNR}s, and emphasize that we are fixing the sample window lengths for each \ac{SNR}.  Whereas \ac{ABT} uses exact knowledge of $\sigma_\theta^2$ along with both the fading coefficient, $\alpha_1(n)$, and noise variance, $\sigma_v^2$.
\begin{figure}[H]
\includegraphics[scale = \appendixfigscaling]{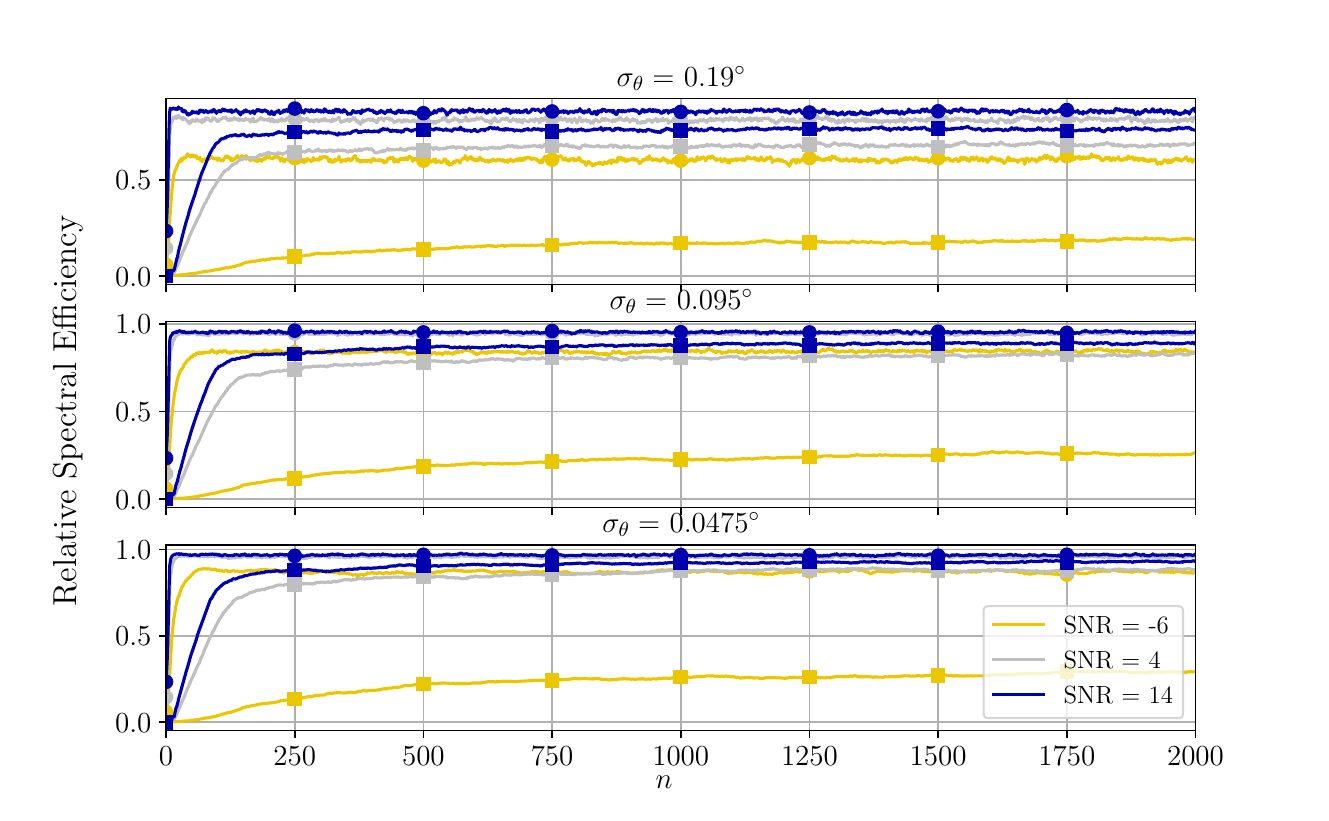}
\centering
\caption{Comparison performance of \ac{DBZ} and \ac{ABT} with discrete Gaussian jumps.}
\label{fig:gaussianjumpsperformance}
\end{figure} 

We elected to use the \ac{DWNA} model in our main simulations in Section \ref{sec:sims} to have both spatially diversity with random variations in \ac{ME} motion over the random path. Essentially, the \ac{DWNA} combines the aspects of the fixed velocity model, \eqref{eq:fixedvmodel}, and the discrete Gaussian jumps model, \eqref{eq:gaussianjumpsmodel}.

\section{Parameters for NYU Sim}\label{sec:nyusimparameters}

We list detailed parameters used in the simulations from Section \ref{sec:sims} for the NYU Sim channels model \cite{sun2017nyusim,ju2018nyusimmotion} in Tables \ref{tab:scenarionyusimparametersT}, \ref{tab:nyusimparametersT}, and \ref{tab:spatialnyusimparametersT} .  We made one modification to the NYU Sim source code to ensure the initial starting position would allow for the \ac{ME} to remain within the beamforming codebook's angular coverage region, $\Phi$.  Depending on the scenario type, i.e., \texttt{RMa}, we vary the initial range of distances at which the \ac{UE} starts.  We set parameters for \texttt{Update Distance} and \texttt{Move Distance}  according to the scenario type as well to reflect practical \ac{BS} employment.  For example, an urban microcell (\texttt{UMi}) will likely service closer proximity \ac{UE}s and require more frequent updates due to a more dynamically changing \ac{LOS} environment \cite{sun2017nyusim}.  The NYU Sim code chooses distance uniformly random between the \texttt{Min Initial Distance} and \texttt{Max Initial Distance}, which we also set to be scenario specific.  We chose a frequency toward the upper limit of NYU Sim supported frequencies ($150$~GHz maximum) and $960$~kHz bandwidth to correspond with a current specified subcarrier spacing for FR-2 in 3GPP 5G Standard Release 17 \cite[Tables 4.3.2-1 and 4.3.2-2]{3GPP_TS_38211_R17}.  We maintain the default value for parameters governing weather, i.e., temperature, humidity, etc.  

\begin{table}[H]
\begin{center}
\caption{Scenario Specific Parameters used in NYU Sim simulations.}
\label{tab:scenarionyusimparametersT}
\centering

\begin{tabularx}{\columnwidth}{ >{\hsize=0.7\hsize\linewidth=\hsize\centering\arraybackslash}X  >{\hsize=0.8\hsize\linewidth=\hsize\centering\arraybackslash}X >{\hsize=0.8\hsize\linewidth=\hsize\centering\arraybackslash}X >{\hsize=0.8\hsize\linewidth=\hsize\centering\arraybackslash}X  }

\textbf{Parameter Name}          & \textbf{RMa}  & \textbf{UMa}      & \textbf{UMi}   \\

\hline\hline

\textbf{Move Distance} & $60$~m & $15$~m & $6$~m\\
\textbf{Max Initial Distance} & $40$~m & $10$~m & $5$~m\\
\textbf{Min Initial Distance} & $40$~m & $20$~m & $10$~m\\
\textbf{Update Distance} & $0.1$~m & $0.025$~m & $0.01$~m\\

\hline
\end{tabularx}
\end{center}
\end{table}

\begin{table}
\begin{center}
\caption{Static Parameters used in NYU Sim simulations.}
\label{tab:nyusimparametersT}
\centering

\begin{tabularx}{\columnwidth}{ >{\hsize=0.7\hsize\linewidth=\hsize\centering\arraybackslash}X  >{\hsize=0.8\hsize\linewidth=\hsize\centering\arraybackslash}X >{\hsize=0.8\hsize\linewidth=\hsize\centering\arraybackslash}X  }

\textbf{Parameter Name}          & \textbf{Value}  & \textbf{Unit}         \\

\hline\hline

Frequency & $142$ & GHz\\
Bandwidth & $960$ & kHz\\
Tx power & $30$ & dBm\\
Environment & LOS & N/A\\
Scenario & Varies & N/A\\
Tx Height & $20$ & m\\
Rx Height & $20$ & m\\
Barometric Pressure & $1013.25$ & mbar\\
Humidity & $50$ & $\%$\\
Temperature & $20$ & deg C\\
Rain Rate & $150$ & mm/hr\\
Polarization & Co-Pol & N/A\\
Foliage Loss & No & N/A\\
Foliage Distance & $0$ & m\\
Foliage Attenuation & $0.4$ & dB\\
Tx Array Type & ULA & N/A\\
Rx Array Type & ULA & N/A\\
Number of Tx Elements & $128$ & Elements\\
Number of Rx Elements & $1$ & Elements\\
Tx Elements per Row & $1$ & Elements\\
Rx Elements per Row & $1$ & Elements\\
Tx Antenna Spacing & $0.5$ & Wavelengths\\
Rx Antenna Spacing & $0.5$ & Wavelengths\\
Tx Azimuth HPBW & $10$ & Degrees\\
Tx Elevation HPBW & $10$ & Degrees\\
Rx Azimuth HPBW & $10$ & Degrees\\
Rx Elevation HPBW & $10$ & Degrees\\

\hline
\end{tabularx}
\end{center}
\end{table}

\begin{table}
\begin{center}
\caption{ Spatial Consistency and Human Blockage Parameters used in NYU Sim simulations.}
\label{tab:spatialnyusimparametersT}
\centering

\begin{tabularx}{\columnwidth}{ >{\hsize=0.7\hsize\linewidth=\hsize\centering\arraybackslash}X  >{\hsize=0.8\hsize\linewidth=\hsize\centering\arraybackslash}X >{\hsize=0.8\hsize\linewidth=\hsize\centering\arraybackslash}X  }

\textbf{Parameter Name}          & \textbf{Value}  & \textbf{Unit}         \\

\hline\hline

Spatial Consistency & On & N/A\\
Correlation Distance Shadow Fading & $60$ & m\\
Correlation Distance LOS Condition & $100$ & m\\
Track Type & Varies & N/A\\
Move Distance & Varies & m\\
Move Direction & $20$ & degrees\\
Update Distance & Varies & m\\
Velocity & $1$ & m/s\\
Human Blockage & On & N/A\\
Human Blockage Default & No & N/A\\
Mean Attenuation & $15.8$ & dB\\
Unshadow to Decay Rate & $0.21$ & Hz\\
Decay to Shadow Rate & $7.88$ & Hz\\
Shadow to Rise Rate & $7.7$ & Hz\\
Rise to Unshadow Rate & $7.67$ & Hz\\

\hline
\end{tabularx}
\end{center}
\end{table}



\end{document}